\documentclass[
    prx,
	twocolumn,
    groupeaddress,
	longbibliography,
	superscriptaddress,
]{revtex4-2}

\usepackage[english]{babel}
\usepackage{siunitx}
\usepackage[utf8]{inputenc}
\usepackage{soul}
\usepackage{rotating}
\usepackage{multirow}
\usepackage{lipsum}
\usepackage{empheq} 
\usepackage{ifthen}
\usepackage{tikz}
\usetikzlibrary{arrows}
\usetikzlibrary {arrows.meta}
\usetikzlibrary{arrows,decorations.pathmorphing}
\usetikzlibrary{calc}
\usepackage{booktabs}

\usepackage{amsmath,amssymb,amsfonts}
\usepackage{color}
\usepackage{graphicx}
\graphicspath{{Images/}}
\usepackage{subcaption}
\usepackage{caption} 
\usepackage{physics}
\usepackage{float}
\usepackage[normalem]{ulem}
\usepackage{tabularx}
\usepackage{bm}
\usepackage{comment}

\renewcommand{\selectlanguage}[1]{} 

\usepackage[
	colorlinks=true,
	linkcolor=blue,
	urlcolor=blue,
	citecolor=blue
]{hyperref}

\emergencystretch=\maxdimen
\newcommand{\dtuelectro}{
    Department of Electrical and Photonics Engineering, Technical University of Denmark,
    2800 Kgs. Lyngby,
    Denmark
}

\newcommand{\dtunanophoton}{
    NanoPhoton - Center for Nanophotonics, Technical University of Denmark, Ørsteds Plads 345A, 2800 Kgs. Lyngby, Denmark}

\begin{document}

\title{
One-dimensional photonic wire as a single-photon source: Implications of cavity QED
to a phonon bath of reduced dimensionality
}

\author{José Ferreira Neto}
\email{jofer@dtu.dk}
\affiliation{\dtuelectro}

\author{Matias Bundgaard-Nielsen}
\affiliation{\dtuelectro}
\affiliation{\dtunanophoton}

\author{Niels Gregersen}
\affiliation{\dtuelectro}

\author{Luca Vannucci}
\affiliation{\dtuelectro}

\date{\today}

\begin{abstract}
While the semiconductor quantum dot placed in a solid-state material allows for deterministic emission of single photons, the photon indistinguishability is strongly influenced by the intrinsic coupling to lattice vibrations, phonons, of the solid-state environment. This work investigates the phonon-induced decoherence for a quantum dot placed in the one-dimensional system of a homogeneous cylindrical nanowire. Such a structure supports multiple longitudinal phonon branches, and we consider both a linear and a quadratic coupling of the emitter to these modes. Under a polaron approach, we initially derive an analytical expression for the 1D pure dephasing rate, which leads to a reduced pure dephasing rate compared with bulk. By implementing these results into a full cavity quantum electrodynamic model, we demonstrate that multimode coupling is necessary to correctly predict the indistinguishability in a 1D system, which may otherwise be significantly underestimated. 
\end{abstract}

\maketitle

\section{Introduction}

The triggered emission of high-quality single photons is a key component of optical quantum technologies \cite{couteau_applications_2023}. Over the past decades, tremendous progress has been made in developing quantum light sources spanning various physical platforms \cite{eisaman_invited_2011}. For applications in quantum information processing, one aims to simultaneously maximize the photon collection efficiency $\varepsilon$ and the photon indistinguishability $\mathcal{I}$. While spontaneous parametric down-conversion offers a viable method for generating indistinguishable single photons, its inherent probabilistic nature severely limits the collection efficiency $\varepsilon$ \cite{tanida_highly_2012}. As an alternative, the semiconductor quantum dot (QD) has emerged as a promising deterministic platform allowing for pure and efficient single-photon generation \cite{somaschi_near-optimal_2016,senellart_high-performance_2017,heindel_quantum_2023}.

Unlike atoms, QDs are inexorably coupled to quantized lattice vibrations, phonons, of the host medium resulting in reduced indistinguishability \cite{iles-smith_phonon_2017}. Even though placing the QD inside an optical cavity can generally increase both $\varepsilon$ and $\mathcal{I}$ through Purcell enhancement and cavity filtering \cite{iles-smith_phonon_2017}, the dimensionality of the hosting material strongly governs the phonon-induced decoherence process \cite{tighineanu_phonon_2018}. This behavior is dictated by the phonon spectral density $J(\omega)$, which generally scales with the phonon frequency to the power of the phonon bath dimensionality, i.e., $J(\omega) \propto \omega^{s}$ \cite{leggett_dynamics_1987,doll_fast_2009,hizhnyakov_zero-phonon_2012,chassagneux_effect_2018,reitz_molecule-photon_2020}.

Seminal work by Besombes \textit{et al.} \cite{besombes_acoustic_2001} first elucidated the subject of non-Lorentzian QD emission and its temperature dependence, attributing these effects to lattice relaxation caused by exciton-acoustic-phonon coupling in bulk media. Subsequently, a microscopic theory of optical transitions in QDs with carrier-phonon interactions was formulated \cite{krummheuer_theory_2002,muljarov_dephasing_2004}. Since then, the theory for electron-acoustic phonon interactions in homogeneous bulk material ($s=3$) has been extensively studied \cite{iles-smith_phonon_2017,tighineanu_phonon_2018,kaer_microscopic_2013,reigue_probing_2017,grange_reducing_2017,gerhardt_intrinsic_2018,morreau_phonon-induced_2019,brash_light_2019,denning_phonon_2020,wang_micropillar_2020,denning_optical_2020,bundgaard-nielsen_non-markovian_2021,gaal_near-unity_2022,brash_nanocavity_2023,pandey_collective_2024}. While the bulk model is widely used for structures like the micropillar single-photon source (SPS) \cite{ding_-demand_2016,wang_micropillar_2020}, its validity becomes questionable for geometries of reduced dimensionality. For instance, we expect the bulk model to break down for structures such as the photonic nanowire \cite{claudon_highly_2010}, the photonic crystal (PhC) waveguide \cite{uppu_scalable_2020}, and the photonic hourglass \cite{gaal_near-unity_2022} due to quantum confinement effects \cite{galland_non-markovian_2008,malic_graphene_2013}, which restrict the motion of electron-hole pairs. On the other hand, phonon signatures for two-dimensional materials ($s=2$) have only recently been investigated \cite{khatri_phonon_2019,denning_quantum_2022,preuss_resonant_2022,svendsen_signatures_2023,mitryakhin_engineering_2024,piccinini_high-purity_2024,vannucci_single-photon_2024}, while a one-dimensional phonon spectral density ($s=1$) has been utilized in \cite{kikas_anomalous_1996,reichman_nonperturbative_1996,liu_dissipation_2020,laferriere_approaching_2023}.

In this work, we address the 1D case ($s=1$) of the photonic nanowire by introducing a fully microscopic theory. We calculate an analytical solution for the quantized vibration modes of an infinite free-standing cylinder. In turn, we introduce mode-dependent exciton-phonon coupling between an embedded on-axis QD and higher-order phonon modes, which we then adopt in a cavity quantum electrodynamic (cQED) model. We determine the two-photon indistinguishability $\mathcal{I}$ for two different configurations: (i) the QD in the bare structure and (ii) a case where the optical cavity effect induced by the nanowire structure is taken into account. The latter is analyzed through a weak phonon master equation (ME) \cite{bundgaard-nielsen_non-markovian_2021}, which is compared with numerically exact tensor network (TN) calculations \cite{pollock_non-markovian_2018,strathearn_efficient_2018,jorgensen_exploiting_2019,gerald_e_fux_tempocollaborationoqupy_2024,cygorek_simulation_2022,cygorek_sublinear_2024,cygorek_ace_2024}. We demonstrate that the Ohmic coupling scenario, where $J(\omega) \propto \omega$, fails to accurately describe the solid-state environment of a 1D photonic wire. Lastly, our results indicate that non-Markovian effects lead to improved photon indistinguishability within this system.

This paper is structured as follows: Sec.~\ref{sec:model} outlines the theoretical model for QD-light emission in a cylindrical nanowire taking into account higher-order mechanical vibration modes. Sec.~\ref{sec:pdrc} covers the calculation of the one-dimensional pure dephasing rate. In Sec.~\ref{sec:indistinguishability}, we present our results for the photon indistinguishability with and without cavity effects. Sec.~\ref{sec:discussion} discusses the key insights of our findings, followed by concluding remarks in Sec.~\ref{sec:conclusions}. The Appendix provides further details on our model, as well as a benchmarking of our tensor network results.

\section{Model} \label{sec:model}
\subsection{Hamiltonian}

\begin{figure}[ht]
    \hspace{1.35em}
    \begin{minipage}{0.120\textwidth}
        \includegraphics[width=0.89\linewidth]{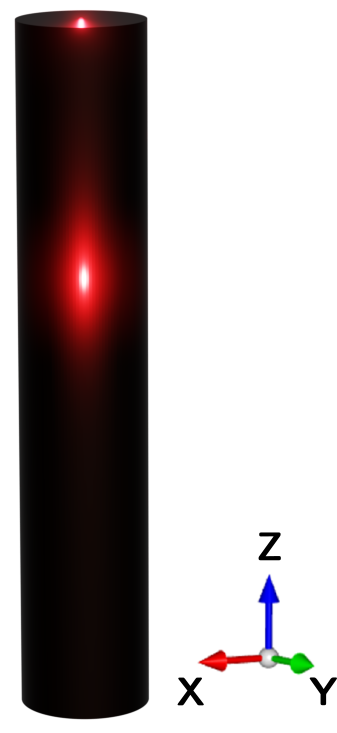}
    \end{minipage}
    \hfill
    \begin{minipage}{0.325\textwidth}
        \centering
        \begin{tikzpicture}
        [
        >=stealth',
        pos=.8,
        photon/.style={decorate,decoration={snake,post length=1mm}}
        ]
        
        \draw [thick,red,arrows = {-Stealth[]}] (4.5,1.5) -- (4.5,0.05);
        \draw [thick,violet,->,photon] (3,1.98) -- (4.2,1.5);
        
        \draw[violet,thin] (3.55,1.15) node[above] {$\hbar\omega_{i,\mathbf{k}}$} node[left=0pt] {} (6.5,1.15);
        \draw[red,thin] (5,0.5) node[above] {$\hbar\omega_{\mathrm{ph}}$} node[left=0pt] {} (6.5,0.5);
        
        \draw[very thick] (3,2) -- node[above] {} node[left=50pt] {$|X\rangle$} (5,2);
        \draw[very thick] (3,0) -- node[above] {} node[left=50pt] {$|G\rangle$} (5,0);
        \draw[gray,ultra thin] (4.2,2.6) -- node[above] {} node[left=30pt] {} (5,2.6);
        \draw[gray,ultra thin] (4.2,2.5) -- node[above] {} node[left=30pt] {} (5,2.5);
        \draw[very thin] (4.2,2.4) -- node[above] {} node[left=30pt] {} (5,2.4);
        \draw[thin] (4.2,2.3) -- node[above] {} node[left=30pt] {} (5,2.3);
        \draw[thin] (4.2,2.2) -- node[above] {} node[left=30pt] {} (5,2.2);
        \draw[thick] (4.2,2.1) -- node[above] {} node[left=30pt] {} (5,2.1);
        \draw[thick] (4.2,1.9) -- node[above] {} node[left=30pt] {} (5,1.9);
        \draw[thin] (4.2,1.8) -- node[above] {} node[left=30pt] {} (5,1.8);
        \draw[thin] (4.2,1.7) -- node[above] {} node[left=30pt] {} (5,1.7);
        \draw[very thin] (4.2,1.6) -- node[above] {} node[left=30pt] {} (5,1.6);
        \draw[gray,ultra thin] (4.2,1.5) -- node[above] {} node[left=30pt] {} (5,1.5);
        \draw[gray,ultra thin] (4.2,1.4) -- node[above] {} node[left=30pt] {} (5,1.4);
        
        \draw[gray,ultra thin] (5.7,2.6) -- node[above] {} node[left=30pt] {} (6.5,2.6);
        \draw[gray,ultra thin] (5.7,2.5) -- node[above] {} node[left=30pt] {} (6.5,2.5);
        \draw[very thin] (5.7,2.4) -- node[above] {} node[left=30pt] {} (6.5,2.4);
        \draw[thin] (5.7,2.3) -- node[above] {} node[left=30pt] {} (6.5,2.3);
        \draw[thin] (5.7,2.2) -- node[above] {} node[left=30pt] {} (6.5,2.2);
        \draw[thick] (5.7,2.1) -- node[above] {} node[left=30pt] {} (6.5,2.1);
        \draw[very thick] (5.7,2) -- node[above] {} node[left=20pt] {...} (6.5,2);
        \draw[thick] (5.7,1.9) -- node[above] {} node[left=30pt] {} (6.5,1.9);
        \draw[thin] (5.7,1.8) -- node[above] {} node[left=30pt] {} (6.5,1.8);
        \draw[thin] (5.7,1.7) -- node[above] {} node[left=30pt] {} (6.5,1.7);
        \draw[very thin] (5.7,1.6) -- node[above] {} node[left=30pt] {} (6.5,1.6);
        \draw[gray,ultra thin] (5.7,1.5) -- node[above] {} node[left=30pt] {} (6.5,1.5);
        \draw[gray,ultra thin] (5.7,1.4) -- node[above] {} node[left=30pt] {} (6.5,1.4);
        \draw[thin] (5.35,2.8) node[above] {$J(\omega)$} node[left=30pt] {} (6.5,2.8);
        
        \end{tikzpicture}
        \end{minipage}
        \caption{Sketch of a cylindrical wire with an embedded QD and its corresponding energy level diagram. Here, $\omega_{i,\mathrm{\mathbf{k}}}$ corresponds to the frequency of the emitted phonon from phonon branch $i$ and wave vector $\mathrm{\mathbf{k}}$, whereas $\omega_{\mathrm{ph}}$ is the frequency of the emitted photon.}
        \label{fig:sketch}
\end{figure}
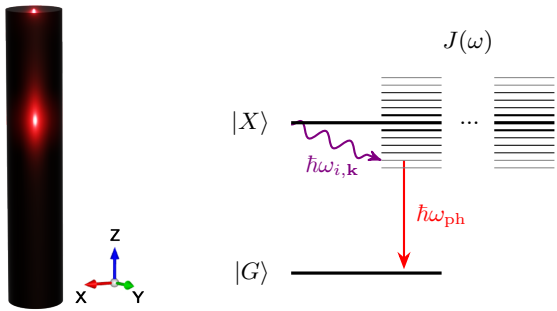

We consider the two-level system of a QD in a 1D nanowire coupled to a continuum of longitudinal acoustic (LA) phonons, sketched in Fig.~\ref{fig:sketch}. The GaAs wire is modeled as a cylindrical structure oriented along the $z$-axis, with a radius denoted by $R$ and length $L$, where $L \gg R$. We assume that the wire maintains its zinc-blende crystal structure unaffected by dimensional confinement effects and that the $z$-axis aligns with the [001]-direction, such that the wire can be treated as isotropic \cite{yu_electronacoustic-phonon_1995}.

In III-V semiconductor materials, two main mechanisms govern the interaction with acoustic phonons: deformation potential and piezoelectric coupling \cite{yu_fundamentals_2010,krummheuer_pure_2005,jahnke_quantum_2012}. The deformation potential mechanism involves changes in the electronic energy states due to variations in crystal volume, primarily coupling electrons to LA phonons. In contrast, piezoelectric coupling arises from electric fields generated in materials lacking inversion symmetry, allowing electrons to interact with both longitudinal and transverse phonons. However, in arsenide-based III-V materials such as GaAs, the deformation potential interaction is the dominant energy-relaxation mechanism, with contributions from piezoelectricity being comparatively minor \cite{krummheuer_theory_2002,ridley_electrons_2009}.

We investigate the impact of phonon-induced transitions on single-photon emission using an independent boson model with the polaron transformation. The total Hamiltonian $\mathcal{H}$ is given by \cite{muljarov_dephasing_2004,bimberg_semiconductor_2008}
\begin{equation}
\label{eqn:hamil}
\mathcal{H}=\mathcal{H}_{0}+\mathcal{H}_{\mathrm{ph}}+\mathcal{H}_{\mathrm{lin}}+\mathcal{H}_{\mathrm{quad}},
\end{equation}
\begin{equation}
\label{eqn:hamil_emitter}
\mathcal{H}_{0}=\hbar\omega_{X}\sigma^{\dag}\sigma,
\end{equation}
\begin{equation}
\label{eqn:h_ph}
\mathcal{H}_{\mathrm{ph}}=\sum_{i,\mathrm{\bf{k}}}\hbar\omega_{i,\mathrm{\bf{k}}}{b^{\dag}_{i,\mathrm{\bf{k}}}}b_{i,\mathrm{\bf{k}}},
\end{equation}
\begin{equation}
\label{eqn:hamil_two}
\mathcal{H}_{\mathrm{lin}}=\sum_{i,\mathrm{\bf{k}}}\hbar g_{i,\mathrm{\bf{k}}}(b_{i,\mathrm{\bf{k}}}^\dag+b_{i,\mathrm{\bf{k}}})\sigma^{\dag}\sigma,
\end{equation}
\begin{equation}
\label{eqn:hamil_three}
\mathcal{H}_{\mathrm{quad}}=\sum_{i,j}\sum_{\mathrm{\bf{k}},\mathrm{\bf{k'}}}\hbar f_{i,j,\mathrm{\bf{k}},\mathrm{\bf{k'}}}(b_{i,\mathrm{\bf{k}}}^\dag+b_{i,\mathrm{\bf{k}}})(b_{j,\mathrm{\bf{k'}}}^\dag+b_{j,\mathrm{\bf{k'}}})\sigma^{\dag}\sigma.
\end{equation}

Here, $\mathcal{H}_{0}$ is the Hamiltonian for a two-level emitter with ground state $|G\rangle$ and excited state (single exciton) $|X\rangle$, $\omega_{X}$ is the bare exciton frequency, $\sigma = |G\rangle\langle X|$ is
the lowering operator, $\mathcal{H}_{\mathrm{ph}}$ represents a multimode acoustic phonon bath with $b_{i,\mathrm{\bf{k}}}^\dag$ ($b_{i,\mathbf{k}}$) the creation (annihilation) operator for the phonon mode $i$
with momentum $\hbar \mathbf{k}$, whereas $\mathcal{H}_{\mathrm{lin}}$ and $\mathcal{H}_{\mathrm{quad}}$ correspond to the linear and quadratic phonon coupling, respectively \cite{muljarov_dephasing_2004,bimberg_semiconductor_2008}. Notably, $\mathcal{H}_{\mathrm{quad}}$ accounts for two-phonon processes involving interactions between all pairs of phonon modes $i, \vb k$ and $j, \vb k'$. The linear and quadratic constants, $g_{i,\mathrm{\bf{k}}}$ and $f_{i,j,\mathrm{\bf{k}},\mathrm{\bf{k'}}}$, are calculated from the matrix elements of the electron-phonon coupling, assuming an isotropic harmonic potential and identical confinement for both electrons and holes in the QD, as discussed in Appendices~\ref{section:lin} and~\ref{section:quad}. 

Under the polaron transformation extended to multiple phonon branches, the linear coupling can be eliminated via $\mathcal{U}_{\rm P} \mathcal{H} \mathcal{U}^\dag_{\rm P}$, with $\mathcal{U}_{\rm P} = \exp [\sigma^\dag \sigma \sum_{i,\vb k} \omega_{i,\vb k}^{-1} g_{i,\vb k} (b^\dag_{i,\vb k} - b_{i,\vb k})]$ \cite{chassagneux_effect_2018}. The transformed Hamiltonian reads $\widetilde{\mathcal{H}} = \hbar \widetilde \omega_X \sigma^\dag \sigma + \mathcal{H}_{\rm ph} + \mathcal{H}_{\rm quad}$, where the polaron-shifted exciton frequency $\widetilde \omega_X = \omega_X - \mathcal{D}$, with $\mathcal{D} = \int_0^{\infty} \dd{\omega} \omega^{-1} J(\omega)$.

Following, for the case of multiple phonon branches, the coupling to phonons is  completely characterized by the phonon spectral density $J(\omega)=\sum\nolimits_{i,\mathbf{k}}\left|g_{i,\mathbf{k}}\right|^2 \delta(\omega-\omega_{i,\mathbf{k}})$, where the index $i$ represents the total number of phonon branches and the summation over $\mathbf{k}$ yields the contribution from different phonon wave vectors \cite{chassagneux_effect_2018}. To assess the coherence evolution of our two-level system as it interacts with the solid-state environment, also referred to as the optical polarization between electrons and holes, we define the physical observable $\langle\sigma_{x}\rangle=\mathrm{Re[e^{\Phi(\tau)}]}$ \cite{nazir_modelling_2016}. The phonon propagator $\Phi(\tau)$ is described by 
\begin{equation}
\label{eqn:7newer}
\Phi(\tau)=\int_{0}^{\infty} \mathrm{d}\omega\ \frac{J(\omega)}{{\omega}^{2}}\left[\frac{\cos(\omega\tau)-1}{\tanh(\beta\hbar\omega/2)}-i\sin(\omega\tau)\right],
\end{equation}

\begin{figure*}[t] 
\centering
\begin{subfigure}{0.4825\textwidth}
  \includegraphics[width=\linewidth]{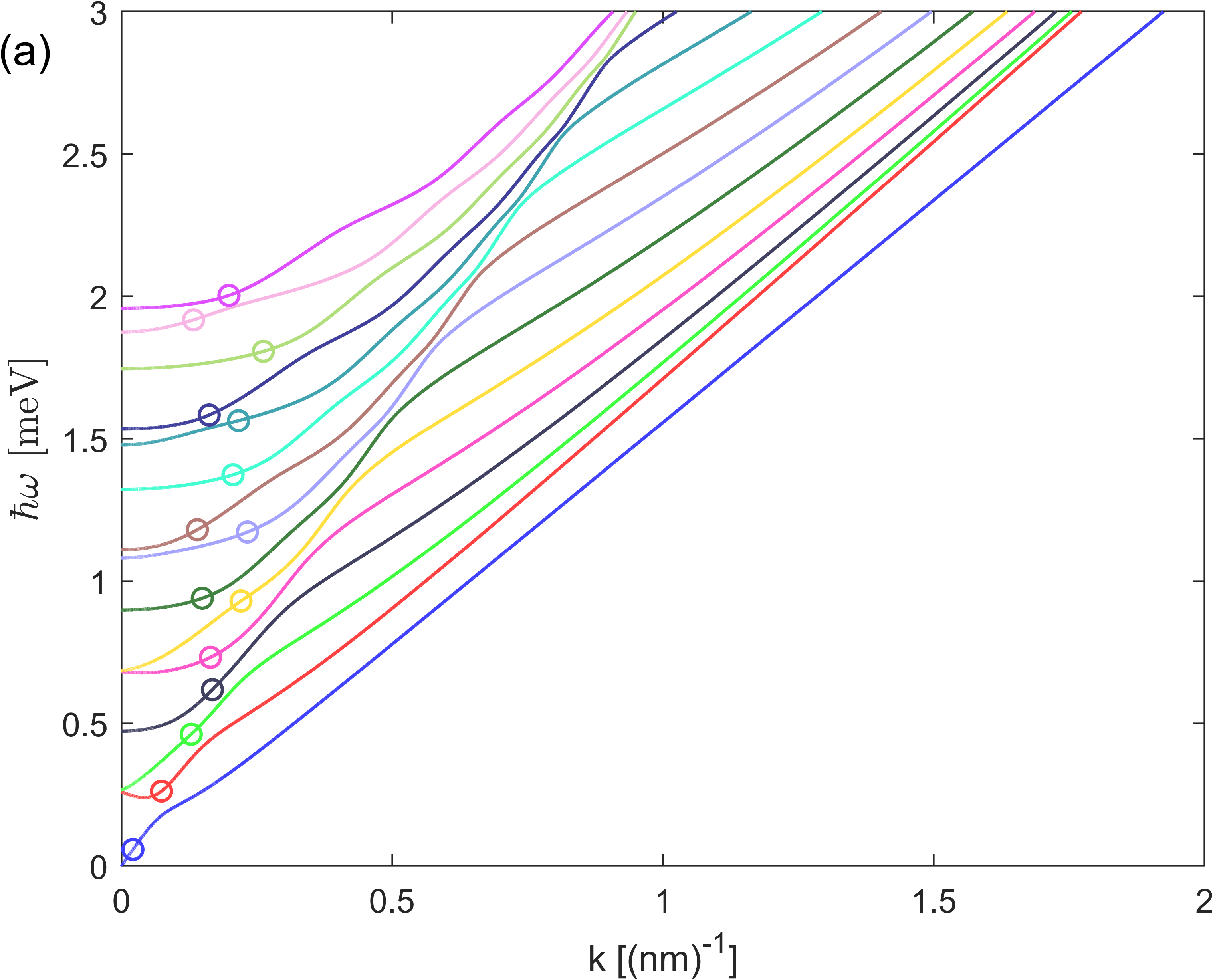}
\end{subfigure}%
\hspace{0.025\textwidth} 
\begin{subfigure}{0.4825\textwidth}
  \includegraphics[width=\linewidth]{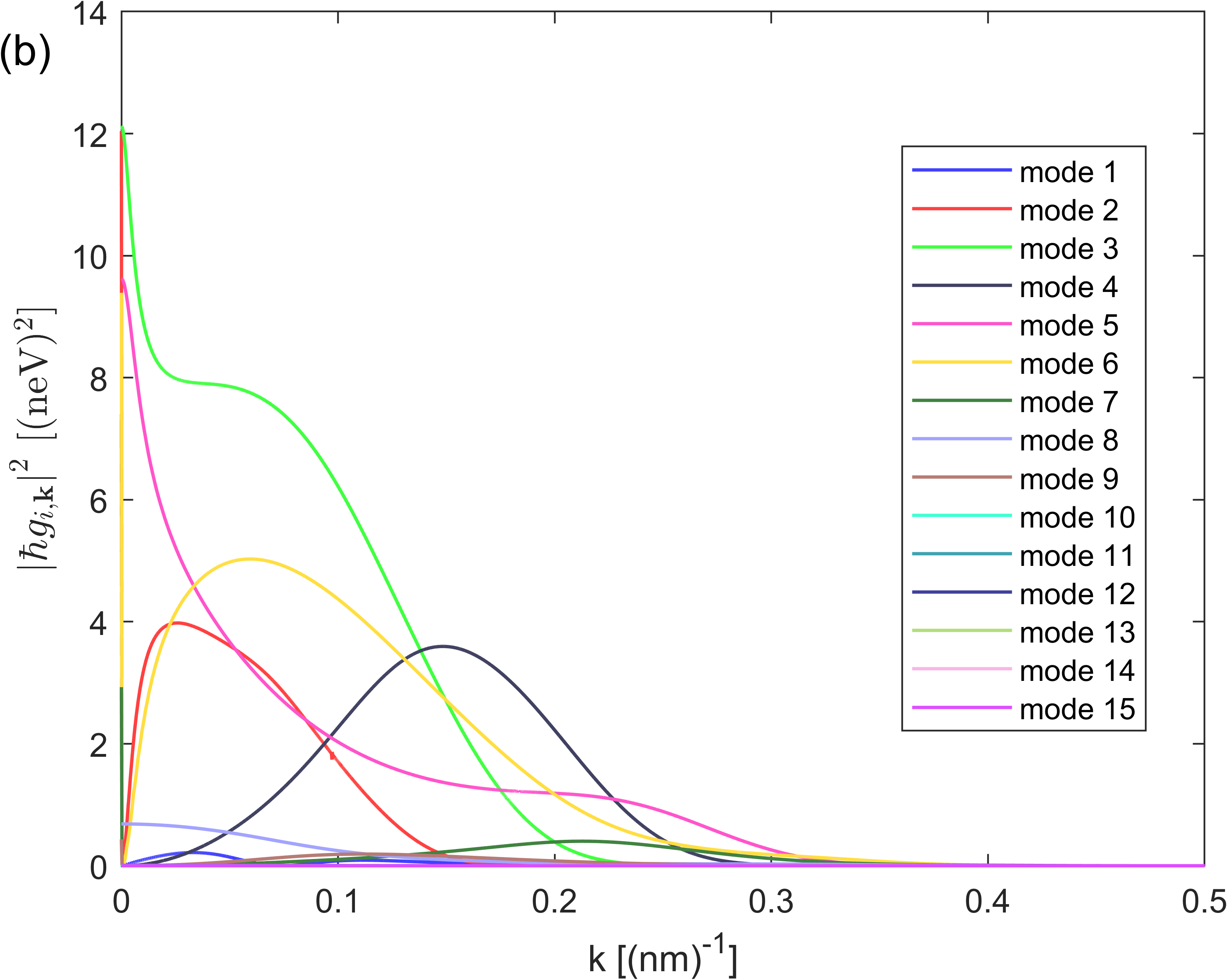}
\end{subfigure}%
\captionsetup{justification=Justified} 
\caption{(a) The first 15 dispersion curves for propagating harmonic waves in a cylindrical rod with $R=25\ \mathrm{nm}$, obtained as solutions of the linear Pochhammer-Chree equation. For later convenience, we indicate with a circle marker the wave vector at which the mode velocity of each mode was extracted for the calculation of the pure dephasing rate presented in Section \ref{sec:pdrc}. (b) The corresponding exciton coupling matrix elements for the first 15 longitudinal modes.}
\label{fig:testandoO}
\end{figure*}

where $\beta=(k_{\mathrm{B}}T)^{-1}$ is the inverse temperature and $k_{\mathrm{B}}$ is the Boltzmann constant. We then straightforwardly obtain that 

\begin{equation}
\label{eqn:7new}
\Phi(\tau)=\sum_{i,\mathbf{k}} \frac{\left|g_{i,\mathbf{k}}\right|^2}{{\omega_{i ,\mathbf{k}}}^{2}}\left[\frac{\cos(\omega_{i,\mathbf{k}}\tau)-1}{\tanh(\beta\hbar\omega_{i,\mathbf{k}}/2)}-i\sin(\omega_{i,\mathbf{k}}\tau)\right].
\end{equation}
 
\subsection{Quantized vibration modes}

For a free-standing isotropic, cubic medium of cylindrical geometry, we obtain the dispersion relation curves of the longitudinal modes predicted by the Pochhammer-Chree equation \cite{stroscio_quantized_1994,sv_pochhammer-chree_2018,royer_elastic_2022}. In Fig.~\ref{fig:testandoO}(a), we show multiple phonon branches obtained for a geometry with a cylindrical cross-section. With this analytical result, we can now evaluate the coupling between the QD and each vibration mode \cite{lindwall_zero-phonon_2007}. We only include longitudinal phonon modes, since a QD situated at the nanowire center is unaffected by flexural vibrations, whereas torsional modes can be disregarded as they do not modulate the bandgap of the two-level system via deformation potential, at least to a first-order approximation \cite{stepanov_large_2016,munsch_resonant_2017,artioli_design_2019}. The results for the coupling constants $g_{i, \vb k}$ are displayed in Fig.~\ref{fig:testandoO}(b). 

\begin{figure}[b] 
  \includegraphics[width=\linewidth]{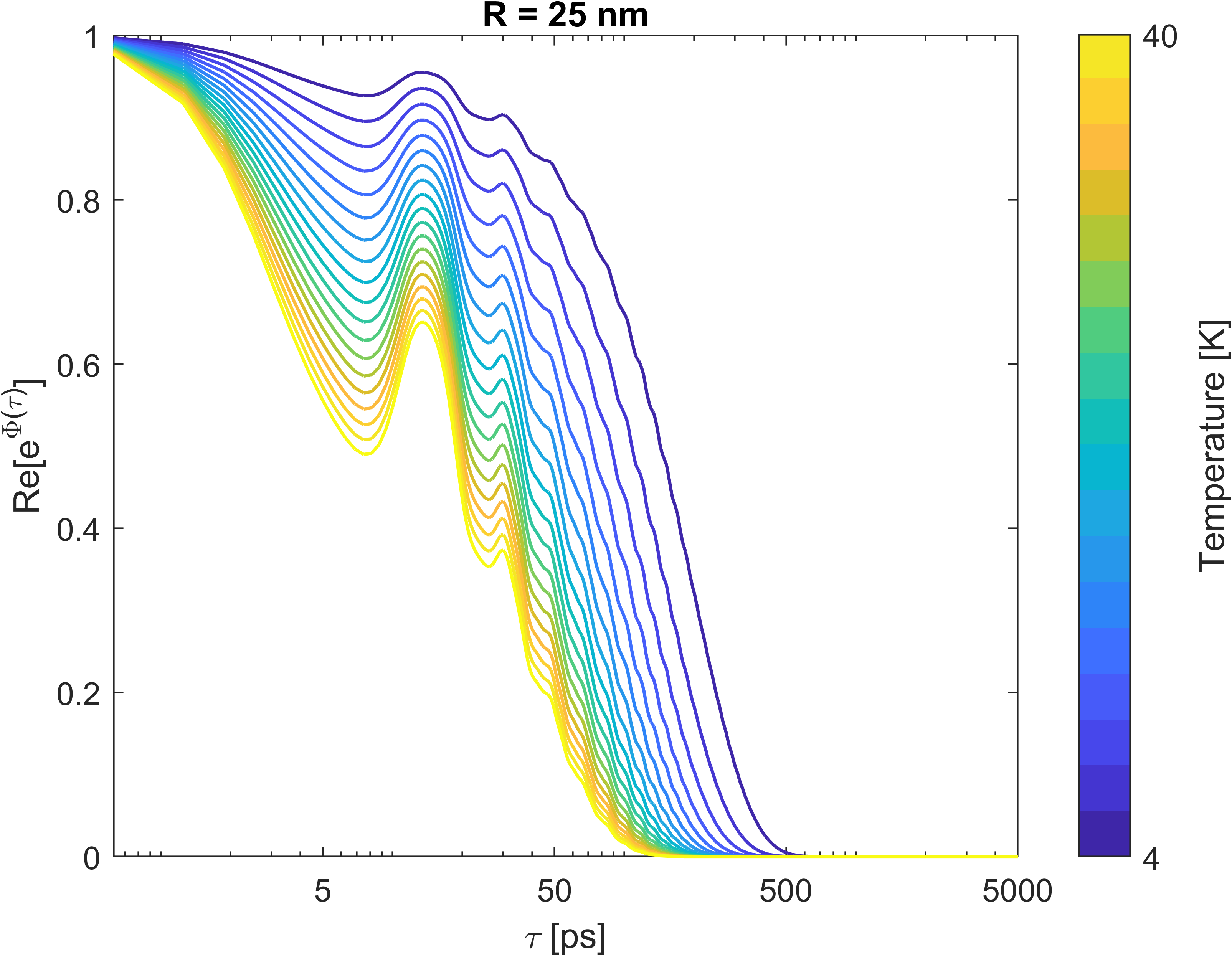}
\caption{Time evolution of the QD coherence for a nanowire radius $R=25\ \mathrm{nm}$ evaluated at different temperatures $T$ $\in$ [\SI{4}{\kelvin}, \SI{40}{\kelvin}] (swept in steps of \SI{2}{\kelvin}).}
\label{fig:coherences}
\end{figure}

We present the coherence dynamics computed from Eq.~\eqref{eqn:7new} for a nanowire radius $R=25\ \mathrm{nm}$ with respect to temperature $T$ in Fig.~\ref{fig:coherences}. The curves exhibit strong signatures of non-Markovianity, as revealed by their accentuated nonmonotonic modulated exponential decay at short timescales, and a vanishing value in the long time limit, i.e., $\lim_{\tau \rightarrow \infty} \mathrm{Re[e^{\Phi(\tau)}]} = 0$. This differs markedly from bulk behavior, which is characterized by a fast initial decay (within picoseconds) followed by a stationary coherence $\ev{\sigma_x} > 0$ \cite{krummheuer_pure_2005,nazir_modelling_2016}.
As a result, the Franck-Condon factor $B^{2} = \exp \qty[-\int_0^{\infty} \mathrm{d}{\omega}\ \omega^{-2} J(\omega) \coth(\beta \hbar \omega / 2)]$ becomes ill-defined for phonon bath dimensionalities $s\leq 2$ \cite{chassagneux_effect_2018}, since the integrand function  $\omega^{-2} J(\omega) \coth(\beta \hbar \omega / 2)$ diverges in such cases. To fully account for the behavior of our system, the phonon propagator evolution $\Phi(\tau)$ should be captured until its coherence vanishes entirely (see complete discussion in Appendix~\ref{app:tn}). While a $R=25\ \mathrm{nm}$ radius is too small to support a guided optical HE$_{11}$ mode, we present the corresponding coherence evolution for the $R=100\ \mathrm{nm}$ radius, similar to the ideal photonic nanowire \cite{claudon_highly_2010}, in the Supplemental Material (SM).

\section{1D pure dephasing rate calculation} \label{sec:pdrc}

Following \cite{reigue_probing_2017}, we generalize the calculation of the pure dephasing rate induced by quadratic phonon coupling to the case of multiple phonon branches. We have
\begin{equation}
\label{eqn:15_novo}
\begin{aligned}
\gamma^{1\mathrm{D}}_{\mathrm{pd}} =\ &\mathrm{Re}\Bigg[\sum_{i,j}\sum_{\mathbf{k,k'}}\int_{0}^{\infty} \mathrm{d}\tau \ |f_{i,j,\mathbf{k,k'}}|^{2} \\
&\hspace{-2.2em}\qquad\times \langle\mathcal{B}_{i,\mathbf{k}}(\tau)\mathcal{B}_{j,\mathbf{k'}}(\tau)\mathcal{B}_{i,\mathbf{k}}(0)\mathcal{B}_{j,\mathbf{k'}}(0)\rangle\Bigg], 
\end{aligned}
\end{equation}
where $\mathcal{B}_{i,\mathbf{k}}(\tau)=b_{i,\mathbf{k}}^\dagger e^{ic_{i}k\tau} + b_{i,\mathbf{k}} e^{-ic_{i}k\tau}$, with $k = |\vb k|$, and the sum is extended to all combinations between phonon branches $i$ and $j$.
Here, $c_{i}$ corresponds to the velocity associated with mode $i$. Whereas only the first phonon mode is taken into account in the bulk model, we now assign mode-dependent velocities to each of the phonon branches. These mode-dependent phase velocities are calculated from the slope of the dispersion relation curves in Fig.~\ref{fig:testandoO}(a) evaluated at the wave vector that maximizes the squared matrix coupling element $|f_{i,j,\mathbf{k,k'}}|^{2}$. This approach will become more evident as we take a closer look into $|f_{i,j,\mathbf{k,k'}}|^{2}$.

Employing Wick's theorem to calculate the multi-operator expectation value in Eq.~\eqref{eqn:15_novo}, we assume that the dominant contribution originates from instances with different phonon wave vectors $\mathbf{k}$ and $\mathbf{k'}$, which allows to factorize the expectation value \cite{reigue_probing_2017}. Applying the commutation relations $\comm{b_{i,\mathbf{k}}}{b_{j,\mathbf{k'}}^\dagger}=\delta_{ij}\delta_{\mathbf{k}\mathbf{k'}}$, and using $\int_0^{\infty} \dd{\tau} e^{\pm i (c_i k - c_j k') \tau} = \pi c_j^{-1} \delta\left(k' - \frac{c_i}{c_j} k\right)$, we are left with the final expression
\begin{equation}
\label{eqn:16_new}
\begin{aligned}
\gamma^{1\mathrm{D}}_{\mathrm{pd}} =\ & \pi\left(\frac{L}{2\pi}\right)^{2}\sum_{i,j}{c_{i}}^{-1}\int \mathrm{d}k\ |f_{i,j,\mathbf{k},c_{ij}\mathbf{k}}|^{2} \\ 
&\hspace{-0.35em}\times \{n_{i}(k)[1+n_{j}(c_{ij}k)] +n_{j}(c_{ij}k)[1+n_{i}(k)]\}, 
\end{aligned}
\end{equation}
with $c_{ij}={c_i}/{c_j}$ the ratio between the mode-dependent velocities and $n_{i}(k)=\{\mathrm{exp}[\beta\hbar\omega_{i}(k)]-1\}^{-1}$ the bosonic occupation number for the phonon branch $i$ with a wave vector module $k$, at temperature $T$.

\begin{figure}[t]
\begin{subfigure}[b]{0.24\textwidth}
  \includegraphics[width=\linewidth]{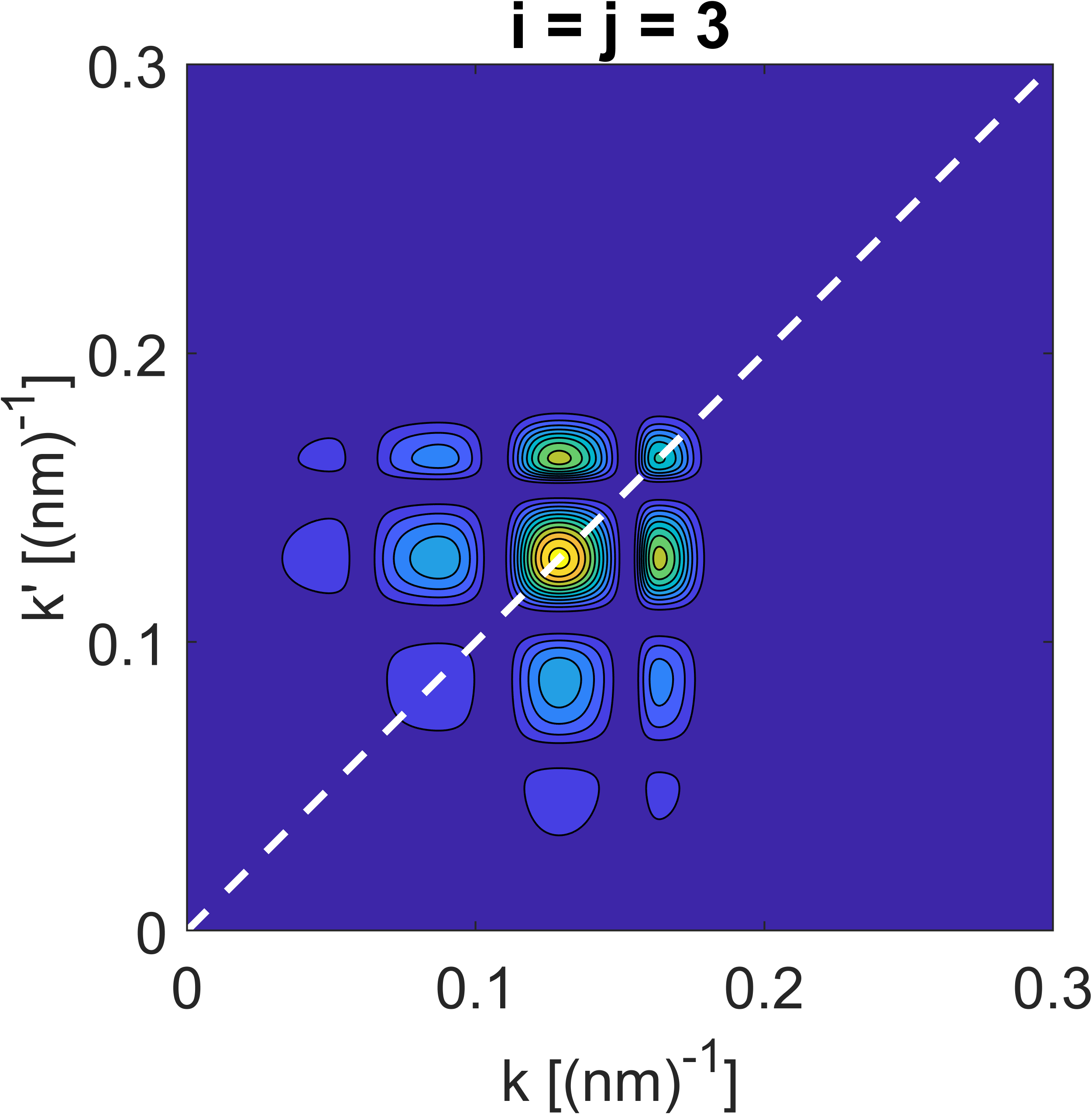}
\end{subfigure}%
\hfill
\begin{subfigure}[b]{0.24\textwidth}
  \includegraphics[width=\linewidth]{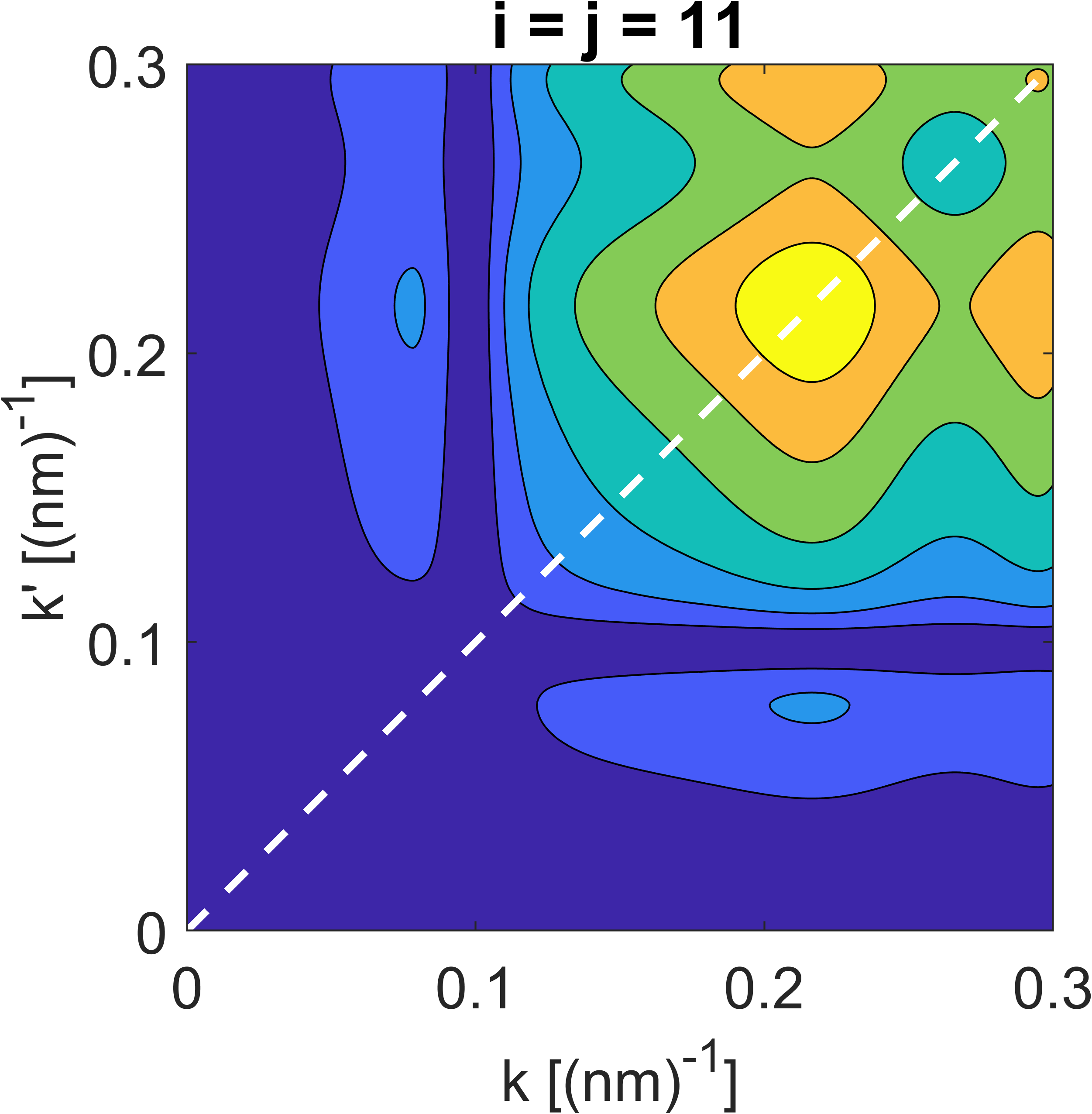}
\end{subfigure}

\vspace{10pt} 

\begin{subfigure}[b]{0.24\textwidth}
  \includegraphics[width=\linewidth]{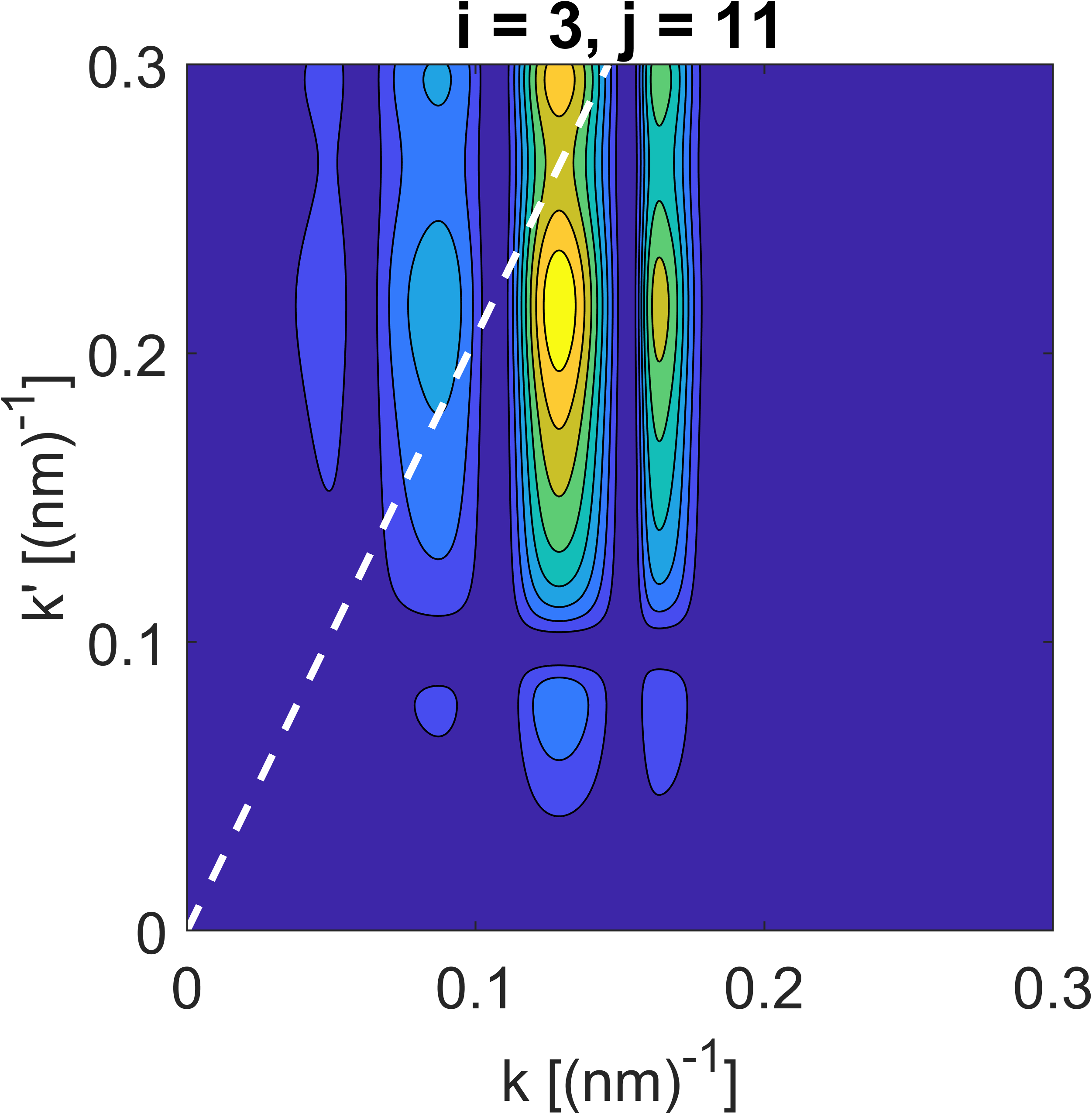}
\end{subfigure}%
\hfill
\begin{subfigure}[b]{0.24\textwidth}
  \includegraphics[width=\linewidth]{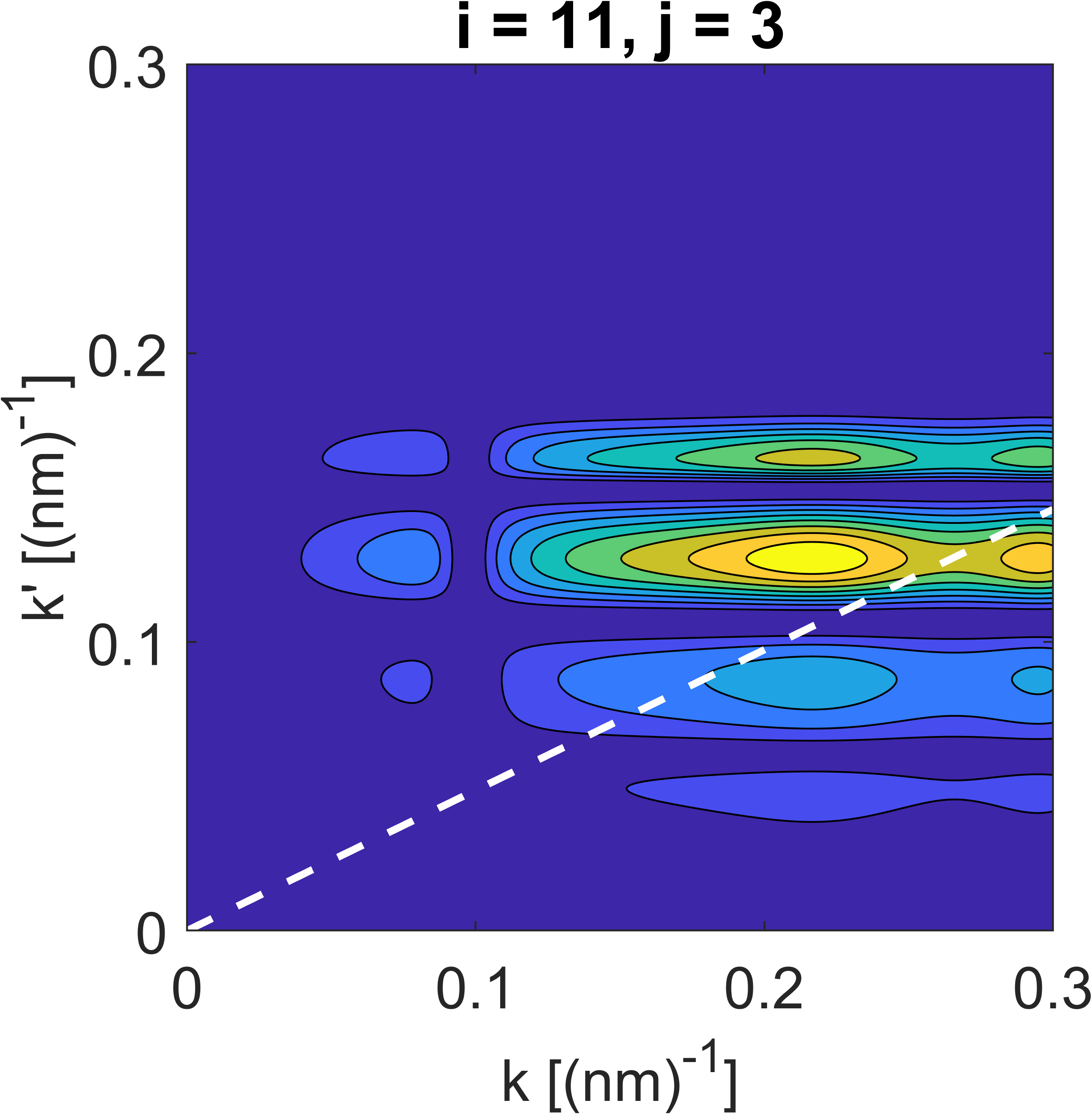}
\end{subfigure}
\caption{Contour plots $|f_{i,j,\mathbf{k,k'}}|^{2}$ for $R=25\ \mathrm{nm}$ for different modes ($i$ and $j$) superimposed with the dashed white lines defined by $k'=\frac{c_{i}}{c_{j}}k$. All plots are normalized with dark blue (yellow) representing 0 (1) arbitrary units.}
\label{fig:withlines}
\end{figure}

We use the following procedure to define the mode-dependent velocities $c_i$, which is based on the inspection of the matrix coupling elements $|f_{i,j,\mathbf{k,k'}}|^{2}$ shown in Fig.~\ref{fig:withlines} and Fig.~\textcolor{blue}{A} in the SM. For $i=j$, we observe that $|f_{i,j,\mathbf{k,k'}}|^{2}$ vanishes almost everywhere except for a small region around $k = k' = k_i$, cf. the exemplary cases $i=3$ and $i=11$ in Fig.~\ref{fig:withlines}.
This suggests to define the mode-dependent velocity $c_i$ as the slope of the phonon branch $i$ at $k_i$, i.e., $c_i = \eval{\dv{\omega_i(k)}{k}}_{k_i}$. The corresponding points are indicated in Fig.~\ref{fig:testandoO}(a) with circle markers, and the values of $c_i$ and $k_i$ are reported in the SM for the first 40 modes.
Similarly, for $i \neq j$, we observe that $|f_{i,j,\mathbf{k,k'}}|^{2}$ is nonzero only in the region around $(k, k') = (k_i, k_j)$, with the same values for $k_i$ and $k_j$ as defined above.
We observe that our choice is consistent with the constraint of $k' = \frac{c_{i}}{c_j} k$ imposed by the delta function $\delta\left(k' - \frac{c_i}{c_j} k\right)$, which is verified in Fig.~\ref{fig:withlines} by tracing a white dashed line.
This fact is also consistent with the interpretation of the phonon-induced pure dephasing in terms of virtual transitions into higher confined states \cite{muljarov_dephasing_2004, reigue_probing_2017, tighineanu_phonon_2018}, restricted here to the first excited electron and hole level, implying that the phonon modes involved must have approximately the same energy $\hbar c_i k \approx \hbar c_j k'$.
Note also that $|f_{i,j,\mathbf{k,k'}}|^{2}$ is symmetric under the simultaneous exchange $i \leftrightarrow j$ and $k \leftrightarrow k'$.

\begin{figure}[hbtp]
\centering
  \includegraphics[width=\linewidth]{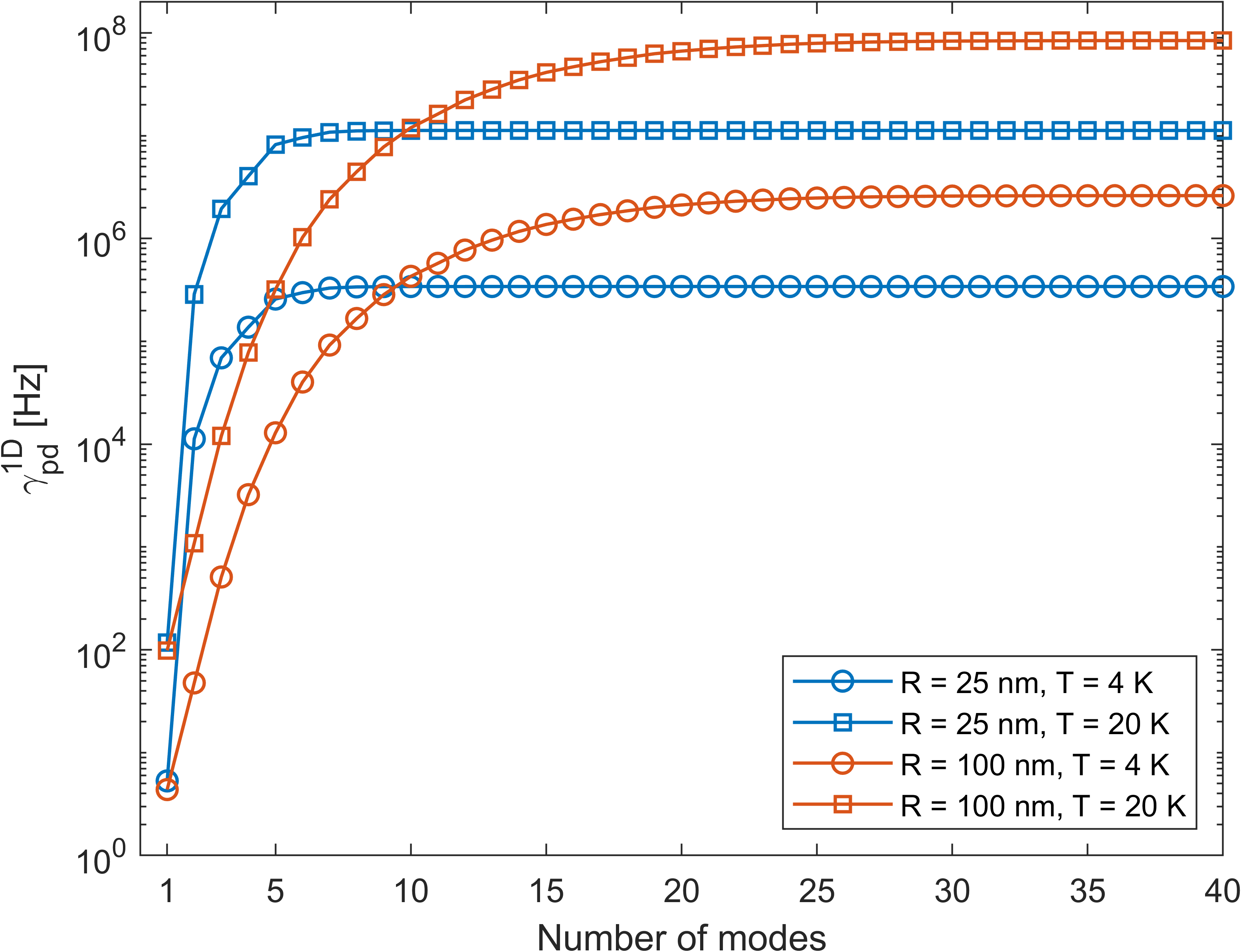}
\caption{Convergence analysis of the 1D pure dephasing rate $\gamma^{1\mathrm{D}}_{\mathrm{pd}}$ as a function of the number of included phonon modes.}
\label{fig:testandooutro}
\end{figure}

The convergence of the pure dephasing rate with the number of phonon modes is presented in Fig.~\ref{fig:testandooutro}, and we observe that $\sim$15 modes are required to achieve convergence for $R=25\ \mathrm{nm}$. An increased nanowire radius requires a higher number of phonon modes for convergence, and for the case of $R=100\ \mathrm{nm}$, we estimate that $\sim$35 modes are needed. In our study, a total of 40 vibration modes were used for both $R=25\ \mathrm{nm}$ and $R=100\ \mathrm{nm}$. 

\begin{figure}[b]
\centering
\includegraphics[width=\linewidth]{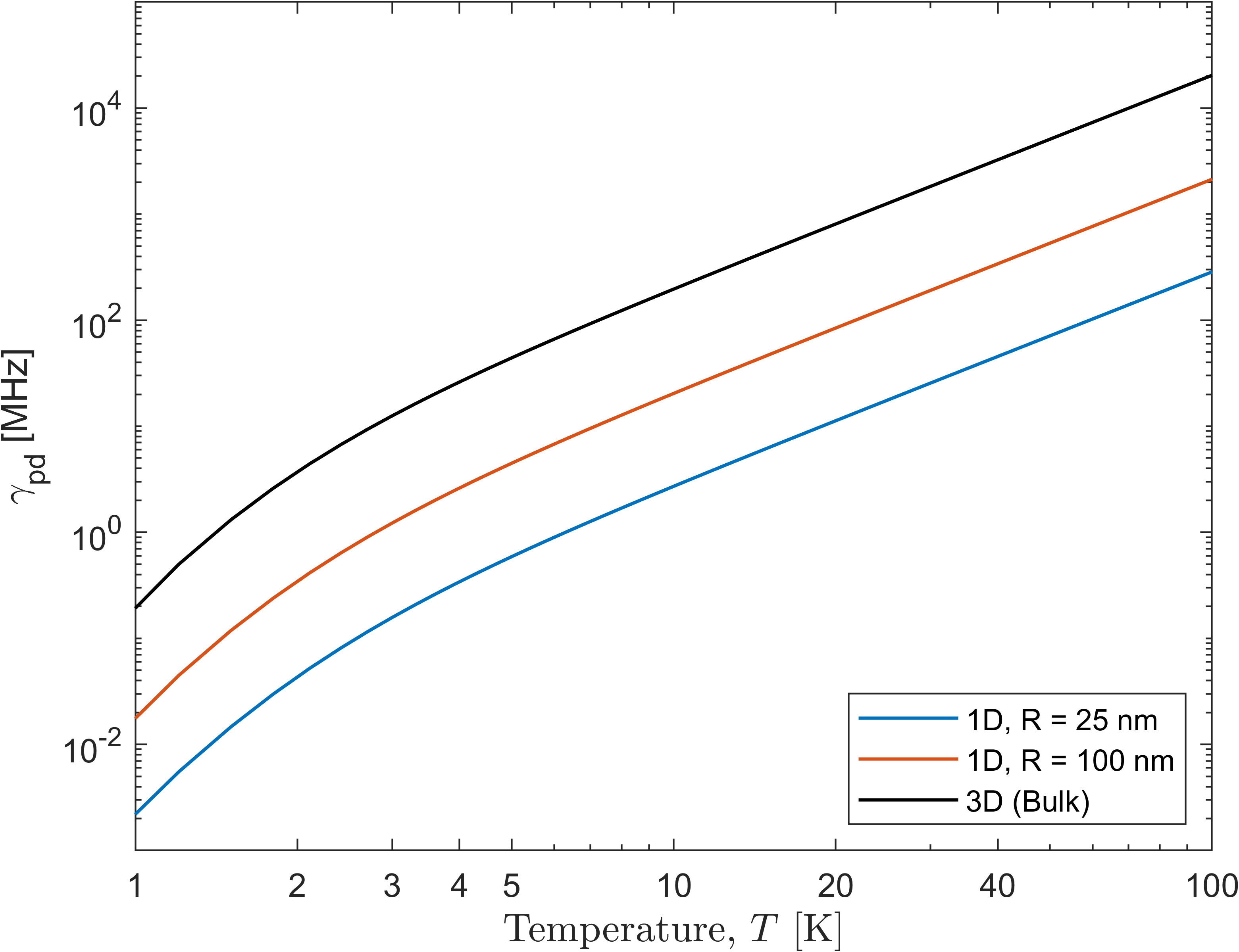}
\caption{Comparison between the pure dephasing rate $\gamma_{\mathrm{pd}}$ calculated for a 1D phonon spectral density, with $R=25\ \mathrm{nm}$ and $R=100\ \mathrm{nm}$, and the bulk prediction.}
\label{fig:loglog}
\end{figure}

Subsequently, we compare our 1D rates with predictions from a 3D bulk model given by \cite{reigue_probing_2017,tighineanu_phonon_2018} 
\begin{equation}
\gamma^{\mathrm{3D}}_{\mathrm{pd}} = \frac{\alpha^{2}\mu}{{\omega_{c}}^{4}}\int\nolimits_{0}^{\infty}\mathrm{d}\omega\ \omega^{10}e^{-2\omega^{2}/\omega_{c}^{2}}\ n(\omega)[n(\omega) + 1], 
\label{eqn:pd}
\end{equation}

where the cutoff frequency $\omega_{c}=\sqrt{2}c_{s}/d$, $d$ is the confinement length of the QD ($d=a_{e}=a_{h}$), and $n(\omega)=[\mathrm{exp}(\beta\hbar\omega)-1]^{-1}$ is the Bose-distributed phonon occupation number. In this bulk model, $\alpha = (4\pi^{2}\hbar\rho_{m} c_{s}^{5})^{-1}(D_{e}-D_{h})^{2}$, $\mu = \pi\hbar^{2}(D_{e}-D_{h})^{-4}(\Delta_{e}^{-1}D_{e}^{2}+\Delta_{h}^{-1}D_{h}^{2})^{2}$, $D_{e}$ ($D_{h}$) is the electron (hole) deformation potential, $\rho_{m}$ is the mass density, and $\Delta_{\lambda}=\hbar^{2}/(a_{\lambda}^{2}m_{\lambda})$, while the subscript $\lambda$ $\in$ \{$e$, $h$\}. Thus, $a_{\lambda}$ and $m_{\lambda}$ denote the confinement length and the effective mass, respectively, for electrons ($e$) and holes ($h$). 

We present the pure dephasing rates in Fig.~\ref{fig:loglog} for both $R=25\ \mathrm{nm}$ and $R=100\ \mathrm{nm}$ as a function of temperature, comparing them to predictions from the bulk model. We observe that the 1D pure dephasing rate tends to approach bulklike behavior as the nanowire radius $R$ increases, in agreement with expectations.
Specific values for $\gamma^{1\mathrm{D}}_{\mathrm{pd}}$ for $T=$ \SI{4}{\kelvin} and $T=$ \SI{20}{\kelvin} for both $R=25\ \mathrm{nm}$ and $R=100\ \mathrm{nm}$ are given in Table~\ref{table:table2}, along with the bulk model predictions. 

\begin{table}[H]
\centering
\begin{tabular}{cccc} 
\hline\hline 
\addlinespace[0.25em]
$T$  & $\gamma_{\mathrm{pd}}^{\mathrm{1D}}$  for $R=25\ \mathrm{nm}$ & $\gamma_{\mathrm{pd}}^{\mathrm{1D}}$  for $R=100\ \mathrm{nm}$ & $\gamma_{\mathrm{pd}}^{\mathrm{3D}}$ (Bulk)  \\ 
\addlinespace[0.125em] \hline
\SI{4}{\kelvin}         & 0.34 MHz        & 2.62 MHz         & 26.15 MHz \\ 
\hline 
\SI{20}{\kelvin}        & 11.26 MHz     &  84.29 MHz       &  806.86 MHz \\  
\hline\hline
\end{tabular}
\caption{1D pure dephasing rates at two different temperatures and predictions from the bulk model.}
\label{table:table2}
\end{table}  

\begin{figure*}[t] 
  \includegraphics[width=\linewidth]{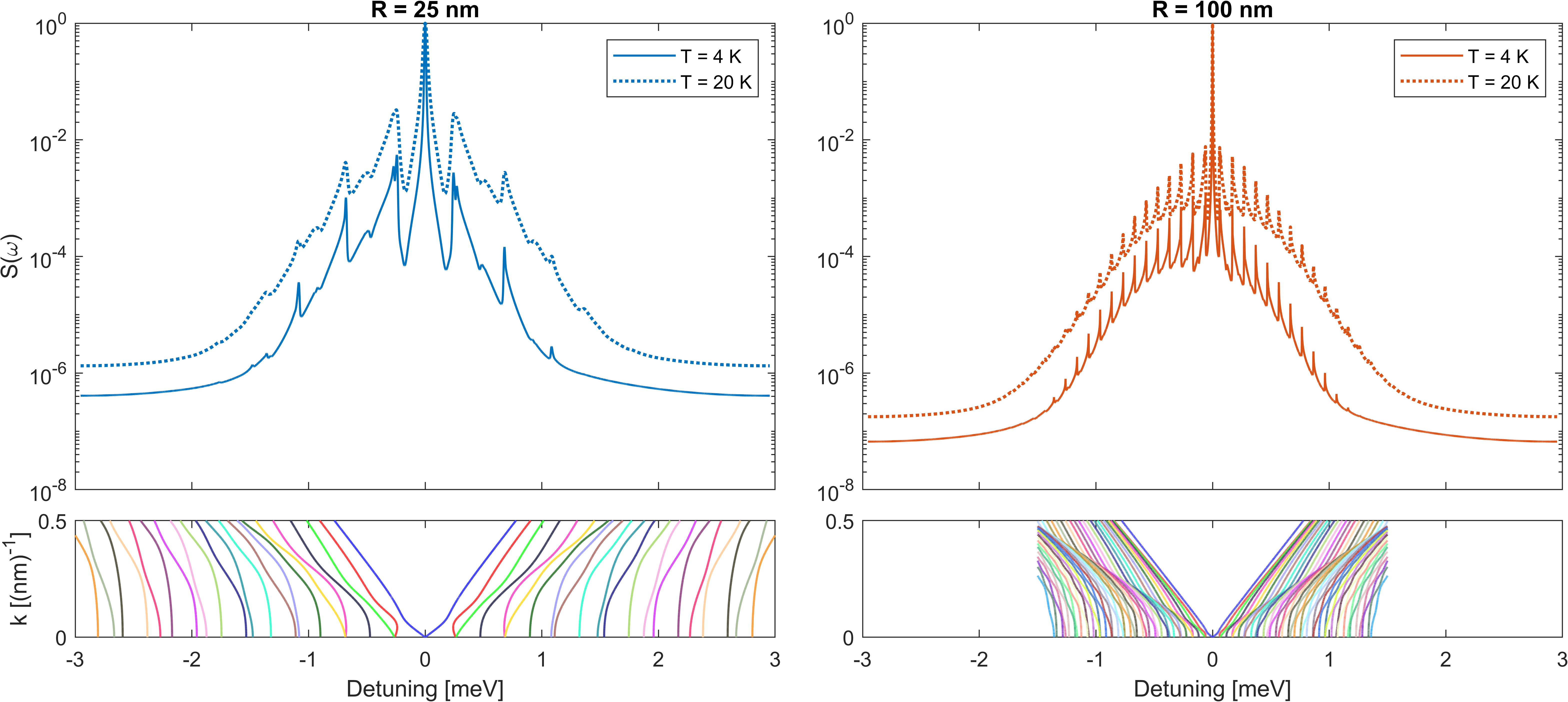}
\caption{Upper panels: Normalized emission spectra for two different temperatures, with $R=25\ \mathrm{nm}$ and $R=100\ \mathrm{nm}$. Lower panels: Dispersion curves for both radii at relevant energies, highlighting the singularities that give rise to the phonon satellite peaks observed in the upper panels. The dispersion is symmetrically replicated on the negative energy side for clarity.}
\label{fig:speccc}
\end{figure*}

\section{Indistinguishability} \label{sec:indistinguishability}
\subsection*{Bare wire calculation}

In the absence of a cavity, the emission spectrum of a QD embedded into a bare wire (see Appendix~\ref{section:bare} for the full derivation and detailed discussion) is given by
\begin{equation}
\label{eqn:20_main}
S(\omega)=\frac{2}{\Gamma_{\mathrm{rad}}}\ \mathrm{Re}\left[\int_{0}^{\infty}\mathrm{d}\tau\ e^{\Phi(\tau)} e^{-\frac{1}{2}(\Gamma_{\mathrm{rad}}+\gamma^{1\mathrm{D}}_{\mathrm{pd}})\tau} e^{-i(\omega-\omega_{0})\tau}\right],
\end{equation}
where $\Gamma_{\mathrm{rad}}$ = 1 GHz is the spontaneous decay rate. The derived emission spectra, shown in the upper panels of Fig.~\ref{fig:speccc}, display remarkable non-Lorentzian features, evidenced by the zero-phonon line (ZPL) accompanied by distinct phonon satellite peaks. The positions of these peaks correspond to the energies where new phonon branches appear \cite{lindwall_zero-phonon_2007}, with these extrema aligning with singularities in the phonon density of states derived from the dispersion curves, as shown in the lower panels of Fig.~\ref{fig:speccc}. Nevertheless, not all phonon modes contribute equally, as their coupling elements differ. For instance, modes that couple weakly to the exciton wave function for $k \rightarrow 0$ result in weaker peaks, while radial modes produce stronger peaks due to their finite exciton coupling matrix element.
For both radii, the emission spectrum at lower temperatures ($T=$ \SI{4}{\kelvin}) shows more pronounced asymmetry when compared to higher temperatures ($T=$ \SI{20}{\kelvin}), and this asymmetry arises from the preferential nature of phonon emission over phonon absorption at low bath temperatures.
Moreover, as the radius increases, the energy separation between the phonon satellite peaks gets increasingly smaller. Consequently, their collective contribution merges into continuous phonon sidebands (PSBs) as the associated phonon spectral density becomes very much bulklike.

\begin{figure}[b]
\centering
\includegraphics[width=\linewidth]{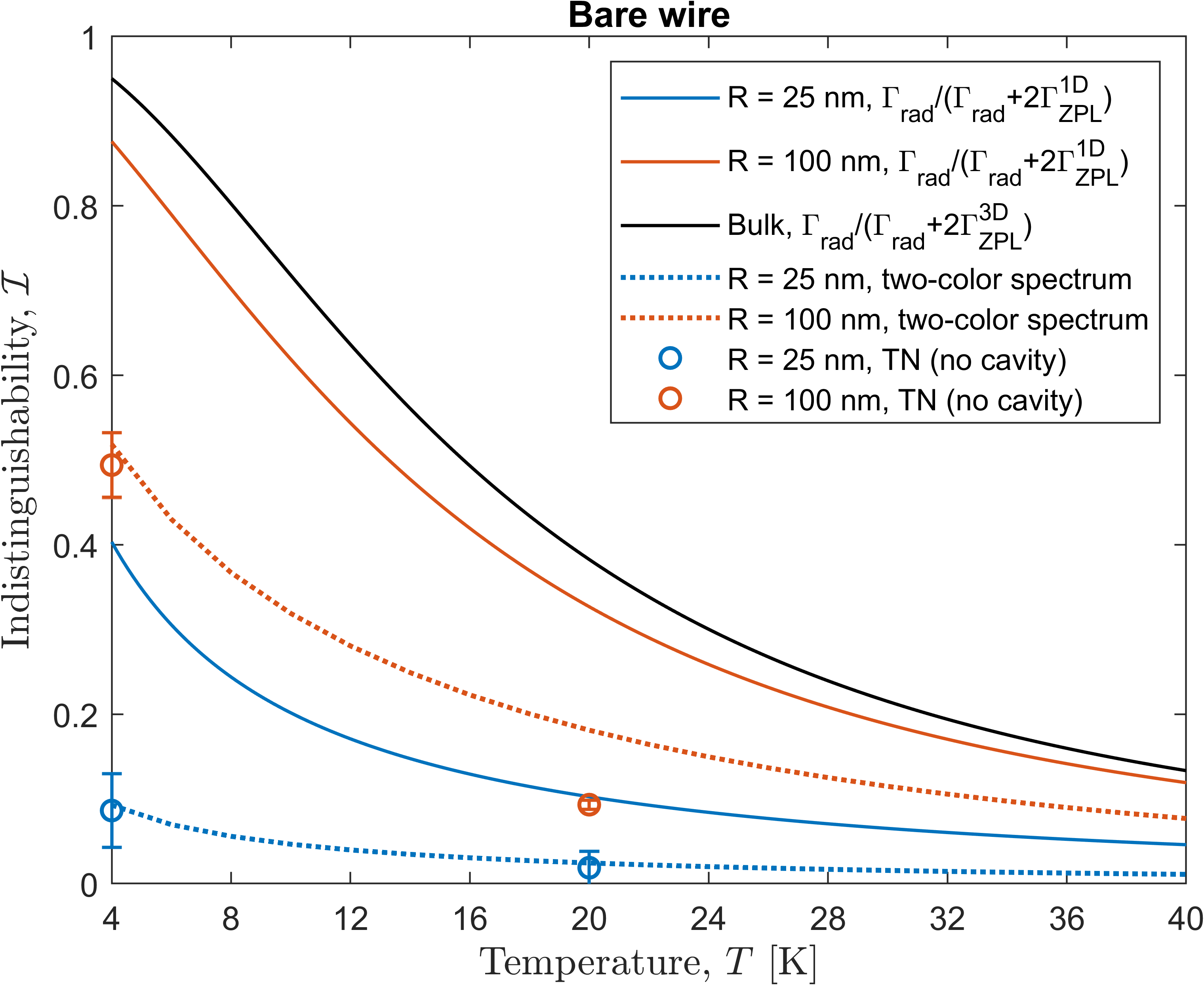}
\caption{Indistinguishability for the bare wire. The solid curves assume complete suppression of the phonon sidebands/satellite peaks and only consider coupling to the first phonon branch $\omega_{1,\mathbf{k}} = c_{s} |\rm{\mathbf{k}}|$, without accounting for cavity coupling effects. The dotted curves correspond to our bare wire calculations determined by Eq.~\eqref{eqn:20_main}, which are compared with tensor network results (empty circles). Details on the error bars for the tensor network results are given in Appendix~\ref{app:tn}.}
\label{fig:indis_25nm}
\end{figure}

In Fig.~\ref{fig:indis_25nm}, we present the indistinguishability computed using three different theoretical models, for the two nanowire radii and for bulk as a function of the temperature. 

Under the assumption of Markovian QD-light interaction and a $\delta$-shaped laser excitation, while considering a perfectly antibunched photon output with stationary noise, the two-photon indistinguishability can be calculated as \cite{kiraz_quantum-dot_2004,tighineanu_phonon_2018,artioli_design_2019}

\begin{equation}
    \mathcal{I}=\Gamma_{\rm rad}\int_{0}^{\infty} \mathrm{d}\tau\ e^{-\Gamma_{\rm rad} \tau}|P(\tau)|^{2},
\label{eq:indi_simple_before}
\end{equation}
where the coherence function $|P(\tau)|=\exp(-\Gamma_{\mathrm{ZPL}}\tau)$ by assuming the complete suppression of the phonon sidebands and coupling to the first phonon branch only \cite{tighineanu_phonon_2018}. Here, $\Gamma_{\mathrm{ZPL}}$ is the total decay rate arising from exciton-phonon interaction, which depends on the phonon bath dimensionality. We then directly obtain that the corresponding photon indistinguishability, plotted using full curves in Fig.~\ref{fig:indis_25nm}, is given by 
\begin{equation}
    \mathcal{I} = \frac{\Gamma_{\mathrm{rad}}}{\Gamma_{\mathrm{rad}} + 2 \Gamma_{\mathrm{ZPL}}}.
\label{eq:indi_simple}
\end{equation}

In the 3D bulk case, the broadening of the ZPL is exclusively due to the quadratic exciton-phonon coupling \cite{muljarov_dephasing_2004}. Consequently, the total decay rate $\Gamma_{\mathrm{ZPL}} = \Gamma^{3\mathrm{D}}_{\mathrm{ZPL}} = \gamma^{\mathrm{3D}}_{\mathrm{pd}}$ is determined solely by the pure dephasing rate owing to virtual phonon transitions, as described in Eq.~\eqref{eqn:pd}. For sufficiently low temperatures, the average phonon energy $k_{\mathrm{B}}T$ is too small to induce virtual phonon transitions, leading to negligible pure dephasing rate of the ZPL, i.e., $\gamma^{\mathrm{3D}}_{\mathrm{pd}}\ll\Gamma_{\mathrm{rad}}$ and consequently $\mathcal{I} \approx 1$. In contrast, in 1D systems, there is additional broadening from linear coupling. This results in $\Gamma^{\mathrm{1D}}_{\mathrm{ZPL}} = \Gamma_{\mathrm{lin}}^{\mathrm{1D}} + \Gamma^{\mathrm{3D}}_{\mathrm{ZPL}}$, where $\Gamma_{\mathrm{lin}}^{\mathrm{1D}} = (2A\rho_{m} c_{s}^{3}\hbar^{2})^{-1}(D_{e} - D_{h})^{2}(1 - 2\nu)^{2}k_{\mathrm{B}}T$ \cite{lindwall_zero-phonon_2007,tighineanu_phonon_2018}, with the wire cross-sectional area $A=\pi R^{2}$. 

\begin{table*}[t]
\centering
\begin{tabular}{ccccccccccc} 
\hline \hline
\addlinespace[0.25em]
$D_{e}$ {[}eV{]} & $D_{h}$ {[}eV{]} & $\rho_{m}$ {[}kg/m$^{3}${]} & $v_{l}$ {[}m/s{]} & $v_{t}$ {[}m/s{]} & $c_{s}$ {[}m/s{]} & $\nu$ & $a_{e}$ {[}$\mathrm{nm}${]} & $a_{h}$ {[}$\mathrm{nm}${]} & $m_{e}$ {[}$m_{0}${]} & $m_{h}$ {[}$m_{0}${]} \\ 
\addlinespace[0.125em] \hline
-15.93           & -8.77            & 5370 & 4780     & 2560 & 4126.14 & 0.2989 & 10               & 10               & 0.067                  & 0.51 \\ 
\hline \hline 
\end{tabular}
\caption{Material parameters for GaAs, used to calculate the electron-phonon coupling parameters, employed throughout this paper \cite{lindwall_zero-phonon_2007,denning_phonon_2020}. Here, $v_{l}$ ($v_{t}$) is the longitudinal (transverse) sound velocity. The bulk sound velocity was calculated from $c_{s}=v_{t}\sqrt{(3v_{l}^{2}-4v_{t}^{2})/(v_{l}^{2}-v_{t}^{2})}$, while $\nu=(v^{2}_{l}-2v^{2}_{t})/(2v^{2}_{l}-2v^{2}_{t})$ is the Poisson ratio, and $m_{0}$ is the free electron mass.}
\label{table:table1}
\end{table*}

Additionally, Fig.~\ref{fig:indis_25nm} presents  results in dotted curves from our own model taking into account multiple mechanical modes. Here, the indistinguishability is calculated with the two-color spectrum (see Appendix~\ref{section:bare}) from Eq.~\eqref{eqn:20_main}. Our predictions take lower values than the results from Eq.~\eqref{eq:indi_simple}, and we attribute this difference to the effect of the PSBs/phonon replicas, which is fully taken into account in our model but not in the model given by Eq.~\eqref{eq:indi_simple}. These initial results appear to confirm previous indications \cite{tighineanu_phonon_2018} that the reduced dimensionality of the 1D wire is indeed strongly detrimental to the indistinguishability. 

Finally, we also present indistinguishability computed using a numerically exact tensor network (empty circles in Fig.~\ref{fig:indis_25nm}). We observe good agreement for $R=25\ \mathrm{nm}$, confirming the validity of our model Eq.~\eqref{eqn:20_main}. However, for $R=100\ \mathrm{nm}$ at $T=$ \SI{20}{\kelvin}, the tensor network predicts lower indistinguishability, and this deviation arises due to the finite memory time of the tensor network as discussed in detail in Appendix~\ref{app:tn}.

\subsection*{Cavity quantum electrodynamic calculation}
Placing the QD in an optical cavity allows for enhancement of the zero-phonon line and suppression of the phonon sidebands through a cavity funneling effect \cite{grange_cavity-funneled_2015,iles-smith_phonon_2017}, thus enhancing the indistinguishability. However, when including a cavity, the known solution to the independent boson model employed in the previous section is no longer valid. Significant effort has been put into developing perturbative master equations \cite{wilson-rae_quantum_2002,iles-smith_phonon_2017,denning_phonon_2020,bundgaard-nielsen_non-markovian_2021}. However, the use of master equations in lower dimensional systems is less explored. While the polaron master equation has been widely used to model non-Markovian phonon effects for QDs in bulk, it is not valid in low dimensional systems due to the vanishing Franck-Condon factor $B^{2}$ \cite{chassagneux_effect_2018}.

Instead, we employ a weak master equation \cite{mccutcheon_general_2011,bundgaard-nielsen_non-markovian_2021} and a numerically exact tensor network \cite{jorgensen_exploiting_2019,strathearn_efficient_2018,pollock_non-markovian_2018,cygorek_sublinear_2024,cygorek_simulation_2022,gerald_e_fux_tempocollaborationoqupy_2024}. The weak master equation is a perturbative approach that neglects the non-Markovian effects of phonons. It establishes a comparison that can point out the nontrivial results of the tensor network, which, on the other hand, offers exact results, albeit with a substantial computational overhead. This is especially true when there is a large separation of timescales between phonon and emitter-cavity dynamics. For a micropillar structure with distributed Bragg reflectors or a photonic crystal cavity, typical light-matter coupling rates and cavity decay rates are $\hbar g=20 - 250$ \textmu$\mathrm{eV}$ and $\hbar \kappa=0.01 - 10 \ \mathrm{meV}$ \cite{denning_phonon_2020}, respectively. 

In this work, we choose $\hbar g=40$ \textmu$\mathrm{eV}$, and the characteristic timescale can be estimated by the two quantities: $\kappa/(4 g^2) = 1.03 - 1028.45\ \mathrm{ps}$ and $\kappa^{-1} = 0.06 - 65.82\ \mathrm{ps}$, indicating that the emitter-cavity timescale is of several hundreds of picoseconds. However, the fast phonon dynamics, with a timescale of $\sim$1 $\mathrm{ps}$, requires very small time steps compared with the long simulation times needed for the emitter-cavity dynamics. We explore this regime conventionally unavailable for tensor network methods by using the code and algorithm developed in Ref.~\cite{cygorek_sublinear_2024}, which utilizes a periodic process tensor structure to reduce the numerical complexity of simulations with many time steps. The cavity-emitter coupling is described by the Jaynes-Cummings Hamiltonian \cite{bundgaard-nielsen_non-markovian_2021}

\begin{equation}
    \mathcal{H}_{\mathrm{S}} = \hbar\omega_{X}\sigma^\dagger\sigma + \hbar\omega_{\rm cav}a^\dagger a + \hbar g(\sigma^\dagger a + a^\dagger \sigma),
    \label{eq:JC}
\end{equation}
where $\omega_{X}$ ($\omega_{\rm cav}$) is the bare exciton (cavity mode) frequency, $a^\dag$ ($a$) is the cavity photon creation (annihilation) operator, and $g$ the light-matter coupling strength introduced above. Next, we will consider the cavity to be on resonance with the phonon-shifted emitter transition $\tilde{\omega}_{X}$. The cavity mode is leaky with a decay rate $\kappa$, included through a Lindblad term. For more details, see Appendix~\ref{app:me}.

\begin{figure}[b]
    \centering
    \includegraphics[width=\linewidth]{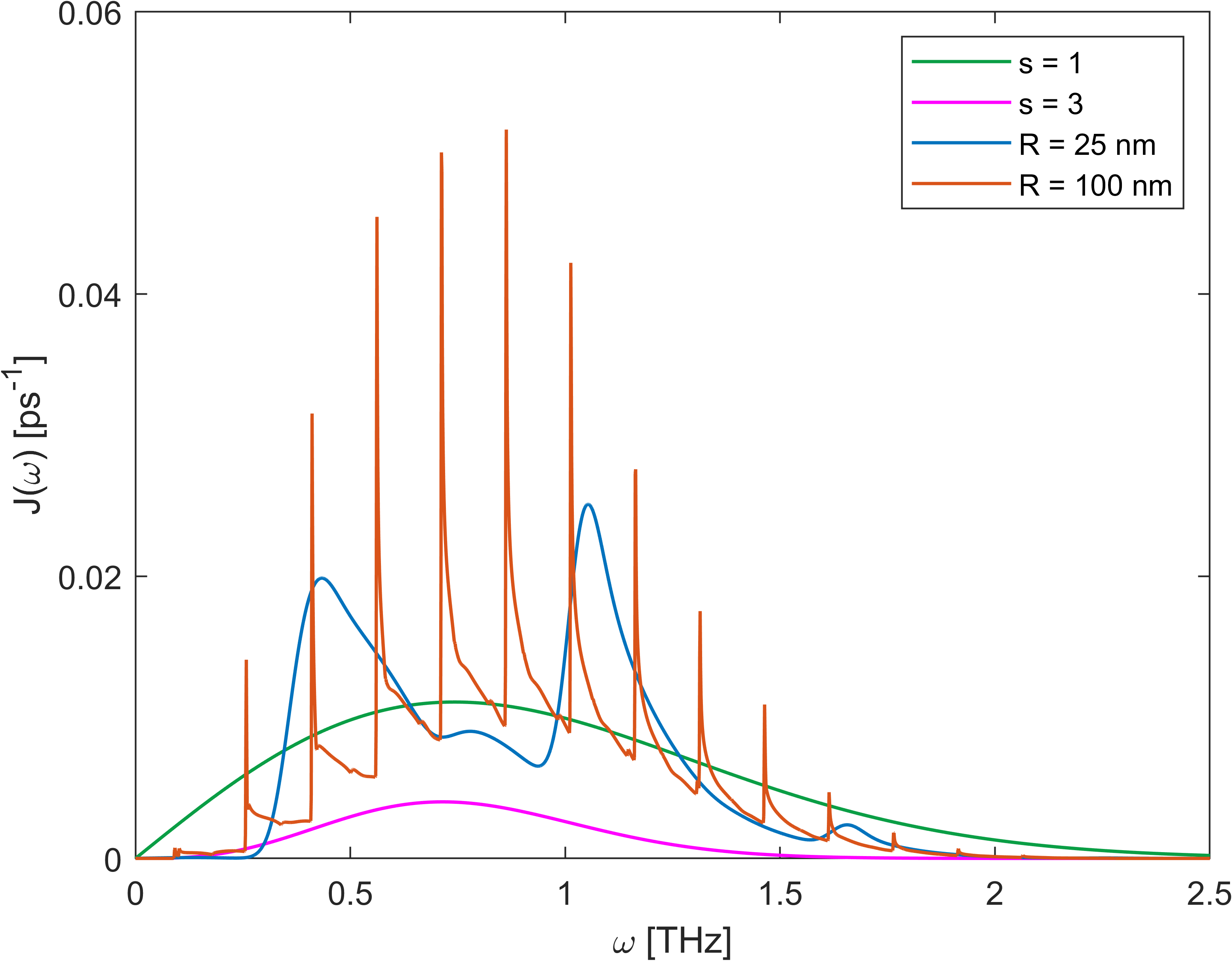}
    \caption{Spectral densities of the phonon environment $J(\omega)$ from different models: Predictions from our multimode 1D phonon model for $R=25 \ \mathrm{nm}$ and $R=100 \ \mathrm{nm}$ and from the expression $J(\omega)=\alpha \omega^s\exp(-\omega^{2}/\omega_c^{2})$ \cite{chassagneux_effect_2018} for $s=\{1,3\}$ with $\omega_c = \{1.05 \ \mathrm{THz}, 0.58 \ \mathrm{THz}\}$ and $\alpha = \{0.012,0.049 \ \mathrm{ps}^{2} \}$, respectively.}
    \label{fig:spectral_density}
\end{figure}

We will calculate the indistinguishability for four different cases. These include: a nanowire of radius (i) $R=25\ \mathrm{nm}$ and (ii) $R=100\ \mathrm{nm}$, where we have calculated the spectral density $J(\omega)$ numerically from the exciton coupling $g_{i,\mathrm{\bf{k}}}$; (iii) the bulk case ($s=3$), where we use the material parameters stated in Table~\ref{table:table1}; and (iv) an Ohmic model where $J(\omega) \propto \omega$ ($s=1$).
The latter is considered here to test the validity of our model, and is obtained with a cutoff frequency $\omega_c = 1.05 \ \mathrm{THz}$ and imposing that the integral $\int J(\omega)\ \mathrm{d}\omega \approx 0.013$ has the same value as the case for the $R=25\ \mathrm{nm}$ nanowire. 
The phonon spectral densities corresponding to these four cases are shown in Fig.~\ref{fig:spectral_density}.

\begin{figure*}[t] 
  \includegraphics[width=\linewidth]{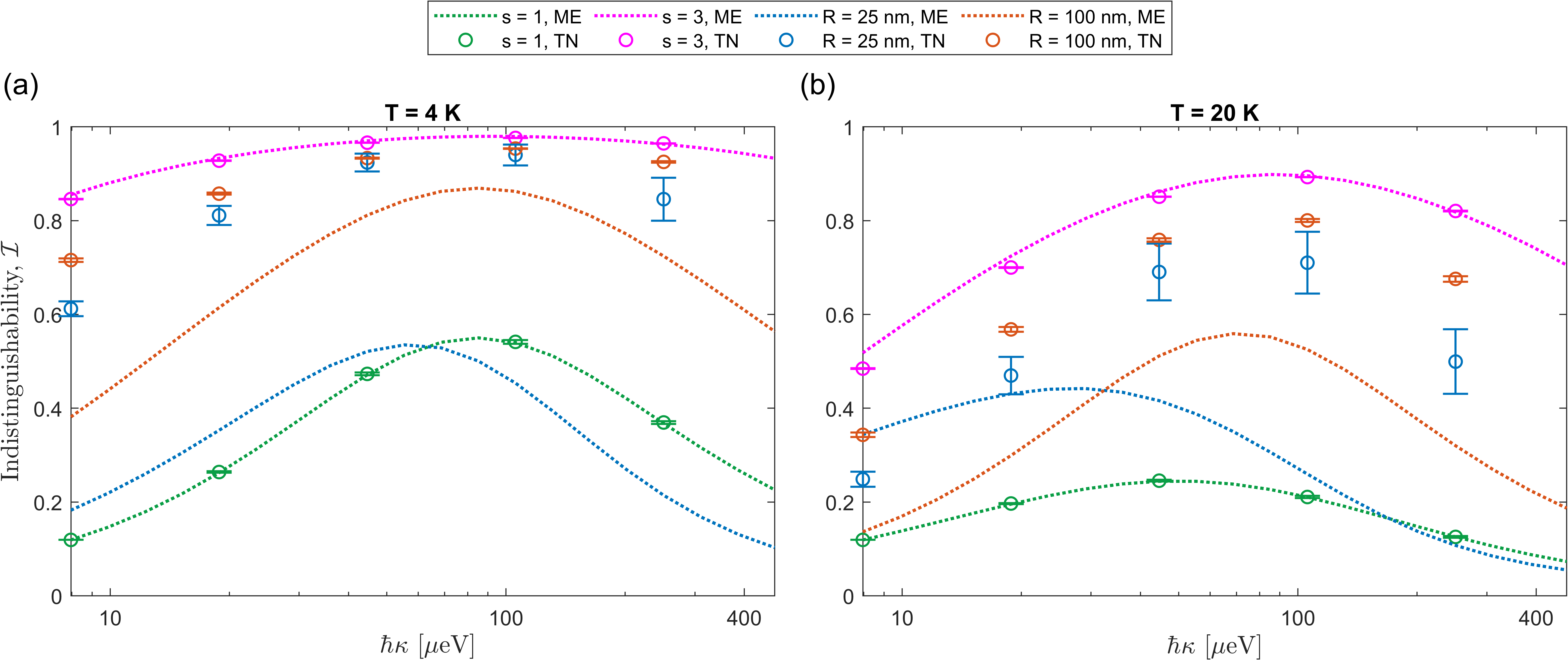}
  \caption{Indistinguishability as a function of the cavity decay rate for $\hbar g=40$ \textmu $\mathrm{eV}$. (a) At a temperature $T=$ \SI{4}{\kelvin} and (b) at $T=$ \SI{20}{\kelvin}. Details on the generation of the error bars can be found in Appendix~\ref{app:tn}.}
\label{fig:indis_all}
\end{figure*}

In Fig.~\ref{fig:indis_all}(a), we present the indistinguishability at \SI{4}{\kelvin} for the four cases as a function of the cavity decay rate predicted using the weak master equation and the tensor network. We use the pure dephasing rates from Fig.~\ref{fig:loglog}, where for $s = 1$, the pure dephasing is the same as for the $R=25\ \mathrm{nm}$ nanowire. We observe that the weak master equation describes the indistinguishability of the $s=1$ and $s=3$ cases in good agreement with the tensor network. For the nanowire with $R=25 \ \mathrm{nm}$, the non-Markovian effects due to long-lived phonons are very strong, and the weak master equation strongly underestimates the indistinguishability. This is in consonance with observations of the quantum regression theorem having significant limitations in this regime, resulting in an overestimated influence of the phonon environment \cite{cosacchi_accuracy_2021}. 

\begin{figure}[b]
    \centering
\includegraphics[width=\linewidth]{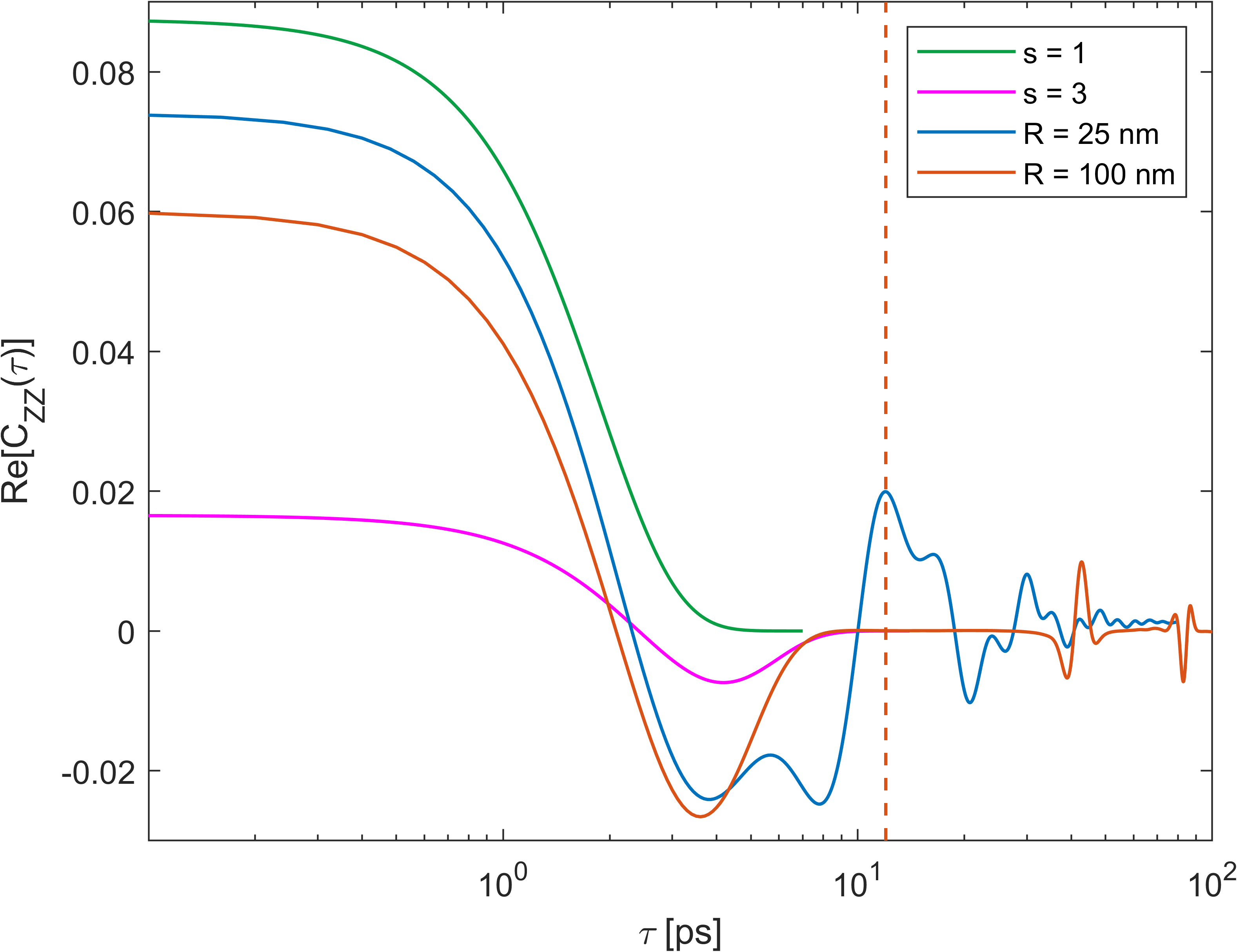}
    \caption{Real component of the environmental correlation function for the four phonon environments considered at $T=4 \ \mathrm{K}$. The dashed vertical line indicates the memory time for $R=100 \ \mathrm{nm}$ (carried out until $\tau=12 \ \mathrm{ps}$), whereas the other correlation functions are plotted over the depicted range of the memory time.} 
    \label{fig:cfunc}
\end{figure}

\begin{figure*}[t] 
  \includegraphics[width=\linewidth]{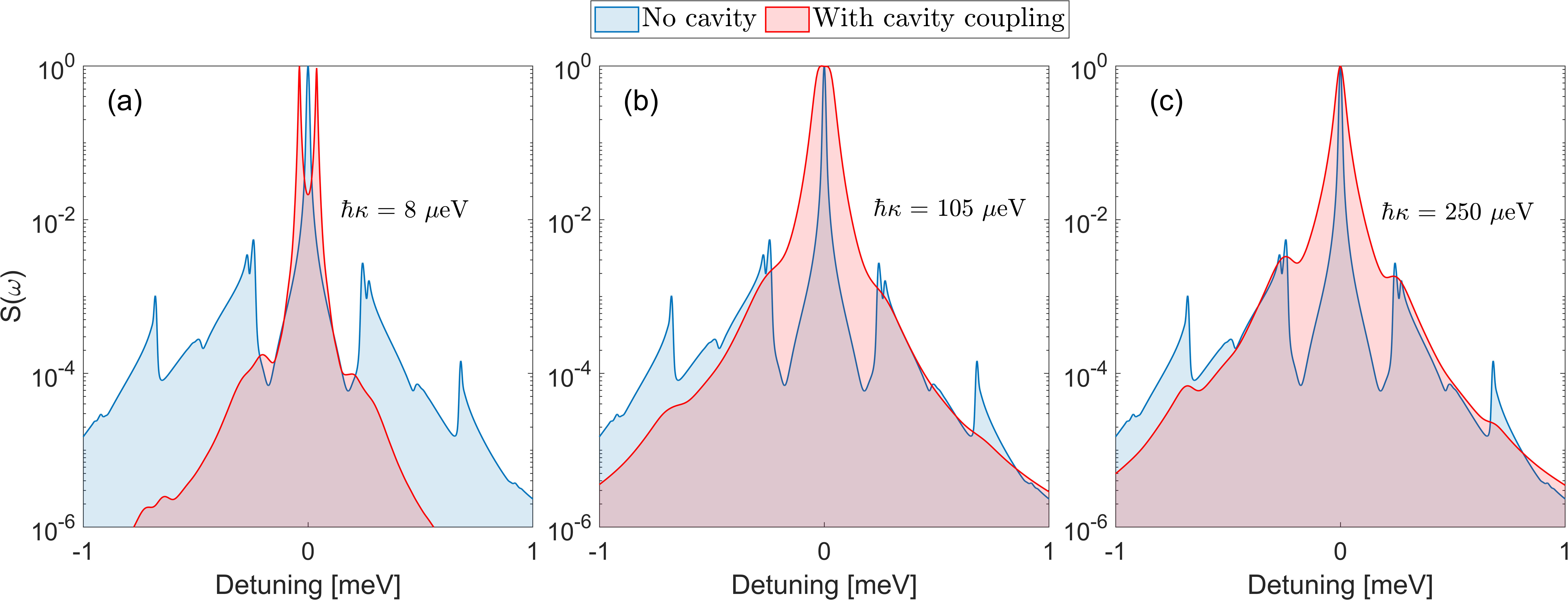}
\caption{Emission spectra obtained from tensor network calculations for different cavity decay rates are contrasted with the bare spectrum (shaded in blue) for a nanowire radius $R=25\ \mathrm{nm}$ at $T=$ \SI{4}{\kelvin}.}
\label{fig:niceee}
\end{figure*}

The presence of long-lived phonon correlations is clearly evidenced by considering the real component of the environmental correlation function $C_{\mathrm{ZZ}}(\tau)$, an analog of the coherence evolution of a two-level system within the independent boson model, also dictated by the phonon spectral density. For the exact expression of $C_{\mathrm{ZZ}}(\tau)$, cf. Appendix~\ref{app:me}. As shown in Fig.~\ref{fig:cfunc}, it is evident that longer-lived correlations are present for both $R=25 \ \mathrm{nm}$ and $R=100 \ \mathrm{nm}$. For $R=100 \ \mathrm{nm}$, the long-lived effects are, however, neglected as they exceed the memory time of the tensor network algorithm. Namely, the longer-lived correlations extend to thousands of picoseconds, which can be associated with the many satellite peaks observed in the spectral density.

The tensor network results for $R=100 \ \mathrm{nm}$ indicate a transition toward bulk behavior, as the indistinguishability approaches the curve corresponding to $s=3$ in Fig.~\ref{fig:indis_all}(a). The finite memory time of the tensor network only allows it to correctly capture the bulklike features, and the transition to bulk behavior is thus not unexpected. Still, it confirms the validity of our 1D nanowire phonon model for large radius. From the corresponding results at \SI{20}{\kelvin} presented in Fig.~\ref{fig:indis_all}(b), we verify that similar physics is observed at a higher temperature despite reduced indistinguishability. However, as the temperature of the bath increases, the weak-coupling treatment becomes an even less accurate approximation due to the increasing relevance of multiphonon effects and the misestimation of the Rabi frequency renormalization \cite{mccutcheon_quantum_2010,nazir_modelling_2016}.

To investigate the effect of the cavity on the emission spectrum, we present example spectra in Fig.~\ref{fig:niceee} from which the reported indistinguishabilities were derived using the tensor network approach. Our focus is on a nanowire with a radius $R$ = 25 $\mathrm{nm}$ at \SI{4}{\kelvin}, as indicated by the empty blue circles in Fig.~\ref{fig:indis_all}(a). When the cavity decay rate $\kappa$ is sufficiently large, we observe that most phonon satellite peaks are recovered, while only finer features of the bare spectrum are suppressed [cf. Fig.~\ref{fig:niceee}(c)]. However, as $\kappa$ decreases, the optical cavity acts as a spectral funnel that enhances emission into the ZPL by reducing contributions from neighboring incoherent phonon peaks \cite{iles-smith_phonon_2017}.

The cQED parameters maximizing the indistinguishability are given by \{$\hbar g$, $\hbar\kappa$, $\hbar \gamma$\} = \{40 \textmu $\mathrm{eV}$, 105 \textmu $\mathrm{eV}$, 0.66 \textmu $\mathrm{eV}$\} for $\gamma=\Gamma_{\mathrm{rad}}=1$ GHz with the corresponding spectrum in Fig.~\ref{fig:niceee}(b). For decreasing $\kappa$, we achieve strong coupling, characterized by $g$ $\gg$ \{$\kappa$, $\gamma$\}. In this regime, phonon modes can mediate transitions between the polariton branches, leading to decoherence in the emission process, as presented in Fig.~\ref{fig:niceee}(a) \cite{denning_phonon_2020}. 

\section{Discussion} \label{sec:discussion}

We introduce a microscopic model for the 1D photonic wire, accounting for exciton coupling to several phonon branches while incorporating cavity coupling effects. Our results highlight the key role of Purcell enhancement and non-Markovian effects in determining the indistinguishability of photons from solid-state quantum emitters embedded in a geometry of reduced dimensionality. The indistinguishability is found to be higher than what is anticipated from models that neglect the cavity coupling effect.

We show that, at the nanoscale, the evolution of QD coherence cannot be accurately captured by uniquely considering the coupling of the exciton to the fundamental longitudinal vibration mode. Thus, to fully describe our quantum emitter interacting with a 1D phonon environment, it becomes crucial to account for multimode phonon coupling, moving beyond the picture of Markovian decay (cf. Fig.~\ref{fig:coherences}).

In Ref. \cite{tighineanu_phonon_2018}, the pure dephasing rate for one-dimensional structures $\gamma^{1\mathrm{D}}_{\mathrm{pd}}$ was assumed to be the same as bulk $\gamma^{3\mathrm{D}}_{\mathrm{pd}}$ due to the significant mismatch between the waveguide size and the QD. However, it is well known that the 1D coherence dynamics strongly differs from the 3D bulk case, making one wary of the validity of the bulk model. Therefore, we deliberately calculated a one-dimensional pure dephasing rate, where we obtained $\gamma^{1\mathrm{D}}_{\mathrm{pd}}(R, T)<\gamma^{3\mathrm{D}}_{\mathrm{pd}}(T)$ for a nanowire radius $R$ at temperature $T$, on account of the reduced number of degrees of freedom. In the limit where $R$ becomes sufficiently big, we can expect to recover the bulk prediction for the pure dephasing rate, as suggested by Fig.~\ref{fig:loglog}. While important, this task is not fully explored in this paper, as our primary interest is in geometries of reduced dimensionality.

One of the challenges in performing a 1D pure dephasing calculation is the need to include a sufficient number of vibration modes $i$ for obtaining accurate results. The computational complexity of calculating $\gamma^{1\mathrm{D}}_{\mathrm{pd}}$ scales with $i^2$. While this quadratic growth is manageable for a moderate number of phonon modes, it becomes cumbersome for larger nanowire radii.

Indistinguishability from a bare wire of radius $R = 25$ nm and $R = 100$ nm is found to be much lower than in bulk (see Fig.~\ref{fig:indis_25nm}), as results derived from Eq.~\eqref{eq:indi_simple} must be regarded as upper bound predictions. This follows from the fact that Eq.~\eqref{eq:indi_simple} is obtained by filtering out incoherent emission from the phonon satellite peaks \cite{phillips_photon_2020}, which increases $\mathcal{I}$ at the expense of reducing the photon collection efficiency $\varepsilon$ \cite{iles-smith_phonon_2017}.
However, for scalable quantum computing with single photons, both $\mathcal{I}$ and $\varepsilon$ must be increased as close as possible to unity simultaneously, making the filtering unpractical.

For the case where cavity effects are exploited, we compared results from a weak coupling master equation with those obtained using tensor network calculations, cf. Fig.~\ref{fig:indis_all}. While the master equation approach successfully captures the overall trend of photon indistinguishability as a function of cavity linewidth at a fixed cavity coupling strength, it underestimates $\mathcal{I}$ in a 1D wire due to its inherent limitations \cite{cosacchi_accuracy_2021,nazir_modelling_2016,mccutcheon_quantum_2010}. Conversely, tensor network calculations, though computationally demanding, offer more reliable predictions.

We thus show that, when making calculations of low-dimensional systems, one has to consider the microscopic details of the phonons. For cylindrical nanowires, approximating the phonons with an Ohmic spectral density ($s = 1$) overestimates the phonon influence. Notably, the nanowire spectral density leads to long-lived phonon interactions, which, through coherent exchange with the emitter, result in greater indistinguishability compared to the Ohmic spectral density, where such long-lived modes and coherent exchange are absent. More specifically, we attribute the physical interpretation of this result to two phenomena. First, the assumption of an Ohmic spectral density leads to complete coherence loss in the long time limit due to the impossibility of benefiting from \textit{coherence trapping}, as the dephasing is effectively Markovian for $s \leq 1$ \cite{addis_coherence_2014}. Second, the long-lived phonon interaction allows the emitter to partially regain coherence lost through an emitted phonon by reabsorbing the phonon at a later time. This \textit{information backflow} is indicative of non-Markovianity, which is reflected by the nonmonotonic behavior of $\langle\sigma_{x}\rangle$ and $\mathrm{Re}[C_{\mathrm{ZZ}}(\tau)]$, presented in this paper. We can verify that for $s=1$ (green curve in Fig.~\ref{fig:cfunc}), the evolution is monotonic, as nonmonotonicity requires  $s>2$ to enter the non-Markovian regime \cite{addis_coherence_2014}.

Using a numerically exact tensor network, we have demonstrated that our model, which incorporates multimode coupling, yields significantly better indistinguishability, providing substantial benefits for SPS applications. Lastly, it should be emphasized that actual indistinguishability measurements are sensitive to losses and decoherence that may occur both during the emitter excitation and emission. We have assumed an instantaneous laser pulse excitation with 100\% state preparation fidelity, thereby initializing the emitter in the excited state at time $t=0$. As a result, only decoherence during the emission process (i.e., via phonon-assisted transitions) was considered.
A detailed modeling of the excitation scheme, which may introduce additional limitations to the indistinguishability through time-jitter or imperfect exciton preparation \cite{gustin_efficient_2020,scholl_crux_2020,wei_tailoring_2022,vannucci_highly_2023}, is beyond the scope of this work.

\section{Conclusions} \label{sec:conclusions}

In summary, by solving the equations for LA phonon modes in a homogeneous cylindrical nanowire, given by their nontrivial dispersion relations, we have derived a closed-form expression for the linear and quadratic electron-phonon coupling, where the latter enabled the development of an analytical expression for the 1D pure dephasing rate  $\gamma^{1\mathrm{D}}_{\mathrm{pd}}$. Assuming that a QD embedded in a free-standing 1D cylindrical waveguide is exclusively coupled to the fundamental LA phonon mode, i.e., only considering the first phonon branch ($\omega_{1,\mathbf{k}} = c_{s} |\rm{\mathbf{k}}|$), Markovian decay is observed over long timescales \cite{lindwall_zero-phonon_2007,tighineanu_phonon_2018}. Conversely, considering the QD to be additionally coupled to higher-order modes, the decay becomes highly non-Markovian. We underscore the importance of going beyond the assumption of Markovian decay in a 1D photonic wire, where memory and backaction effects in the reservoir can increase photon indistinguishability \cite{bylicka_non-markovianity_2014}. Our results hint at the crossover between the physics of bulk media and nanoscale structures, facilitating the development of solid-state quantum technologies with enhanced coherence properties. In future follow-up work, it would be of interest to investigate phonon modeling in inhomogeneous media \cite{grosse_electron-phonon_2007} and explore how reduced dimensionality impacts the deformation potential \cite{murphy-armando_deformation_2010}, thereby influencing single-photon characteristics. 

\section*{Acknowledgements}
J.F.N. is thankful to Devashish Pandey and Resul Al for the fruitful discussions at the early stage of this work. This paper was supported by the European Union’s Horizon 2020 Research and Innovation Programme under the Marie Skłodowska-Curie Grant (Agreement No. 861097), and the Danish National Research Foundation through the NanoPhoton-Center for Nanophotonics, Grant No. DNRF147. The authors also acknowledge support from the European Research Council (ERC-CoG ``Unity", Grant No. 865230).

\appendix
\section{Linear electron-phonon coupling}
\label{section:lin}
The diagonal matrix element for linear electron-phonon coupling via deformation potential is given by \cite{takagahara_theory_1999,bruus_manybody_2004}
\begin{equation}
\label{eqn:8_before}
M^{0,0}_{\lambda,i,\mathrm{\bf{k}}}=\frac{D_\lambda}{\sqrt{N}}\langle\psi_{\lambda,0}|\mathbf{\boldsymbol{\nabla}} \cdot \mathbf{u}_{i,\mathrm{\bf{k}}}(\mathbf{r})|\psi_{\lambda,0}\rangle,
\end{equation}
where the subscript $\lambda$ $\in$ \{$e$, $h$\} refers to electrons ($e$) and holes ($h$), $|\psi_{\lambda,0}\rangle$ corresponds to the ground state wave function, $D_{\lambda}$ is the associated deformation potential, $\mathbf{u}_{i,\mathrm{\bf{k}}}(\mathbf{r})$ is the normalized displacement field of the phonon mode $i$, and $N$ corresponds to the total number of ions in the one-dimensional lattice. In cylindrical coordinates and with the quantized wire modes, the coupling matrix element becomes
\begin{equation}
\label{eqn:10}
M^{0,0}_{\lambda,i,\mathrm{\bf{k}}}= D_\lambda \frac{R}{2}\sqrt{\frac{\hbar}{V \rho_{m}\ \omega_{i,\mathrm{\bf{k}}}}}\frac{\chi_{i,\mathrm{\bf{k}}}}{\sqrt{\sigma_{i,\mathrm{\bf{k}}}}}(k^2+{k_{l}}^2)\ F_{\lambda,i,\mathrm{\bf{k}}}^{0,0}.
\end{equation}
\\

Here, $V\rho_{m} = \mathcal{M}N$, where the volume of the cylindrical wire is given by $V = \pi R^{2}L$, recall that $\rho_{m}$ denotes the mass density, $\mathcal{M}$ is the ion mass, and $F_{\lambda,i,\mathrm{\bf{k}}}^{0,0}$ is the form factor
\begin{equation}
    F_{\lambda,i,\mathrm{\bf{k}}}^{0,0} =  \mel{\psi_{\lambda,0}}{J_{0}(k_{l} \sqrt{x^2 + y^2})\ e^{ikz}}{\psi_{\lambda,0}}.
\end{equation}
It follows from the normalization condition of acoustic-phonon modes in quantum wires that \cite{stroscio_quantized_1994,stroscio_phonons_2001}
\begin{widetext}
\begin{flalign}
\label{eqn:9_new}
\sigma_{i,\mathrm{\bf{k}}} = \Bigg\{ & \frac{k_{t}^{4}}{k^{2}}\frac{R^{2}}{2}[J_{1}^{2}(k_{t}R)+J_{0}^{2}(k_{t}R)]
- 2k_{t}^{2}\chi_{i,\mathrm{\bf{k}}}\frac{R}{k_{l}^{2}-k_{t}^{2}}[k_{l}J_{1}(k_{l}R)J_{0}(k_{t}R) - k_{t}J_{1}(k_{t}R)J_{0}(k_{l}R)] \nonumber \\ 
&\hspace{-1.025em}+ 2k_{t}k_{l}\chi_{i,\mathrm{\bf{k}}}\frac{R}{k_{l}^{2}-k_{t}^{2}}[k_{t}J_{0}(k_{t}R)J_{1}(k_{l}R) - k_{l}J_{0}(k_{l}R)J_{1}(k_{t}R)] + k^{2}\chi_{i,\mathrm{\bf{k}}}^{2}\frac{R^{2}}{2}[J_{1}^{2}(k_{l}R)+J_{0}^{2}(k_{l}R)] \nonumber \\
&\hspace{-1.025em}+ k_{t}^{2}\frac{R^{2}}{2}[J_{1}^{2}(k_{t}R)-J_{0}(k_{t}R)J_{2}(k_{t}R)] + k_{l}^{2}\chi_{i,\mathrm{\bf{k}}}^{2}\frac{R^{2}}{2}[J_{1}^{2}(k_{l}R)-J_{0}(k_{l}R)J_{2}(k_{l}R)] \Bigg\},
\end{flalign}
\end{widetext}
where $J_{0}$ and $J_{1}$ correspond to the lowest- and first-order Bessel functions of the first kind, respectively. Also, for the case of free-surface boundary condition (FSBC), i.e., unrestricted displacements at the boundary between the cylindrical wire and the vacuum, $\chi_{i,\mathrm{\bf{k}}}$ is given by   
\begin{equation}
\label{eqn:9_1}
\chi_{i,\mathrm{\bf{k}}}=\frac{k_{t}(k_{t}^2-k^2)}{2k^2k_{l}}\frac{J_{1}(k_{t}R)}{J_{1}(k_{l}R)}. 
\end{equation}
In addition, $k_{l,t}$ is defined as
\begin{equation}
\label{eqn:9_2}k_{l,t}=\sqrt{\frac{\omega_{i,\mathrm{\bf{k}}}^{2}}{v_{l,t}^2}-k^2},
\end{equation}
where, as a reminder, $v_{l,t}$ represents the longitudinal and transverse sound velocities, respectively. The longitudinal waves are coupled axial and radial modes with the associated quantized wave vectors $k_{l,t}$. 

Considering an isotropic harmonic potential for electrons and holes, with the confinement length $a_{\lambda}$, the ground state wave function is 
\begin{equation}
\label{eqn:whatever}
\psi_{\lambda,0}(\mathbf{r})=\frac{1}{\pi^{3/4}{a_{\lambda}}^{3/2}}\ \mathrm{exp}\left(-\frac{x^2 + y^2 + z^2}{2{a_{\lambda}}^{2}}\right).
\end{equation}

In cylindrical coordinates $(\varrho, \varphi, z)$, the form factor is then
\begin{equation}
\label{eqn:12}
\begin{aligned}
F_{\lambda,i,\vb k}^{0,0} =\ &\frac{1}{\pi^{3/2}{a_{\lambda}}^{3}}\ \int_{0}^{2\pi}\mathrm{d}\varphi \int_{0}^{\infty} \varrho\ \mathrm{exp}\left(\frac{-\varrho^2}{{a_{\lambda}}^{2}}\right)\ J_{0}(k_{l} \varrho)\ \mathrm{d} \varrho\ \\
&\hspace{-0.2em}\times \int_{-\infty}^{\infty}\mathrm{exp}\left(\frac{-z^{2}}{{a_{\lambda}}^{2}}+ikz\right)\ \mathrm{d}z,
\end{aligned}
\end{equation}
where the radial integral results in
\begin{equation}
\label{eqn:12_other}
\begin{aligned}
    \int_{0}^{\infty} \varrho\ \mathrm{exp}\left(\frac{-\varrho^2}{{a_{\lambda}}^{2}}\right)\ J_{0}(k_{l} \varrho)\ \mathrm{d} \varrho  = \frac{a_{\lambda}^2}{2}\ \exp\left(\frac{-a_{\lambda}^2 k_{l}^2}{4}\right),
\end{aligned}
\end{equation}
and the integral over the vertical coordinate $z$ corresponds to
\begin{equation}
\label{eqn:12_3}
\int_{-\infty}^{\infty}\mathrm{exp}\left(\frac{-z^{2}}{{a_{\lambda}}^{2}}+ikz\right)\ \mathrm{d}z = a_{\lambda}\sqrt{\pi}
\ \mathrm{exp}\left(\frac{-a_{\lambda}^2 k^2}{4}\right).
\end{equation}
Combining the results above, the form factor $F_{\lambda,i,\mathrm{\bf{k}}}^{0,0}$ is given by
\begin{equation}
\label{eqn:12_4}
\begin{aligned}
    F_{\lambda,i,\mathrm{\bf{k}}}^{0,0}  = \exp\left(\frac{-a_{\lambda}^2 k_{l}^2}{4}\right)\ \exp\left(\frac{-a_{\lambda}^2 k^2}{4}\right).
\end{aligned}
\end{equation}
Following, making use of the definition from Eq.~\eqref{eqn:9_2}, we are left with
\begin{equation}
\label{eqn:12_5}
\begin{aligned}
    F_{\lambda,i,\mathrm{\bf{k}}}^{0,0}  = \exp\left(\frac{-a_{\lambda}^2 \omega_{i,\mathrm{\mathbf{k}}}^2}{4v_{l}^{2}}\right).
\end{aligned}
\end{equation}
Ultimately, the difference between electron and hole contributions results in the exciton coupling matrix elements, i.e., $\hbar g_{i,\mathbf{k}}$ = $M^{0,0}_{e,\mathrm{\bf{k}}}-M^{0,0}_{h,\mathrm{\bf{k}}}$. These results are displayed in Fig.~\ref{fig:testandoO}(b) in the main text. 

\section{Quadratic electron-phonon coupling}
We begin by defining the coupling constants $f_{i,j,\mathbf{k,k'}}$ between phonon modes $i$ and $j$ as

\begin{equation}
\label{eqn:15_1}
f_{i,j,\mathbf{k,k'}}=\frac{1}{\hbar}\sum_{\lambda=e,h}\sum_{\nu\neq 0} \frac{M_{\lambda,i,\mathrm{\mathbf{k}}}^{0,\nu}M_{\lambda,j,\mathrm{\mathbf{k'}}}^{\nu,0}}{\Delta_{\lambda}},
\end{equation}
with 
\begin{equation}
\label{eqn:15_3}
\Delta_{\lambda}=E_{\lambda,\nu}-E_{\lambda,0}=\frac{\hbar^{2}}{a_{\lambda}^{2}m_{\lambda}},
\end{equation}
where $E_{\lambda,\nu}$ is the energy of the $\nu$-th state, and the sum involves virtual transitions to higher-lying electronic states, whereas $m_{\lambda}$ is the effective electron/hole mass.

The coupling matrix element between the excitonic ground state and the first excited state is
\begin{equation}
\label{eqn:15_2_before}
M_{\lambda,i,\mathrm{\mathbf{k}}}^{0,\nu}=\frac{D_{\lambda}}{\sqrt{N}}\langle\psi_{\lambda,0}|\mathbf{\boldsymbol{\nabla}}\cdot \mathbf{u}_{i,\mathbf{k}}(\mathbf{r})|\psi_{\lambda,\nu}\rangle,
\end{equation}
where the superscript $\nu$ $\in$ \{0,1,...\} indicates the bound states in the QD confinement potential.
Analogously to \eqref{eqn:10}, we have 
\begin{equation}
\label{eqn:15_6}
M_{\lambda,i,\mathrm{\mathbf{k}}}^{0,\nu} = D_\lambda \frac{R}{2}\sqrt{\frac{\hbar}{V \rho_{m}\ \omega_{i,\mathrm{\bf{k}}}}}\frac{\chi_{i,\mathrm{\bf{k}}}}{\sqrt{\sigma_{i,\mathrm{\bf{k}}}}}(k^2+{k_{l}}^2)\ F_{\lambda,i,\mathrm{\bf{k}}}^{0,\nu},
\end{equation}
\label{section:quad}
with the form factor between two different electronic states
\begin{equation}
\label{eqn:12_newmesmo}
F_{\lambda,i,\mathrm{\bf{k}}}^{0,\nu} =  \mel{\psi_{\lambda,0}} {J_{0}(k_{l} \sqrt{x^2 + y^2}) e^{i k z}} {\ \psi_{\lambda,\nu}}.
\end{equation}
Thereafter, we restrict the sum on $\nu$ to the first excited state \cite{reigue_probing_2017}.
For a 3D isotropic harmonic oscillator, the latter is the 3-fold degenerate state $\nu = (1, m)$, with $m \in \{-1, 0, +1\}$, and the wave functions correspond to
\begin{equation}
\label{eqn:15}
\psi_{\lambda,(1,m)}(\mathbf{r})=\frac{2\sqrt{2}}{\sqrt{3}\pi^{1/4}{a_{\lambda}}^{5/2}}\ |\mathbf{r}|\ \mathrm{exp}\left(\frac{-|\mathbf{r}|^{2}}{2{a_{\lambda}}^{2}}\right)\ Y^{m}_{l=1}(\theta,\phi),
\end{equation}
where the spherical harmonics for $l=1$ read
\begin{equation}
\label{eqn:15_new}
Y_{l=1}^{m=-1}(\theta,\phi)=\frac{1}{2}\sqrt{\frac{3}{2\pi}}e^{-i\phi}\sin(\theta)=\frac{1}{2}\sqrt{\frac{3}{2\pi}}\frac{x-iy}{r},
\end{equation}
\begin{equation}
\label{eqn:15_new2}
Y_{l=1}^{m=0}(\theta,\phi)=\frac{1}{2}\sqrt{\frac{3}{\pi}}\cos(\theta)=\frac{1}{2}\sqrt{\frac{3}{\pi}}\frac{z}{r},
\end{equation}
\begin{equation}
\label{eqn:15_new3}
Y_{l=1}^{m=+1}(\theta,\phi)=-\frac{1}{2}\sqrt{\frac{3}{2\pi}}e^{i\phi}\sin(\theta)=-\frac{1}{2}\sqrt{\frac{3}{2\pi}}\frac{x+iy}{r},
\end{equation}
with $r=|\mathrm{\mathbf{r}}|=\sqrt{x^2+y^2+z^2}$. One can verify that only $F_{\lambda,i,\mathrm{\bf{k}}}^{0,(1,m=0)}$ has a non-vanishing contribution given by 
\begin{equation}
\label{eqn:15_4}
\begin{aligned}
F_{\lambda,i,\mathrm{\bf{k}}}^{0,(1,m=0)} =\ &\frac{\sqrt{2}}{\pi^{3/2}{a_{\lambda}}^{4}}\int_{0}^{\infty} \varrho \, \exp\left(-\frac{\varrho^2}{{a_{\lambda}}^{2}}\right)\ J_{0}(k_{l} \varrho)\ \dd{\varrho} \\
&\hspace{-0.2em}\times \int_{-\infty}^{\infty} \ z\ \mathrm{exp}\left(\frac{-z^{2}}{{a_{\lambda}}^{2}}+ikz\right)\ \mathrm{d}z\int_{0}^{2\pi}\mathrm{d}\varphi.
\end{aligned}
\end{equation}
The radial integral remains precisely the same as given by Eq.~\eqref{eqn:12_other}, whereas the integral over the vertical coordinate $z$ reads
\begin{equation}
\label{eqn:15_5}
\int_{-\infty}^{\infty}z\ \mathrm{exp}\left(\frac{-z^{2}}{{a_{\lambda}}^{2}}+ikz\right)\ \mathrm{d}z = \frac{i}{2}\sqrt{\pi}\ a_{\lambda}^{3}\ k\  \mathrm{exp}\left(\frac{-a_{\lambda}^2 k^2}{4}\right).
\end{equation}
The form factor $F_{\lambda,i,\mathrm{\bf{k}}}^{0,(1,m=0)}$ is then
\begin{equation}
\label{eqn:15_other}
F_{\lambda,i,\mathrm{\bf{k}}}^{0,(1,m=0)}  = \frac{i\sqrt{2}}{2} a_{\lambda} k\ F_{\lambda,i,\mathrm{\bf{k}}}^{0,0},
\end{equation}
with $F_{\lambda,i,\mathrm{\bf{k}}}^{0,0}$ as defined in \eqref{eqn:12_5}.

\section{Bare wire calculation}
\label{section:bare}
Formally, one can define the emission power spectrum
\begin{equation}
\label{eqn:17one}
\begin{aligned}
S(\omega)&=\langle\sigma^{\dag}(\omega)\sigma(\omega)\rangle \\
&=\int_{-\infty}^{\infty}\mathrm{d}t\int_{-\infty}^{\infty}\mathrm{d}\tau\ e^{-i\omega\tau}\ \langle\sigma^{\dag}(t+\tau)\sigma(t)\rangle.
\end{aligned}
\end{equation}
We can equivalently write
\begin{equation}
\label{eqn:18}
S(\omega)=2\ \mathrm{Re}\left[\int_{0}^{\infty}\mathrm{d}t\int_{0}^{\infty}\mathrm{d}\tau\ e^{-i\omega\tau}\ \langle\sigma^{\dag}(t+\tau)\sigma(t)\rangle\right],
\end{equation}
where
\begin{equation}
\label{eqn:18_otheR}
\langle\sigma^{\dag}(t+\tau)\sigma(t)\rangle=g^{(1)}(t,\tau)=\Delta(\tau)\ \tilde{g}^{(1)}(t,\tau),
\end{equation}
which corresponds to the first-order dipole correlation function. Plugging this definition in Eq.~\eqref{eqn:18}, we obtain
\begin{equation}
\label{eqn:19}
S(\omega)=2\ \mathrm{Re}\left[\int_{0}^{\infty}\mathrm{d}t\int_{0}^{\infty}\mathrm{d}\tau\ \Delta(\tau)\ \tilde{g}^{(1)}(t,\tau)e^{-i\omega\tau}\right],
\end{equation}
where $\Delta(\tau)=e^{\Phi(\tau)}$ is the free phonon correlation function. Using the quantum regression theorem, while assuming that the quantum emitter is initially excited at time $t=0$, the correlation function in the polaron frame can be written as 
\begin{equation}
\label{eqn:20_otheR}
\tilde{g}^{(1)}(t,\tau)=e^{-\Gamma_{\mathrm{rad}}t} e^{-\frac{1}{2}(\Gamma_{\mathrm{rad}}+\gamma^{1\mathrm{D}}_{\mathrm{pd}})\tau} e^{i\omega_{0}\tau},
\end{equation}
with $\Gamma_{\mathrm{rad}}$ the rate of spontaneous emission (radiative rate) of the QD exciton. Then, the emission spectrum takes the form 
\begin{equation}
\label{eqn:20}
S(\omega)=\frac{2}{\Gamma_{\mathrm{rad}}}\ \mathrm{Re}\left[\int_{0}^{\infty}\mathrm{d}\tau\ e^{\Phi(\tau)} e^{-\frac{1}{2}(\Gamma_{\mathrm{rad}}+\gamma^{1\mathrm{D}}_{\mathrm{pd}})\tau} e^{-i(\omega-\omega_{0})\tau}\right],
\end{equation}
as given by Eq.~\eqref{eqn:20_main} in the main text. Additionally, we can conveniently define $G(\omega)$ as 
\begin{equation}
\label{eqn:20_neww} 
G(\omega) = \int_{0}^{\infty}\mathrm{d}\tau\ e^{\Phi(\tau)} e^{-\frac{1}{2}(\Gamma_{\mathrm{rad}}+\gamma^{1\mathrm{D}}_{\mathrm{pd}})\tau} e^{-i(\omega-\omega_{0})\tau}. 
\end{equation}
Ultimately, we can compactly write the emission power as
\begin{equation}
\label{eqn:21} 
S(\omega)=\frac{2}{\Gamma_{\mathrm{rad}}}\ G_{\mathrm{R}}(\omega),
\end{equation}
where $G_{\mathrm{R}}(\omega)$ stands for the real part of $G(\omega)$.
\par
We can write the bare spectrum without loss of generality as \cite{iles-smith_phonon_2017}
\begin{equation}
\label{eqn:22}
S(\omega,\nu)=F(\omega,\nu)+F(\nu,\omega)^{\ast},
\end{equation}
with 
\begin{equation}
\label{eqn:23}
F(\omega,\nu)=\int_{0}^{\infty}\mathrm{d}t\int_{0}^{\infty}\mathrm{d}\tau\ e^{i(\nu-\omega)t}e^{-i\omega \tau} \langle\sigma^{\dag}(t+\tau)\sigma(t)\rangle.
\end{equation}
Here, however, we are not allowed to separate the contribution from the zero-phonon line and the phonon sidebands due to the assumed dimensionality of the phonon bath ($s=1$). Once again, making use of Eq.~\eqref{eqn:18_otheR} and that $\Delta(\tau)=e^{\Phi(\tau)}$, we obtain
\begin{equation}
\label{eqn:24}
\begin{aligned}
F(\omega,\nu) =&\ \int_{0}^{\infty}\mathrm{d}t\ e^{i(\nu-\omega)t} e^{-\Gamma_{\mathrm{rad}}t} \\
&\hspace{+0.2em}\times\int_{0}^{\infty}\mathrm{d}\tau\ e^{\Phi(\tau)} e^{-\frac{1}{2}(\Gamma_{\mathrm{rad}}+\gamma^{1\mathrm{D}}_{\mathrm{pd}})\tau} e^{-i(\omega-\omega_{0})\tau}.
\end{aligned}
\end{equation}
The first integral above is trivial. By resorting to Eq.~\eqref{eqn:20_neww}, we can now conveniently rewrite our expression as
\begin{equation}
\label{eqn:25}
\begin{aligned}
F(\omega,\nu)=\frac{1}{\Gamma_{\mathrm{rad}}-i(\nu-\omega)}\ G(\omega). 
\end{aligned}
\end{equation}
Next, by taking the complex conjugate of both sides of Eq.~\eqref{eqn:25} and interchanging the variables $\omega$ and $\nu$, we arrive at
\begin{equation}
\label{eqn:28}
F(\nu,\omega)^{\ast}=\frac{1}{\Gamma_{\mathrm{rad}}-i(\nu-\omega)}\ G(\omega)^{\ast}.
\end{equation}
Plugging Eqs.~\eqref{eqn:25} and \eqref{eqn:28} into Eq.~\eqref{eqn:22}, we are left with 
\begin{align}
\label{eqn:combined_new}
S(\omega,\nu) &= \frac{G(\omega) + G(\omega)^{\ast}}{\Gamma_{\mathrm{rad}} - i(\nu - \omega)} = \frac{2G_{\mathrm{R}}(\omega)}{\Gamma_{\mathrm{rad}} - i(\nu - \omega)},
\end{align}
and by setting $\omega=\nu$
\begin{equation}
\label{eqn:31}
S(\omega,\omega)=\frac{2}{\Gamma_{\mathrm{rad}}}\ G_{\mathrm{R}}(\omega).
\end{equation}
Comparing Eq.~\eqref{eqn:21} and Eq.~\eqref{eqn:31}, one can retrieve the emission power spectrum
\begin{equation}
\label{eqn:32}
S(\omega,\omega)=S(\omega).
\end{equation}
Now that we have unequivocally introduced the variables above, we can calculate the indistinguishability $\mathcal{I}$ of two-photon interference in frequency space. To do so, we employ the definition from the two-color spectrum \cite{iles-smith_phonon_2017}
\begin{equation}
\label{eqn:33}
\mathcal{I}=P_{\mathrm{D}}^{-2}\int_{-\infty}^{\infty}\mathrm{d}\omega\int_{-\infty}^{\infty}\mathrm{d}\nu\ |S(\omega,\nu)|^{2},
\end{equation}
where the frequency-integrated spectral power $P_{\mathrm{D}}$ reads
\begin{equation}
\label{eqn:34}
P_{\mathrm{D}}=\int_{-\infty}^{\infty}\mathrm{d}\omega\ S(\omega,\omega).
\end{equation}

\section{Cavity quantum electrodynamic calculation \label{app:me}}
In the main text, we employ a weak master equation for comparison with the results obtained from the tensor network. In this Appendix, we provide details on the utilized master equation. We derive a second-order master equation for the density operator $\rho_{\mathrm{S}}$ of the emitter-cavity system \cite{bundgaard-nielsen_non-markovian_2021}. The system Hamiltonian is given in Eq.~\eqref{eq:JC}, which includes a temperature-dependent 1D pure dephasing rate due to the quadratic phonon interaction $\gamma^{1\mathrm{D}}_{\mathrm{pd}}(T)$ and a cavity decay rate $\kappa$ leading to
\begin{equation}
\label{eqn:35}
\dot{\rho_{\mathrm{S}}}=-\frac{i}{\hbar}[\mathcal{H}_{\mathrm{S}},\rho_{\mathrm{S}}]+2\gamma^{1\mathrm{D}}_{\mathrm{pd}}(T)\mathcal{L}_{\sigma^\dag\sigma}[\rho_{\mathrm{S}}]+\kappa\mathcal{L}_{a}[\rho_{\mathrm{S}}]+\mathcal{K}[\rho_{\mathrm{S}}],
\end{equation}
where $\sigma=|G\rangle\langle X|$ is the lowering operator as previously defined in the main text, $\mathcal{L}_O[\rho_{\mathrm{S}}] = O \rho_{\mathrm{S}} O^\dag - \frac 1 2 \acomm{O^\dag O}{\rho_{\mathrm{S}}}$ is the Lindblad dissipator and $\mathcal{K}[\rho_{\mathrm{S}}]$ corresponds to the phononic dissipator given by \cite{nazir_modelling_2016,bundgaard-nielsen_non-markovian_2021}
\begin{equation}
\label{eqn:dissipator}
\begin{aligned}
\mathcal{K}[\rho_{\mathrm{S}}] =&\
-\int_0^{\infty} \mathrm{d} \tau\ \left(C_{\mathrm{ZZ}}(\tau)\left[O_{\mathrm{Z}}, \tilde{O}_{\mathrm{Z}}(-\tau) \rho_{\mathrm{S}}(t)\right]\right. \\
&\hspace{+0.3em}\left. +C_{\mathrm{ZZ}}(-\tau)\left[\rho_{\mathrm{S}}(t) \tilde{O}_{\mathrm{Z}}(-\tau), O_{\mathrm{Z}}\right]\right),
\end{aligned}
\end{equation}

with $O_{\mathrm{Z}} = \sigma^\dagger \sigma$ and $\tilde{O}_{\mathrm{Z}}(\tau)=\mathrm{e}^{i\mathcal{H}_{\mathrm{S}}\tau/\hbar}O_{\mathrm{Z}}\mathrm{e}^{-i\mathcal{H}_{\mathrm{S}}\tau/\hbar}$. The environmental correlation function is described by \cite{nazir_modelling_2016,bundgaard-nielsen_non-markovian_2021}
\begin{equation}
\label{eqn:36}
C_{\mathrm{ZZ}}(\tau)=\int_{0}^{\infty}\mathrm{d}\omega\ J(\omega)\left[\frac{\cos(\omega\tau)}{\tanh(\beta\hbar\omega/2)}-i\sin(\omega\tau)\right],
\end{equation}
where $\beta=(k_{\mathrm{B}}T)^{-1}$, with $k_{\mathrm{B}}$ the Boltzmann constant. The indistinguishability $\mathcal{I}$ is calculated as in Appendix~\ref{section:bare} by replacing the two-time emitter correlation with the cavity two-time correlation $\expval{a^\dagger(t+\tau)a(t)}$, considering an initially excited emitter.

\section{Tensor network \label{app:tn}}

In this Appendix, we provide details on the accuracy of the tensor network. We emphasize that this work is based on a tensor network representation of the environment as implemented in the ACE algorithm. We refer to the literature for further details ~\cite{cygorek_sublinear_2024,cygorek_ace_2024}. In the tensor network, three numerical parameters determine the accuracy: the time step $\Delta t$, the Singular Value Truncation threshold $\epsilon$ of the Matrix Product Operator compression, and the memory time $\tau_{\mathrm{mem}}$. The memory time $\tau_{\mathrm{mem}}$ should be chosen to capture all environment features. In reality, it means that the real part of the environmental correlation function $\mathrm{Re}[C_{\mathrm{ZZ}}(\tau)]$ is zero after $\tau_{\mathrm{mem}}$. The environmental correlation function is given as in Eq.~\eqref{eqn:36} \cite{cygorek_sublinear_2024}. In Fig.~\ref{fig:cfunc} in the main text, we show the correlation functions of the four phonon environments plotted over the range of their memory time $\tau_{\mathrm{mem}}$, which are $\tau_{\mathrm{mem}} = \{7 \ \mathrm{ps},14 \ \mathrm{ps},80 \ \mathrm{ps},12 \ \mathrm{ps}$\} for $s = \{1,3\}$, nanowire radius $R=25 \ \mathrm{nm}$, and $ R=100\ \mathrm{nm}$, respectively. 

For $R=100\ \mathrm{nm}$, we have made an exception since we plot it over $\tau=100 \ \mathrm{ps}$, but the cutoff is $\tau_{\mathrm{mem}}=12 \ \mathrm{ps}$ as indicated by the vertical line. To estimate the impact of the truncation parameter $\epsilon$, we do each calculation twice using first a larger value $\epsilon_{\mathrm{C}}$ and secondly a smaller $\epsilon_{\mathrm{F}}$, where the subscripts denote the result of the coarse (C) or fine (F) calculation. Comparing the two indistinguishabilities, we thus estimate the relative error as $\delta_{\epsilon} = \abs{\mathcal{I}_{\mathrm{F}}-\mathcal{I}_{\mathrm{C}}}/\mathcal{I}_{\mathrm{F}}$. For $s=\{1,3\}$ and the nanowire radius $R={100 \ \mathrm{nm}}$, we use $\epsilon_{\mathrm{C}} = 10^{-7}$ and $\epsilon_{\mathrm{F}} = 10^{-8}$, whereas for $R=25 \ \mathrm{nm}$ we use $\epsilon_{\mathrm{C}} = 10^{-6}$ and $\epsilon_{\mathrm{F}} = 5 \times 10^{-7}$.

When calculating the nanowire spectral density, it should be noted that we have employed the numerical trick of $J(\omega)=\sum_{i,\mathrm{\mathbf{k}}} |g_{i,\mathrm{\mathbf{k}}}|^{2}\delta(\omega-\nobreak\omega_{i,\mathrm{\mathbf{k}}}) \approx \sum_{i,\mathrm{\mathbf{k}}} |g_{i,\mathrm{\mathbf{k}}}|^{2}(\sqrt{\pi} \sigma)^{-1}\exp[(\omega-\omega_{i,\mathrm{\mathbf{k}}})^{2}/\sigma^{2}]$, the delta function is thus approximated as a Gaussian with a finite width $\sigma$. We do this to smoothen the spectral density. This is much like the method employed in Ref.~\cite{svendsen_signatures_2023}. The smoothening is done to make the phonon coherence finite, thus rendering the calculations with the numerical tensor network feasible. However, unlike in \cite{svendsen_signatures_2023}, $\sigma$ is not a physical parameter but rather just a numerical trick.  

As depicted in Fig.~\ref{fig:indis_25nm} in the main text, we confirmed that this smoothening still allows for an accurate description of the nanowire phonons by comparing the bare wire calculations with the tensor network \cite{jorgensen_exploiting_2019,strathearn_efficient_2018,pollock_non-markovian_2018,cygorek_sublinear_2024,cygorek_simulation_2022,gerald_e_fux_tempocollaborationoqupy_2024} and obtaining very good agreement between the two methods. The deviation for $R=100\ \mathrm{nm}$ at $T=$ \SI{20}{\kelvin} stems from the finite memory of the tensor network and not the smoothening. The finite memory time of the tensor network makes it only capture the bulklike features of the spectral density, whereas the longer-lived thin peaks in the spectral density are not well described. 

We estimate the error due to the smoothening of the spectral density in a similar fashion as for the truncation threshold $\epsilon$ by considering $\sigma_{\mathrm{C}} = 0.1 \ \mathrm{THz}$ and $\sigma_{\mathrm{F}} = 0.05 \ \mathrm{THz}$, and estimating the relative error $\delta_{\mathrm{smooth}}$. For the nanowire simulations, we thus estimate the tensor network error as $\delta_{\mathrm{tot}} = \sqrt{\delta_{\epsilon}^2 +  \delta_{\mathrm{smooth}}^2}$, whereas for $s=\{1,3\}$ it is simply given by $\delta_{\mathrm{tot}} = \delta_\epsilon$. These errors are shown as the error bars in Figs.~\ref{fig:indis_25nm} and~\ref{fig:indis_all} with a size defined by the product $\mathcal{I} \delta_{\mathrm{tot}}$. Finally, estimating the error due to the finite discretization $\Delta t$ is more challenging. Changing $\Delta t$ also changes the requirement for convergence on the truncation cutoff $\epsilon$ \cite{cygorek_simulation_2022}. We thus assured in an initial study that we observed no significant changes for $\Delta t = 0.4 \ \mathrm{ps}$ when a small enough $\epsilon$ was chosen.

\bibliography{biblio}
\newpage
\onecolumngrid
\section*{Supplemental Material}






\begin{center}
\textbf{Mode-dependent velocities}
\end{center}
\label{section:tables}

\renewcommand{\thetable}{\Alph{table}}
\setcounter{table}{0} 

\begin{table}[H]
\begin{minipage}{.5\textwidth}
  \centering
  \caption{For $R=25\ \mathrm{nm}$.}
  \begin{tabular}{ccc}
    \hline \hline
    \addlinespace[0.25em]
    Mode number & $k$ [$(\mathrm{nm})^{-1}$] & Mode velocity [m/s] \\
    \addlinespace[0.125em] \hline
    1 & 0.0212 & 4029.79 \\ \hline
    2 & 0.0740 & 2228.36 \\ \hline
    3 & 0.1289 & 2699.92 \\ \hline
    4 & 0.1681 & 2936.31 \\ \hline
    5 & 0.1645 & 1422.98 \\ \hline
    6 & 0.2208 & 1995.43 \\ \hline
    7 & 0.1495 & 1014.86 \\ \hline
    8 & 0.2327 & 1157.07 \\ \hline
    9 & 0.1403 & 1615.65 \\ \hline
    10 & 0.2061 & 900.59 \\ \hline
    11 & 0.2163 & 751.15 \\ \hline
    12 & 0.1621 & 1231.65 \\ \hline
    13 & 0.2620 & 858.65 \\ \hline
    14 & 0.1333 & 933.79 \\ \hline
    15 & 0.1986 & 962.11 \\ \hline
    16 & 0.2154 & 450.09 \\ \hline
    17 & 0.1533 & 941.42 \\ \hline
    18 & 0.2563 & 966.05 \\ \hline
    19 & 0.2605 & 453.92 \\ \hline
    20 & 0.1409 & 766.76 \\ \hline
    21 & 0.1972 & 373.50 \\ \hline
    22 & 0.3036 & 455.57 \\ \hline
    23 & 0.1609 & 799.26 \\ \hline
    24 & 0.2231 & 337.47 \\ \hline
    25 & 0.2474 & 299.03 \\ \hline
    26 & 0.1413 & 613.78 \\ \hline
    27 & 0.2574 & 334.69 \\ \hline
    28 & 0.2749 & 345.31 \\ \hline
    29 & 0.1256 & 433.28 \\ \hline
    30 & 0.2540 & 284.50 \\ \hline
    31 & 0.1640 & 621.33 \\ \hline
    32 & 0.1878 & 344.21 \\ \hline
    33 & 0.2448 & 238.23 \\ \hline
    34 & 0.1486 & 512.62 \\ \hline
    35 & 0.2228 & 290.44 \\ \hline
    36 & 0.2647 & 231.94 \\ \hline
    37 & 0.1458 & 414.04 \\ \hline
    38 & 0.2441 & 242.63 \\ \hline
    39 & 0.2647 & 214.81 \\ \hline
    40 & 0.1407 & 390.07 \\
    \hline\hline 
  \end{tabular}
  \label{table:25nm}
\end{minipage}%
\begin{minipage}{.5\textwidth}
  \centering
  \caption{For $R=100\ \mathrm{nm}$.}
  \begin{tabular}{ccc}
    \hline \hline
    \addlinespace[0.25em]
    Mode number & $k$ [$(\mathrm{nm})^{-1}$] & Mode velocity [m/s] \\
    \addlinespace[0.125em] \hline
    1 & 0.0324 & 1675.37 \\ \hline
    2 & 0.0243 & 3293.10 \\ \hline
    3 & 0.0361 & 2927.45 \\ \hline
    4 & 0.0525 & 3316.89 \\ \hline
    5 & 0.0706 & 3610.03 \\ \hline
    6 & 0.0891 & 3765.23 \\ \hline
    7 & 0.1080 & 3871.78 \\ \hline
    8 & 0.1272 & 3943.57 \\ \hline
    9 & 0.1194 & 3197.17 \\ \hline
    10 & 0.1400 & 3369.49 \\ \hline
    11 & 0.1604 & 3477.39 \\ \hline
    12 & 0.1426 & 2919.16 \\ \hline
    13 & 0.1650 & 3053.05 \\ \hline
    14 & 0.1370 & 2628.07 \\ \hline
    15 & 0.1947 & 2128.02 \\ \hline
    16 & 0.2325 & 2305.48 \\ \hline
    17 & 0.1641 & 1807.18 \\ \hline
    18 & 0.2081 & 2059.09 \\ \hline
    19 & 0.1289 & 1396.18 \\ \hline
    20 & 0.1761 & 1804.67 \\ \hline
    21 & 0.2219 & 2010.03 \\ \hline
    22 & 0.1383 & 1428.03 \\ \hline
    23 & 0.1885 & 1759.48 \\ \hline
    24 & 0.2356 & 2074.39 \\ \hline
    25 & 0.1399 & 1428.03 \\ \hline
    26 & 0.1808 & 1759.48 \\ \hline
    27 & 0.2175 & 2074.39 \\ \hline
    28 & 0.1519 & 1428.03 \\ \hline
    29 & 0.1950 & 1759.48 \\ \hline
    30 & 0.2325 & 2074.39 \\ \hline
    31 & 0.1640 & 1428.03 \\ \hline
    32 & 0.2081 & 1759.48 \\ \hline
    33 & 0.1289 & 2074.39 \\ \hline
    34 & 0.1761 & 1428.03 \\ \hline
    35 & 0.2219 & 1759.48 \\ \hline
    36 & 0.1383 & 2074.39 \\ \hline
    37 & 0.1885 & 1428.03 \\ \hline
    38 & 0.2356 & 1759.48 \\ \hline
    39 & 0.1399 & 2074.39 \\ \hline
    40 & 0.1808 & 1428.03 \\ 
    \hline \hline
  \end{tabular}
  \label{table:100nm}
\end{minipage}
\end{table}

Next, we display the first 40 contour plots from which the data in Table~\ref{table:25nm} was generated. Analogously, we could also present the corresponding contour plots for Table~\ref{table:100nm} (with $R=100\ \mathrm{nm}$). However, for conciseness, these were not included in the SM as they do not significantly differ from the behavior depicted in Fig.~\ref{fig:coolplots_25nm_1}.

\newpage

\renewcommand{\thefigure}{\Alph{figure}}
\setcounter{figure}{0} 

\begin{figure}[H]
\centering
\begin{subfigure}{.235\textwidth}
  \includegraphics[width=\linewidth]{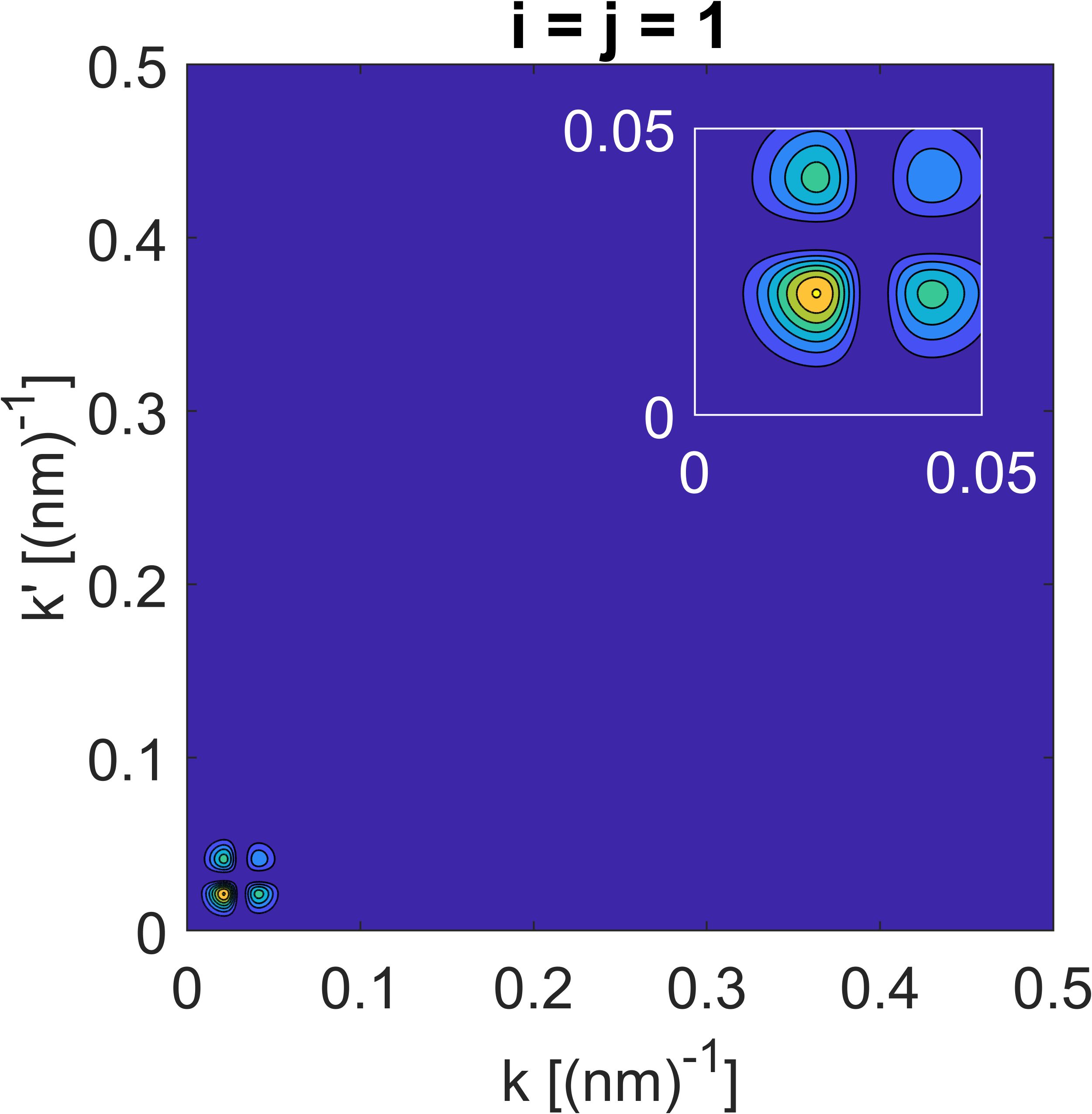}
\end{subfigure}
\begin{subfigure}{.235\textwidth}
  \includegraphics[width=\linewidth]{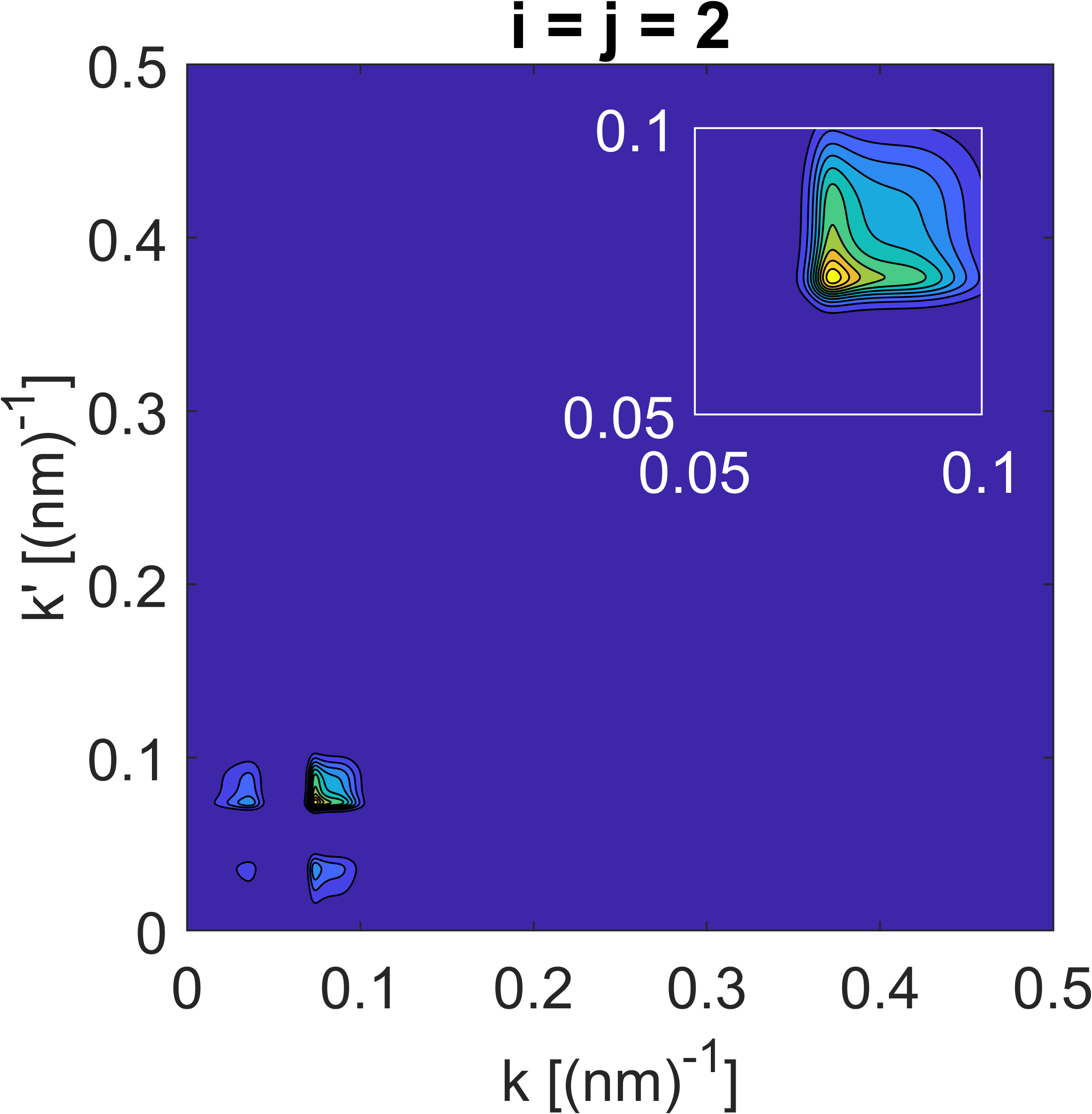}
\end{subfigure}
\begin{subfigure}{.235\textwidth}
  \includegraphics[width=\linewidth]{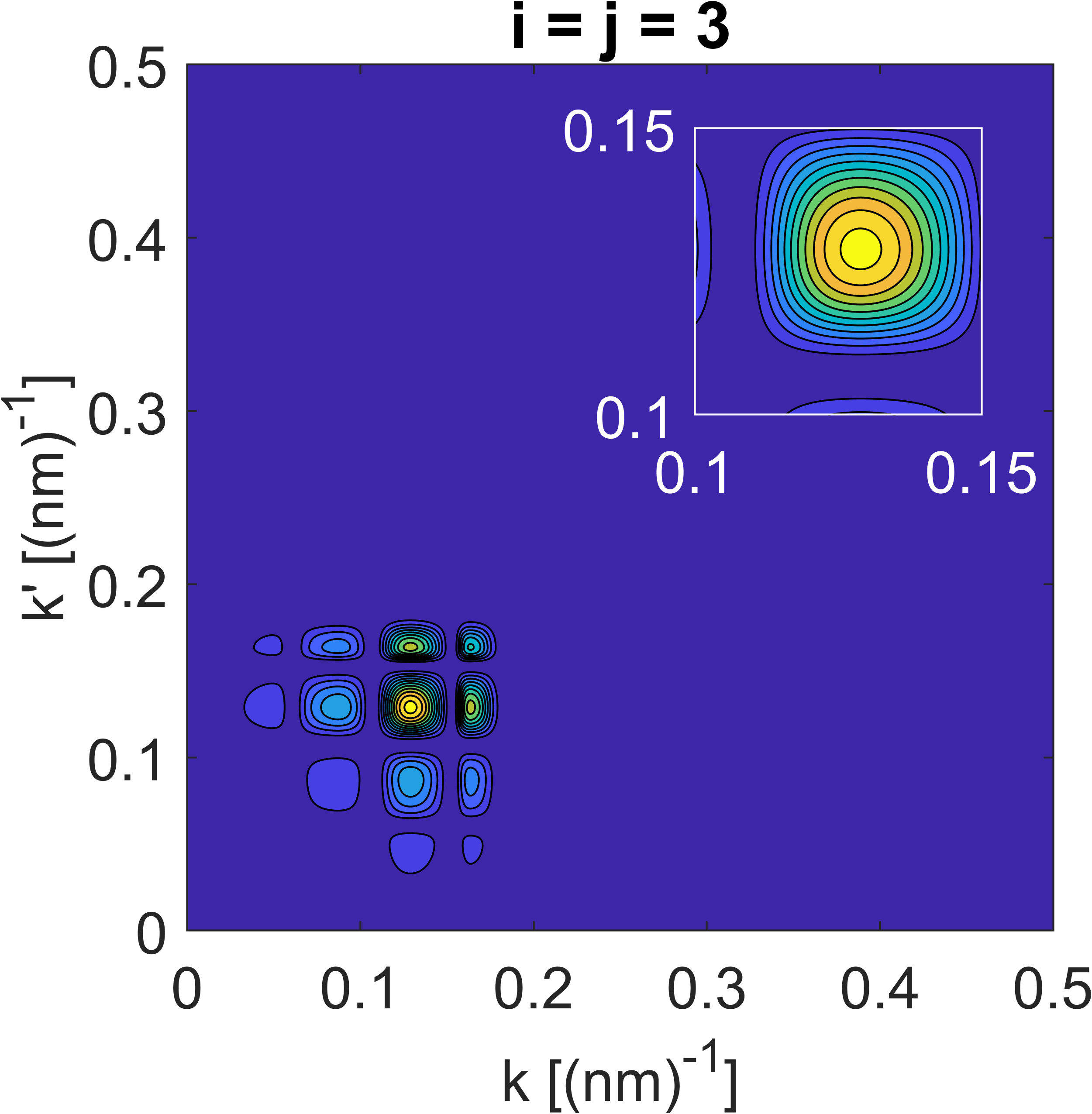}
\end{subfigure}
\begin{subfigure}{.235\textwidth}
  \includegraphics[width=\linewidth]{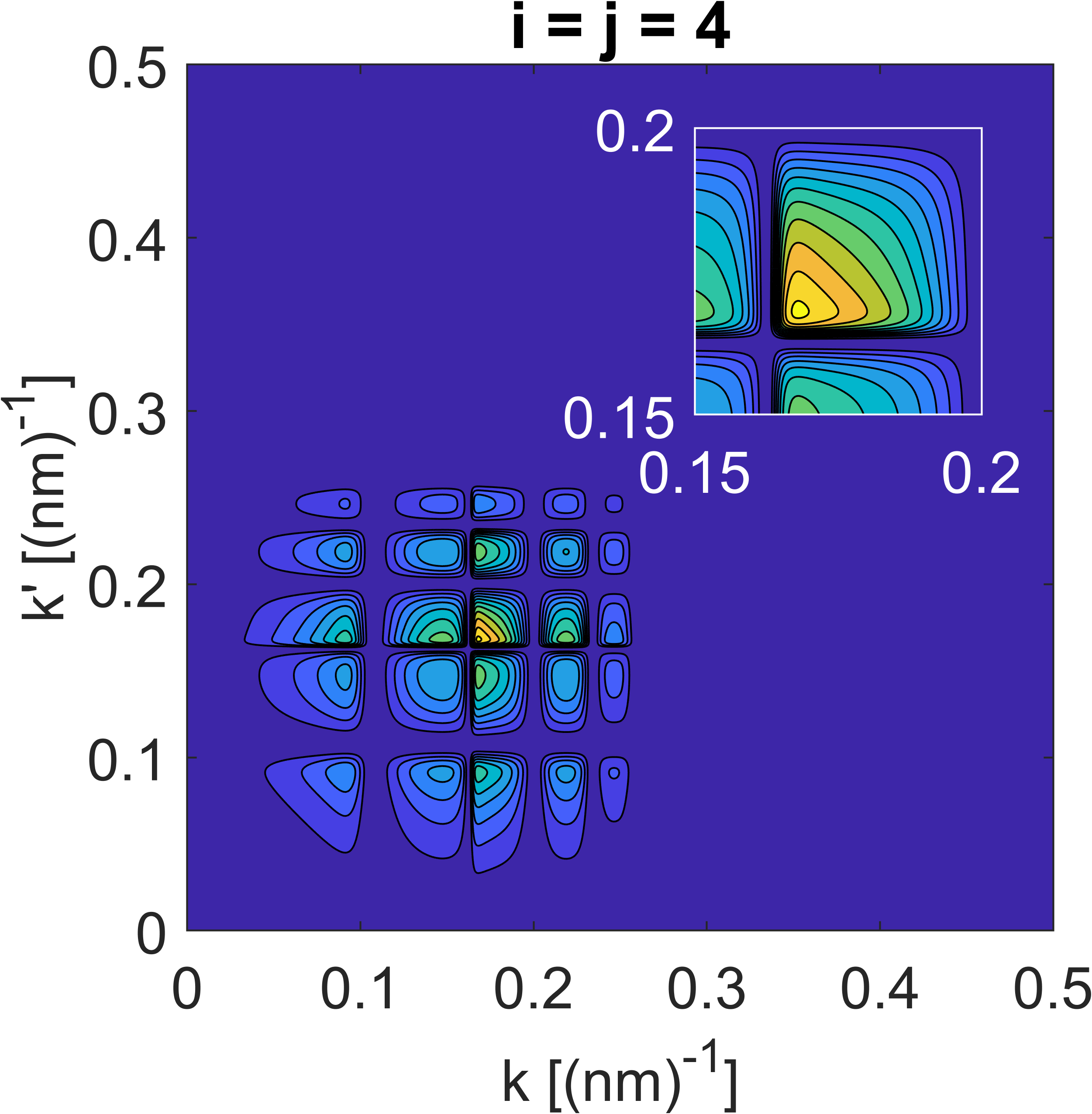}
\end{subfigure}
\newline
\newline
\begin{subfigure}{.235\textwidth}
  \includegraphics[width=\linewidth]{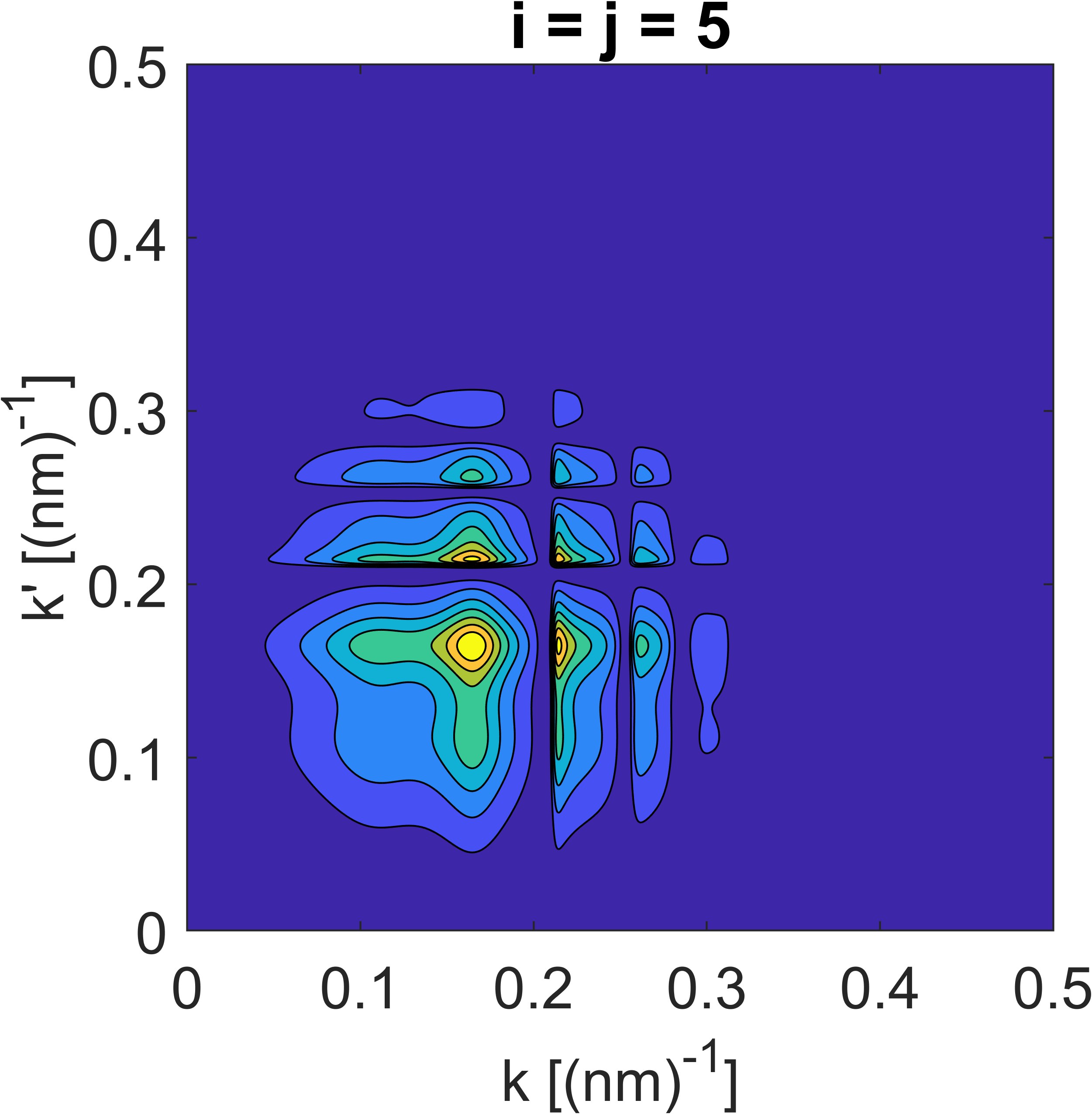}
\end{subfigure}%
\begin{subfigure}{.235\textwidth}
  \includegraphics[width=\linewidth]{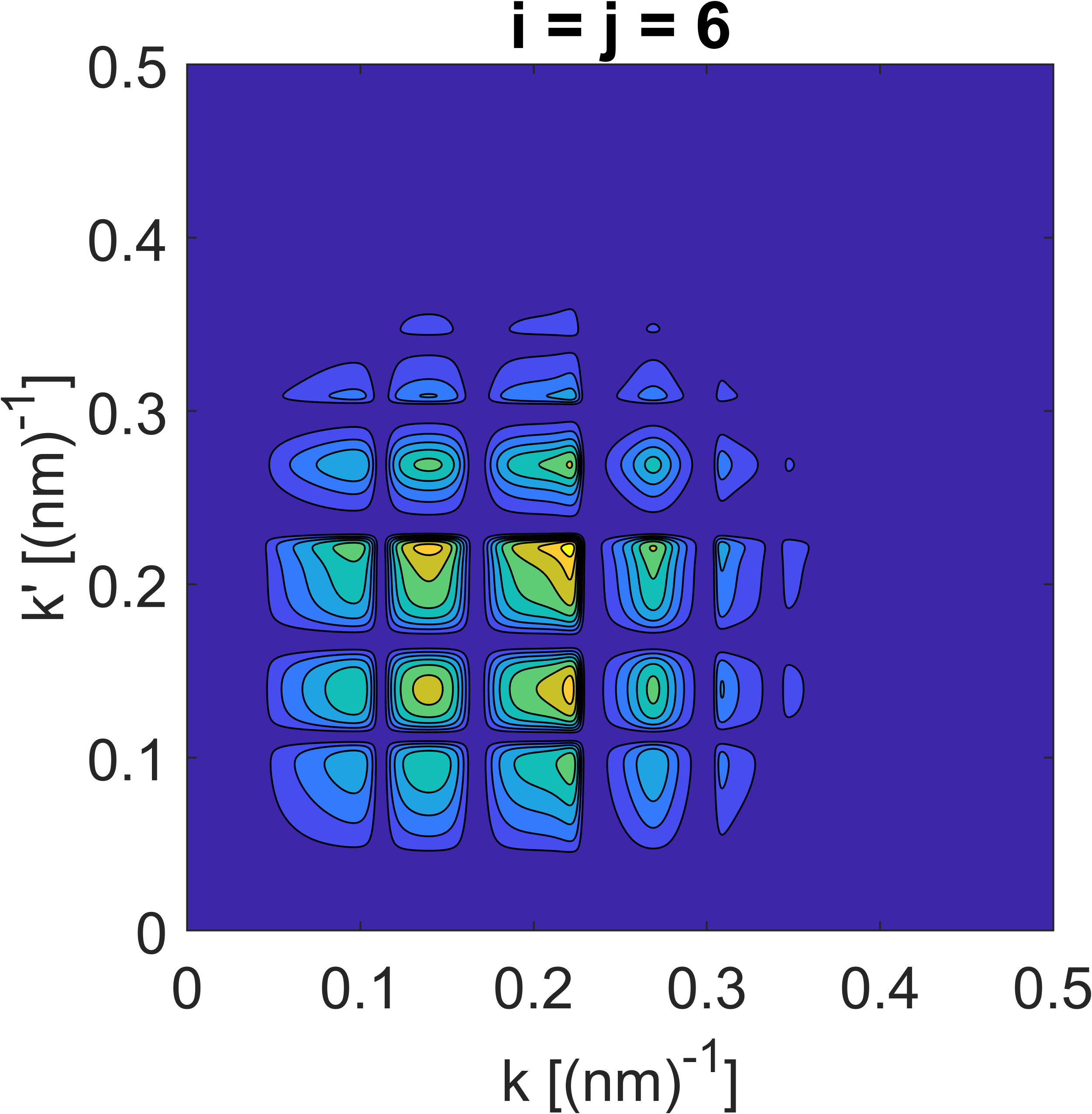}
\end{subfigure}
\begin{subfigure}{.235\textwidth}
  \includegraphics[width=\linewidth]{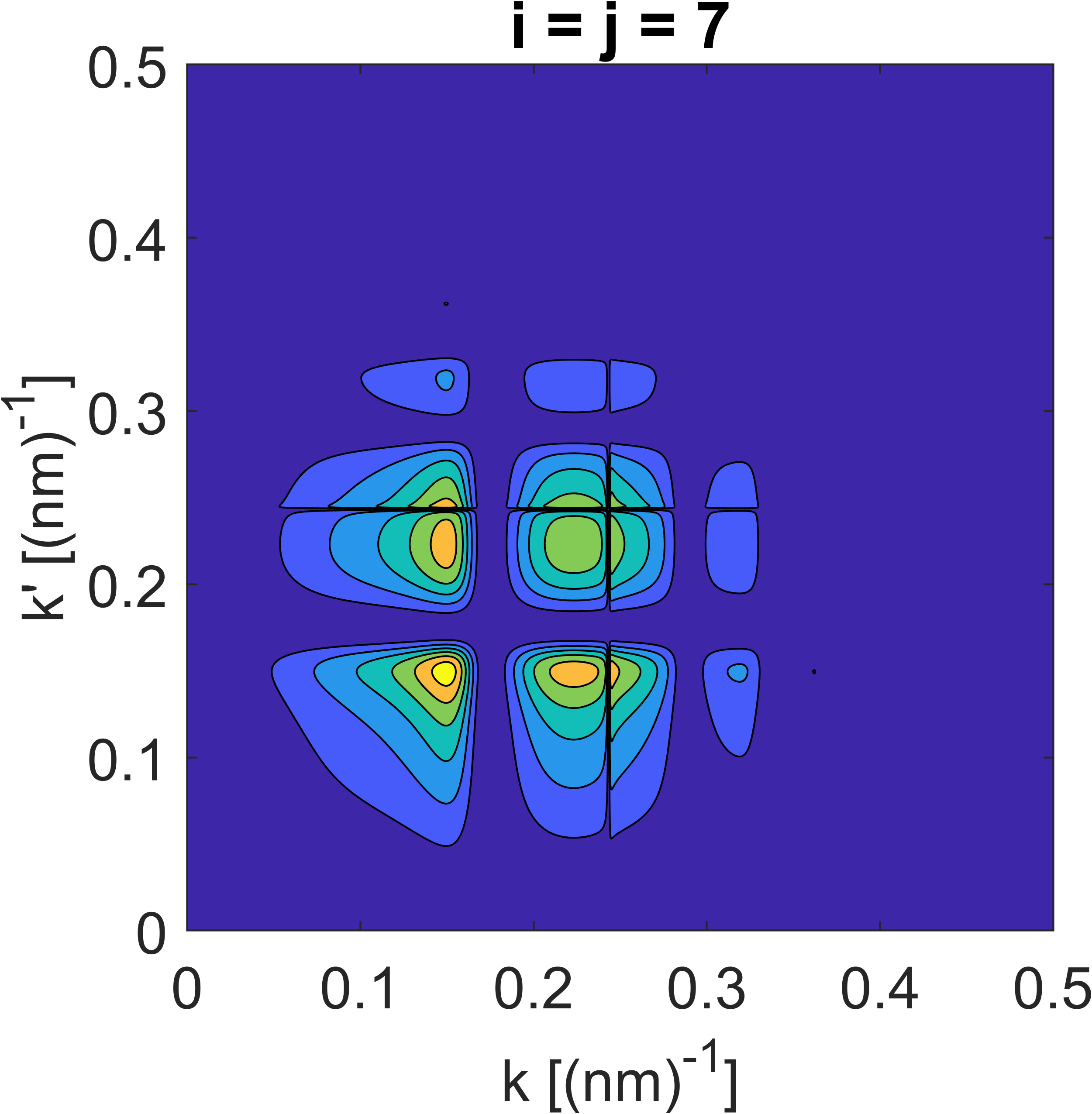}
\end{subfigure}
\begin{subfigure}{.235\textwidth}
  \includegraphics[width=\linewidth]{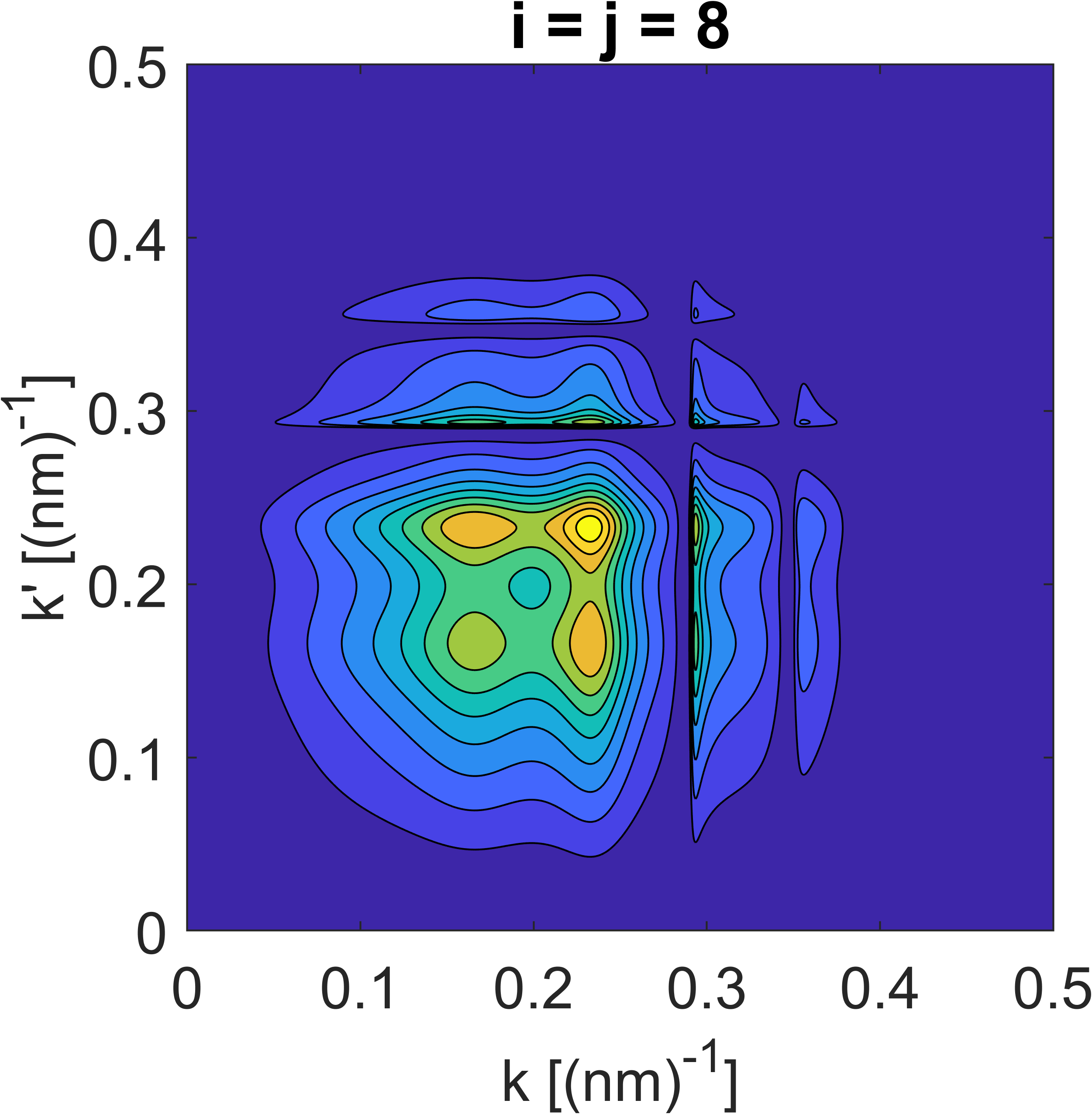}
\end{subfigure}
\newline
\newline
\begin{subfigure}{.235\textwidth}
  \includegraphics[width=\linewidth]{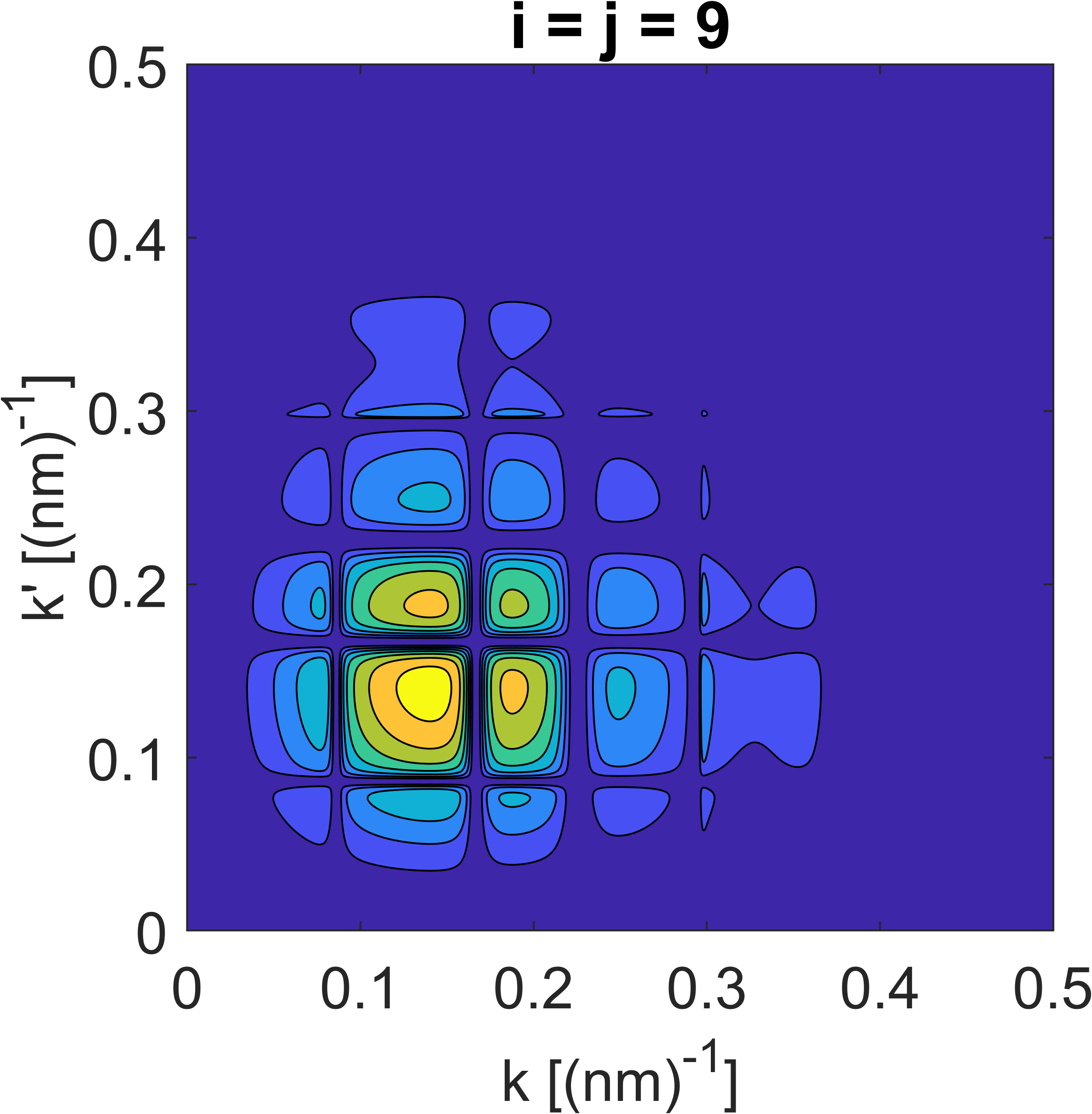}
\end{subfigure}%
\begin{subfigure}{.235\textwidth}
  \includegraphics[width=\linewidth]{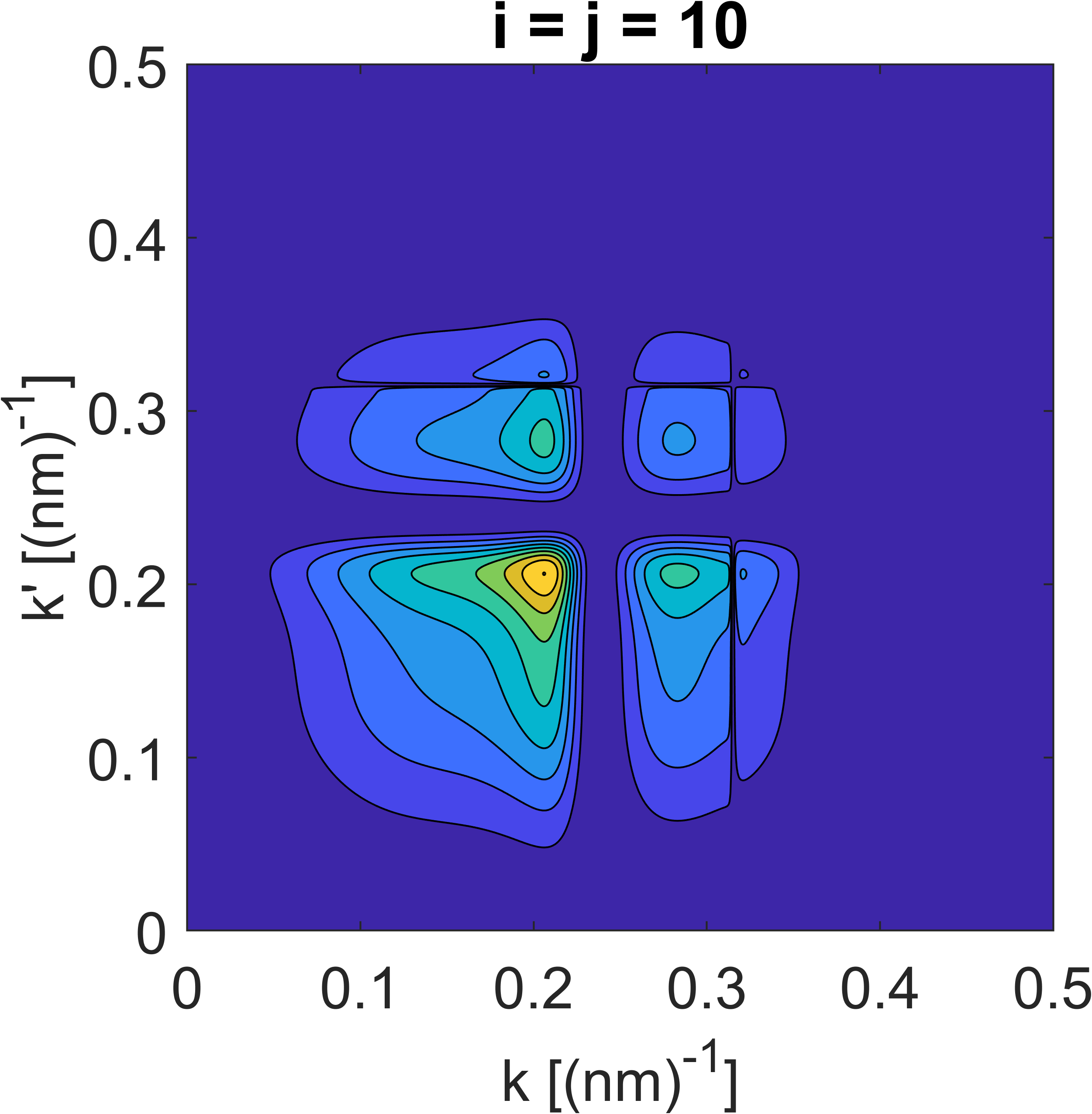}
\end{subfigure}
\begin{subfigure}{.235\textwidth}
  \includegraphics[width=\linewidth]{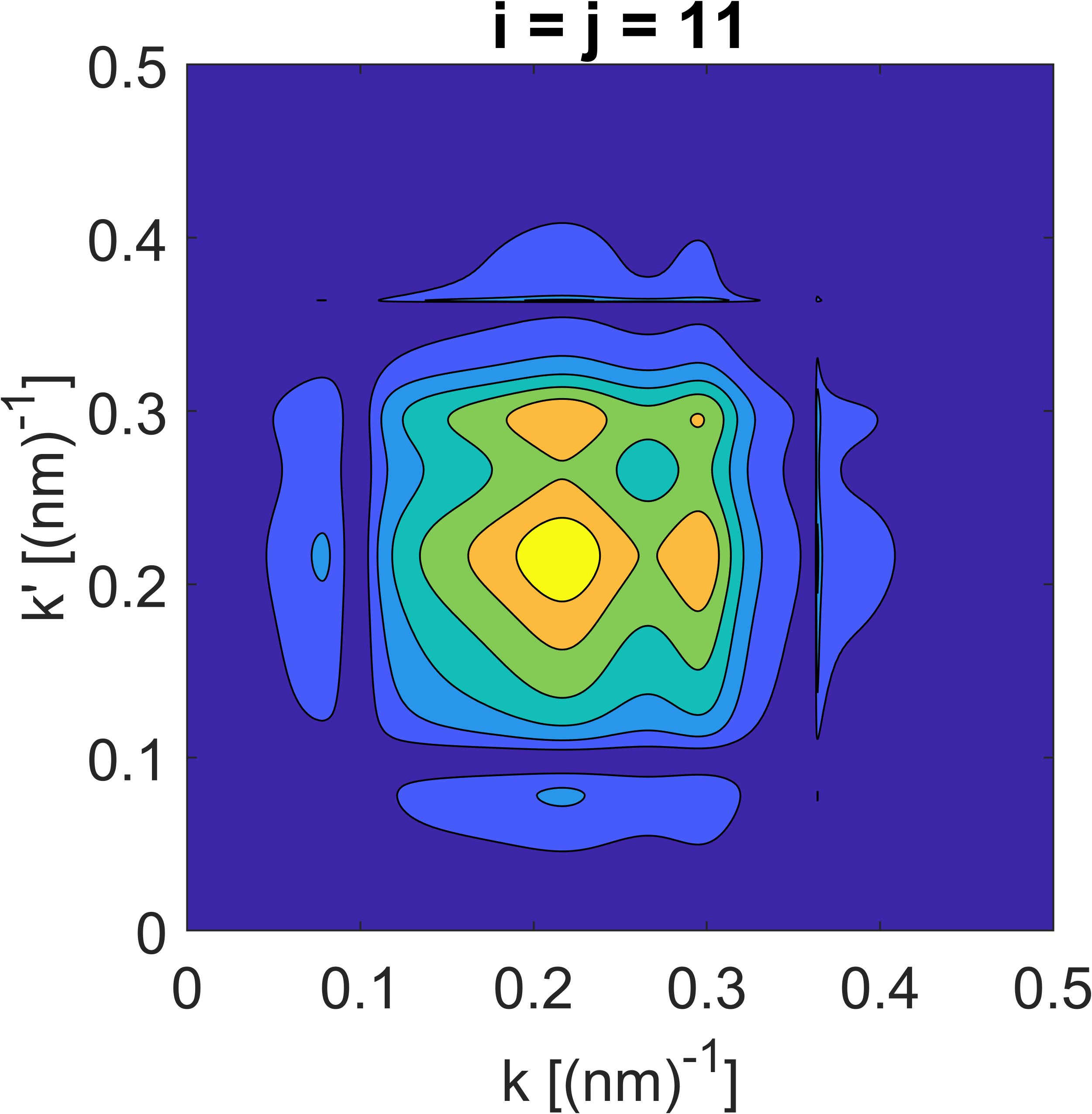}
\end{subfigure}
\begin{subfigure}{.235\textwidth}
  \includegraphics[width=\linewidth]{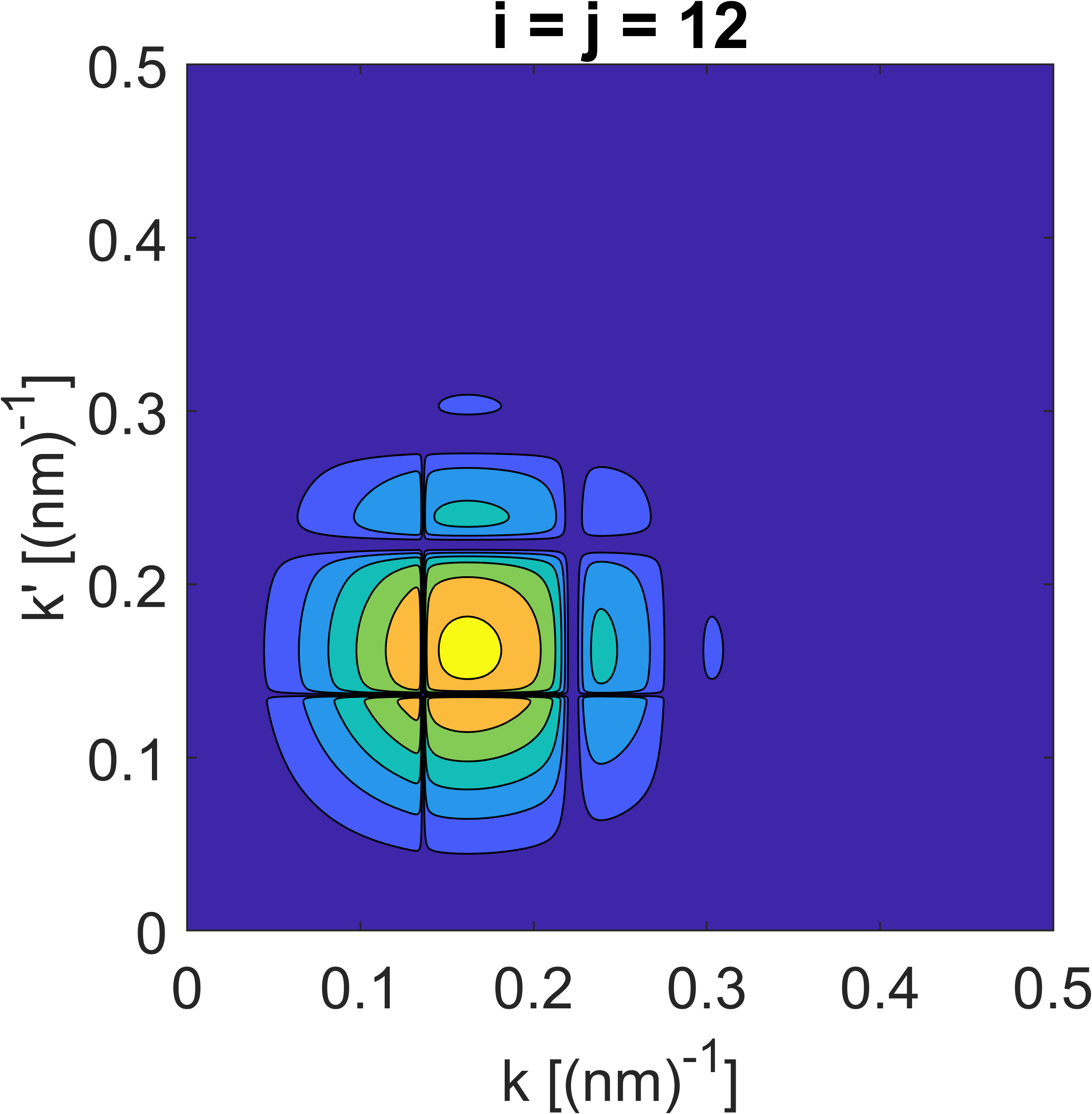}
\end{subfigure}
\newline
\newline
\begin{subfigure}{.235\textwidth}
  \includegraphics[width=\linewidth]{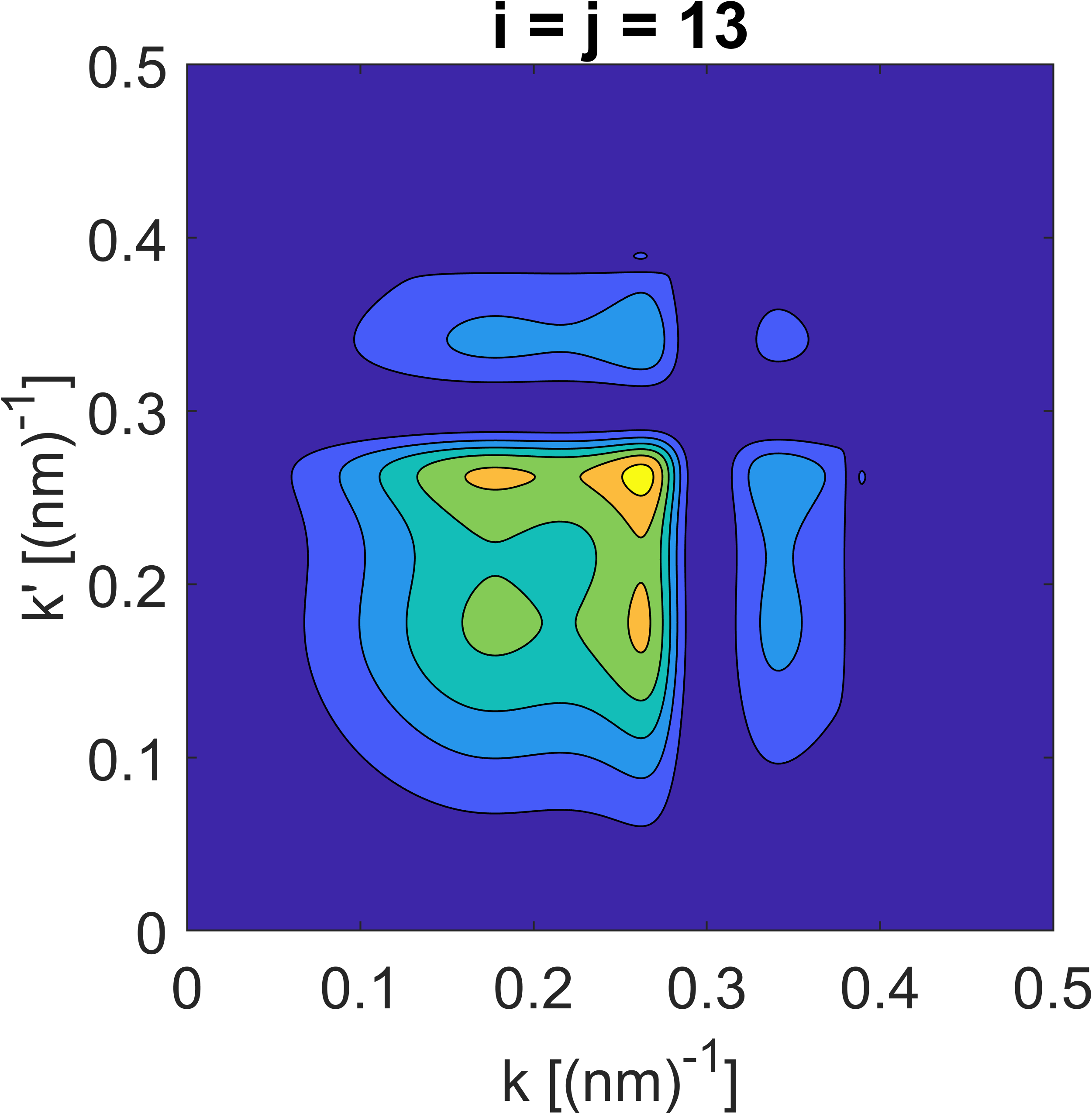}
\end{subfigure}%
\begin{subfigure}{.235\textwidth}
  \includegraphics[width=\linewidth]{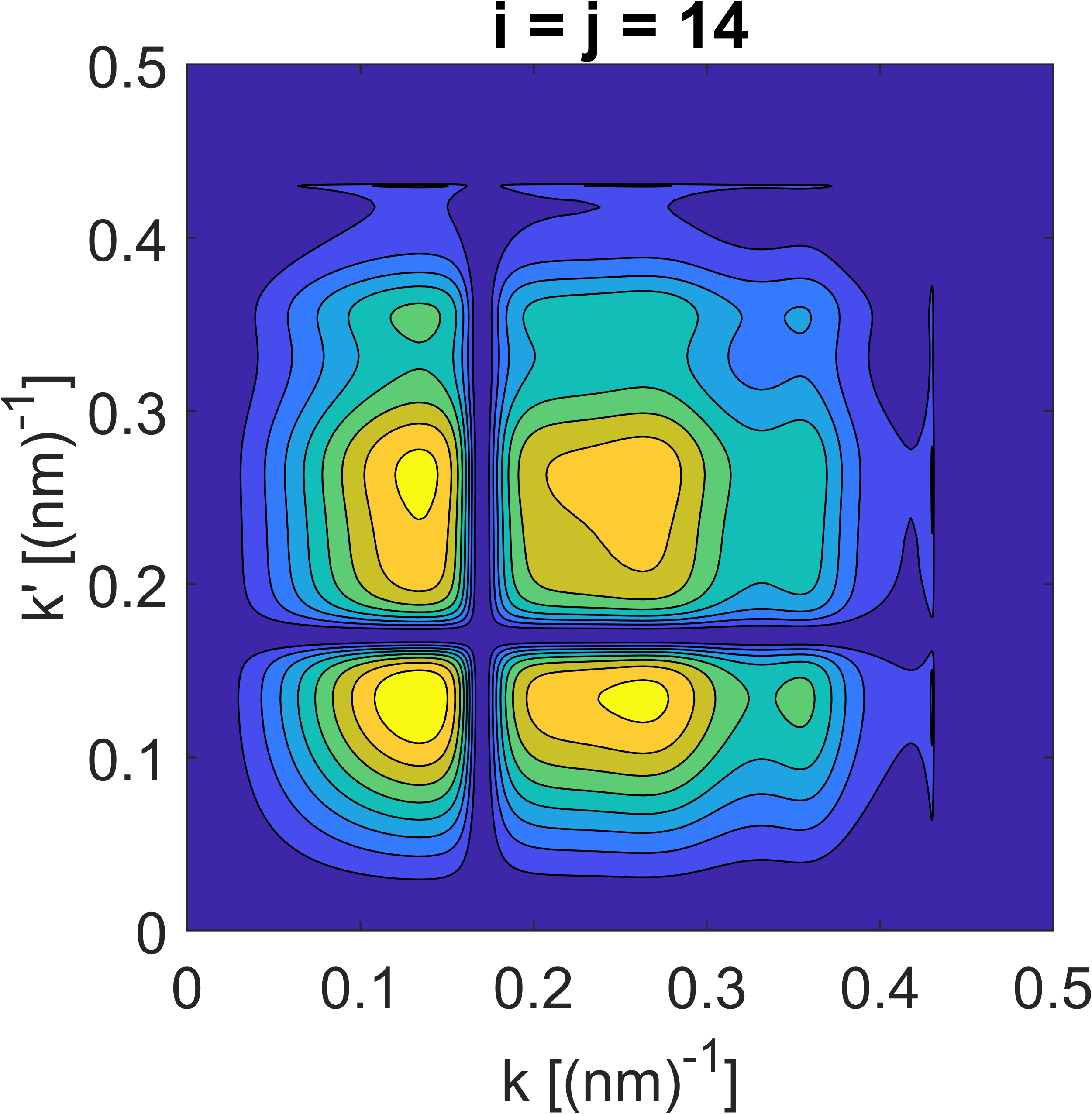}
\end{subfigure}
\begin{subfigure}{.235\textwidth}
  \includegraphics[width=\linewidth]{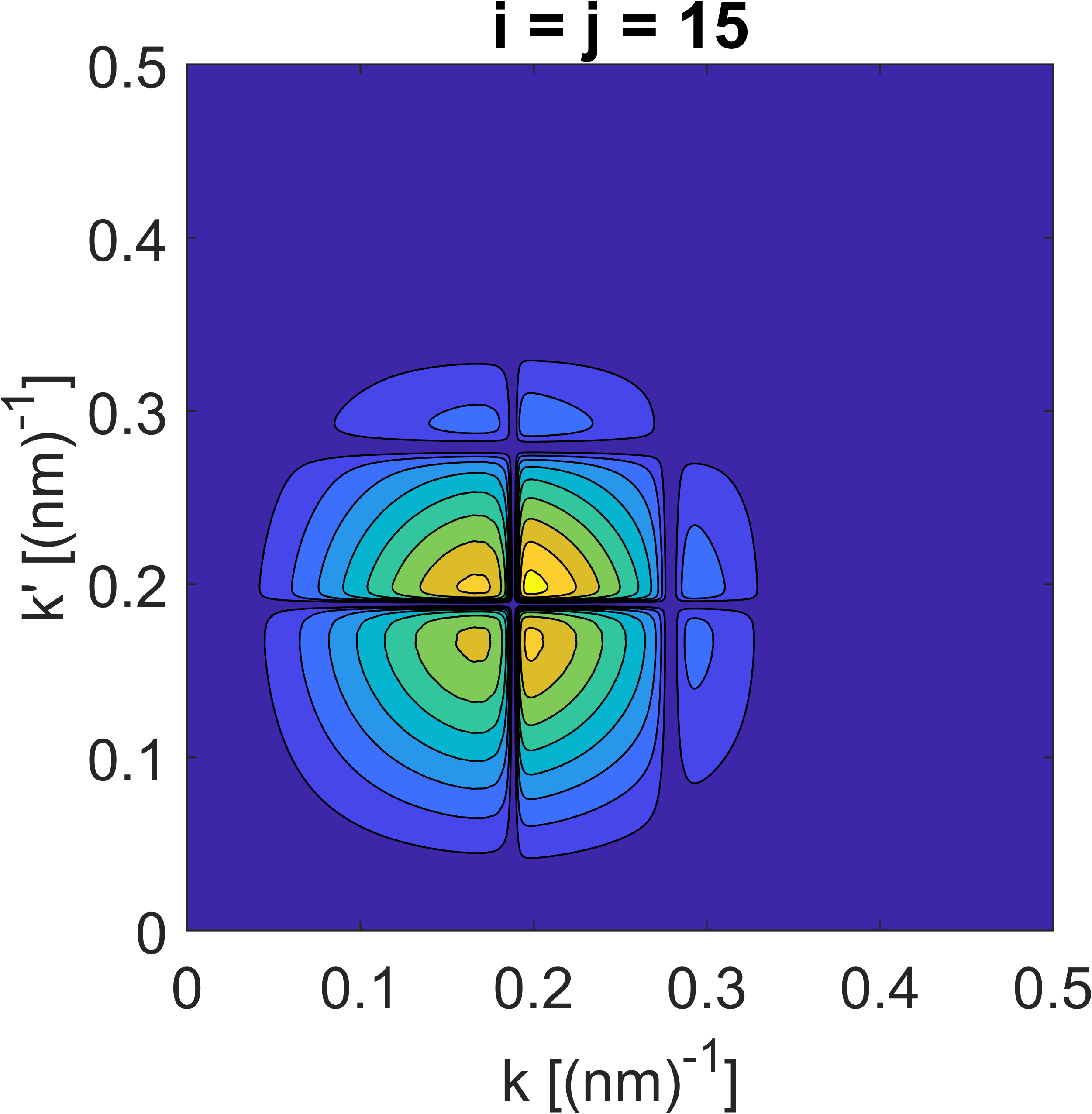}
\end{subfigure}
\begin{subfigure}{.235\textwidth}
  \includegraphics[width=\linewidth]{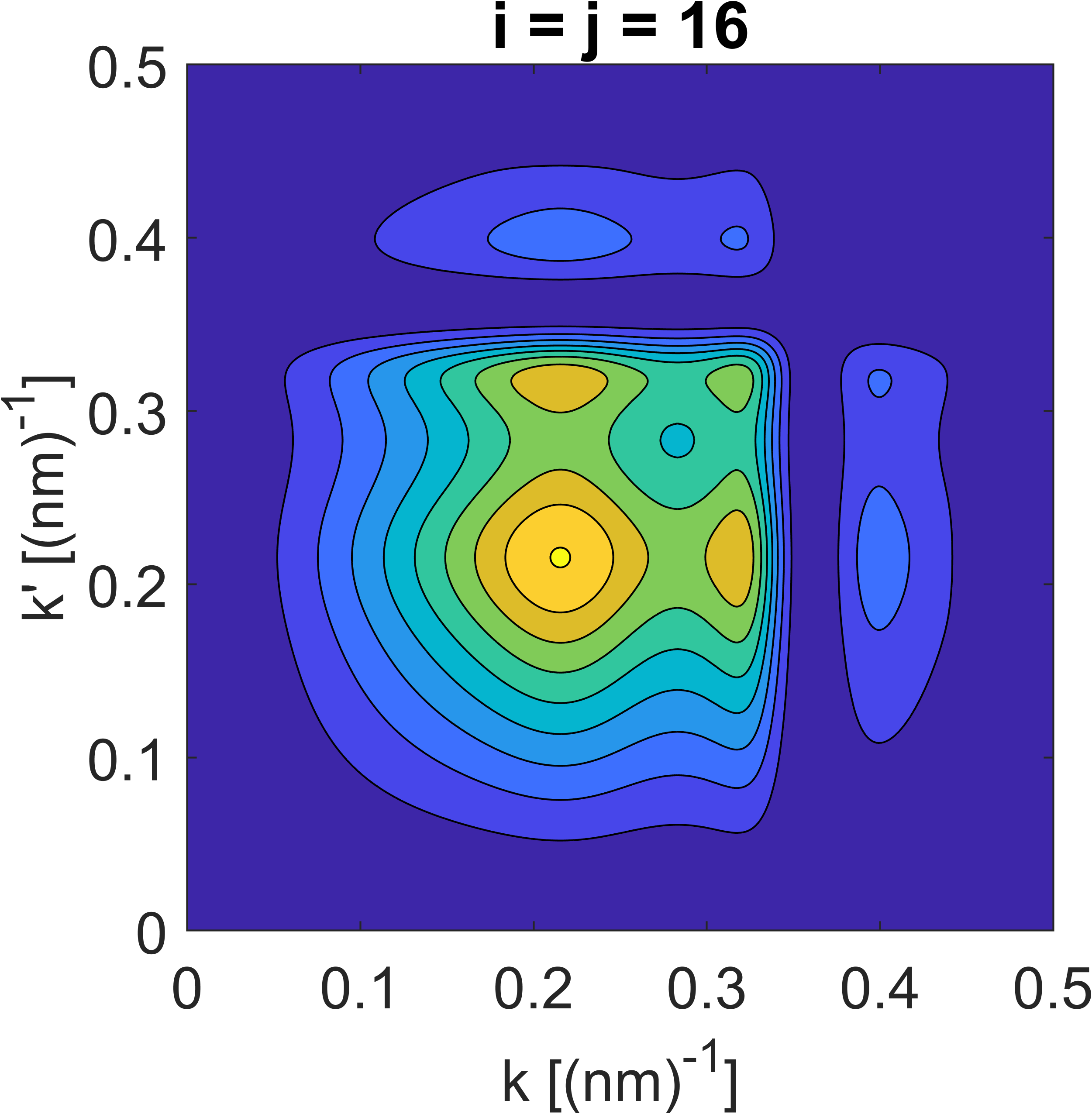}
\end{subfigure}
\newline
\newline
\begin{subfigure}{.235\textwidth}
  \includegraphics[width=\linewidth]{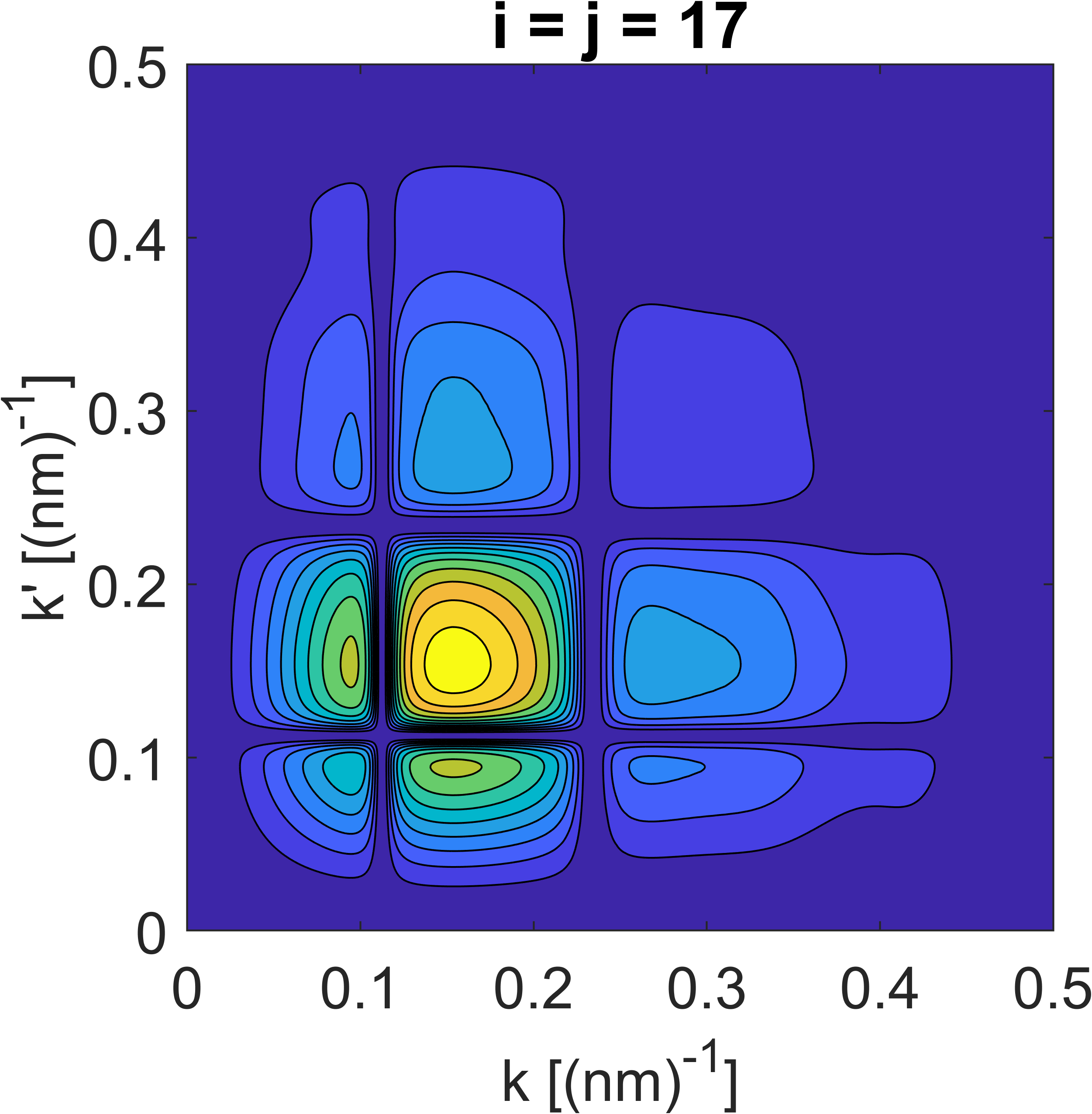}
\end{subfigure}%
\begin{subfigure}{.235\textwidth}
  \includegraphics[width=\linewidth]{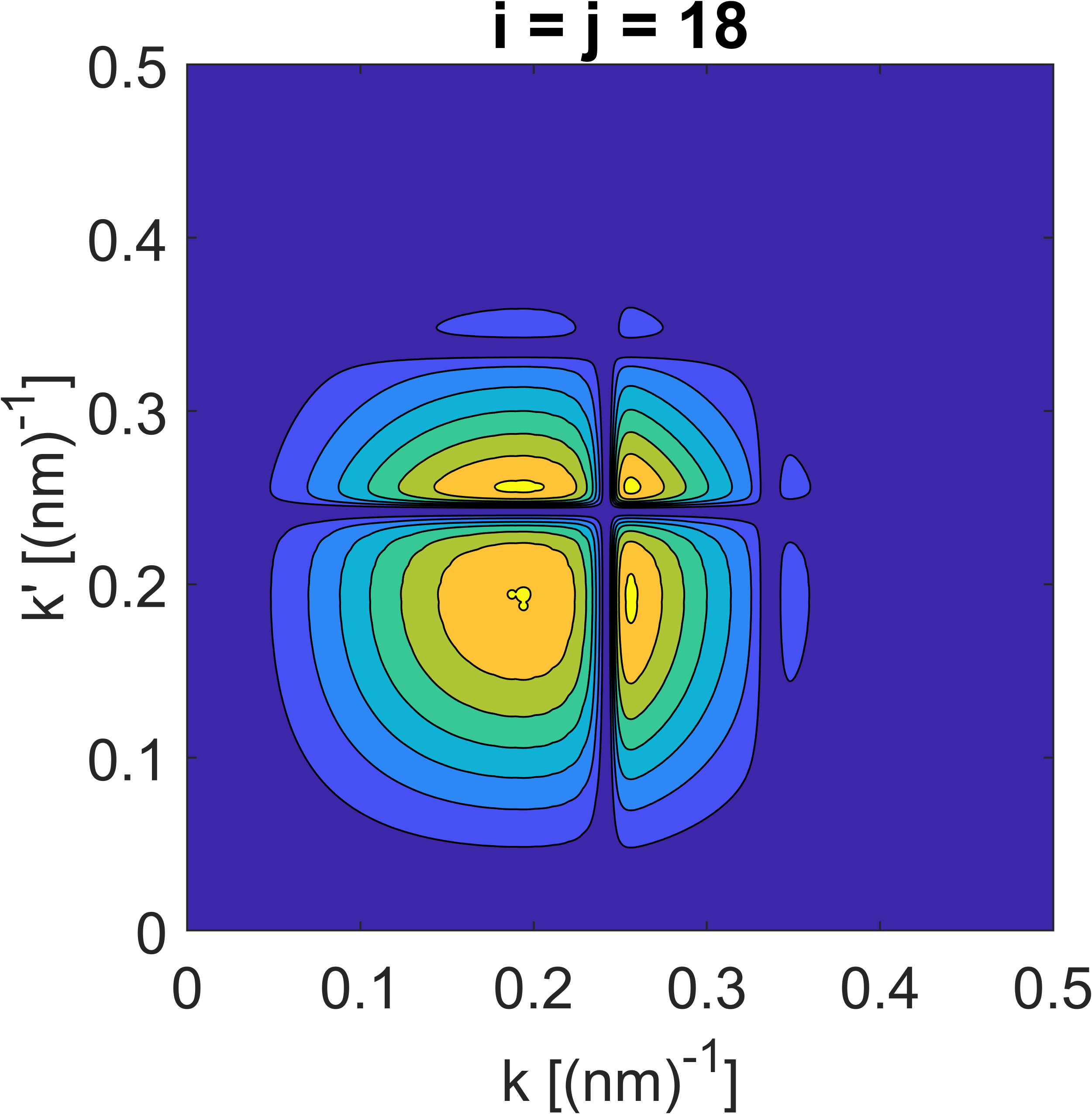}
\end{subfigure}
\begin{subfigure}{.235\textwidth}
  \includegraphics[width=\linewidth]{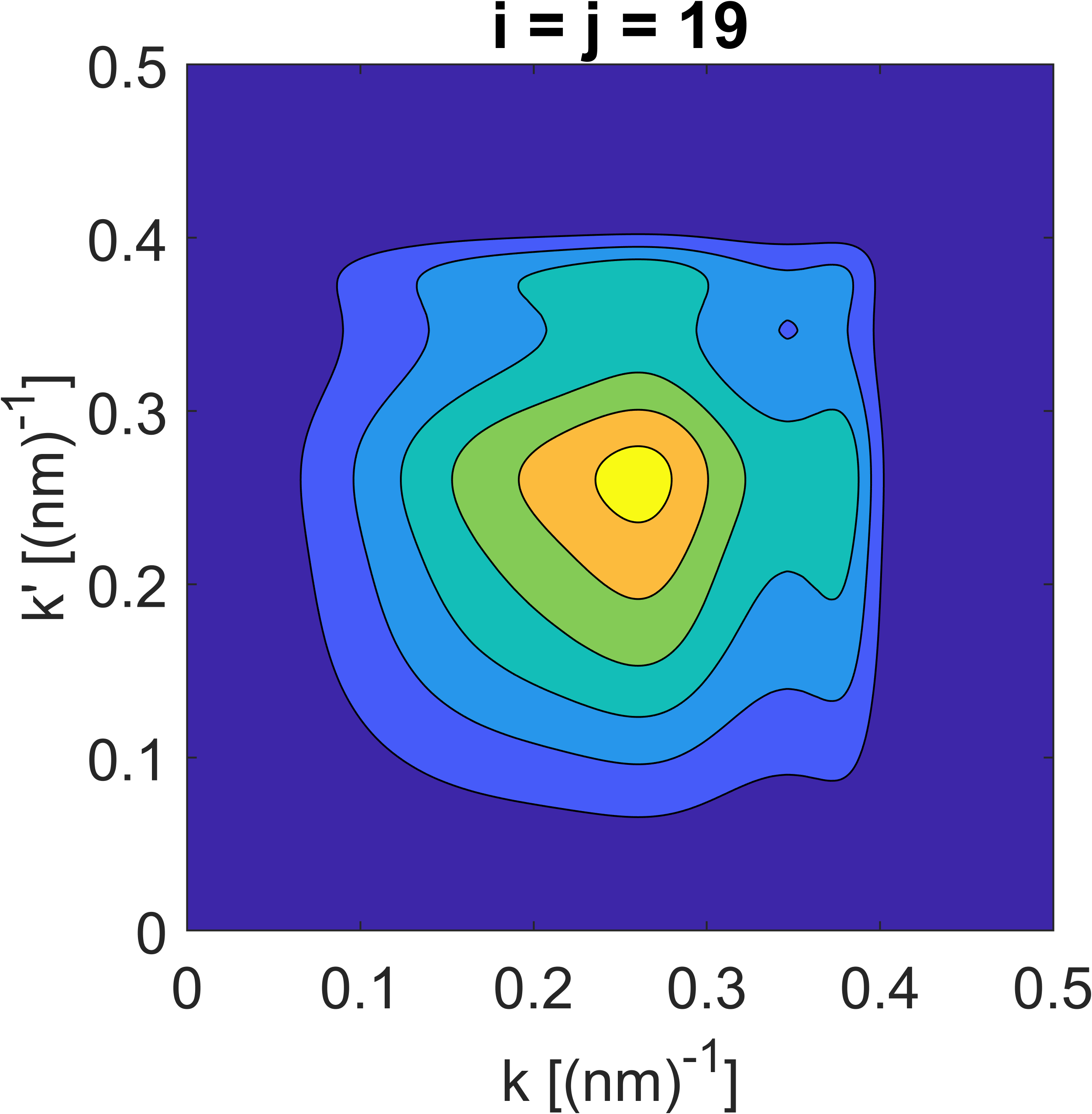}
\end{subfigure}
\begin{subfigure}{.235\textwidth}
  \includegraphics[width=\linewidth]{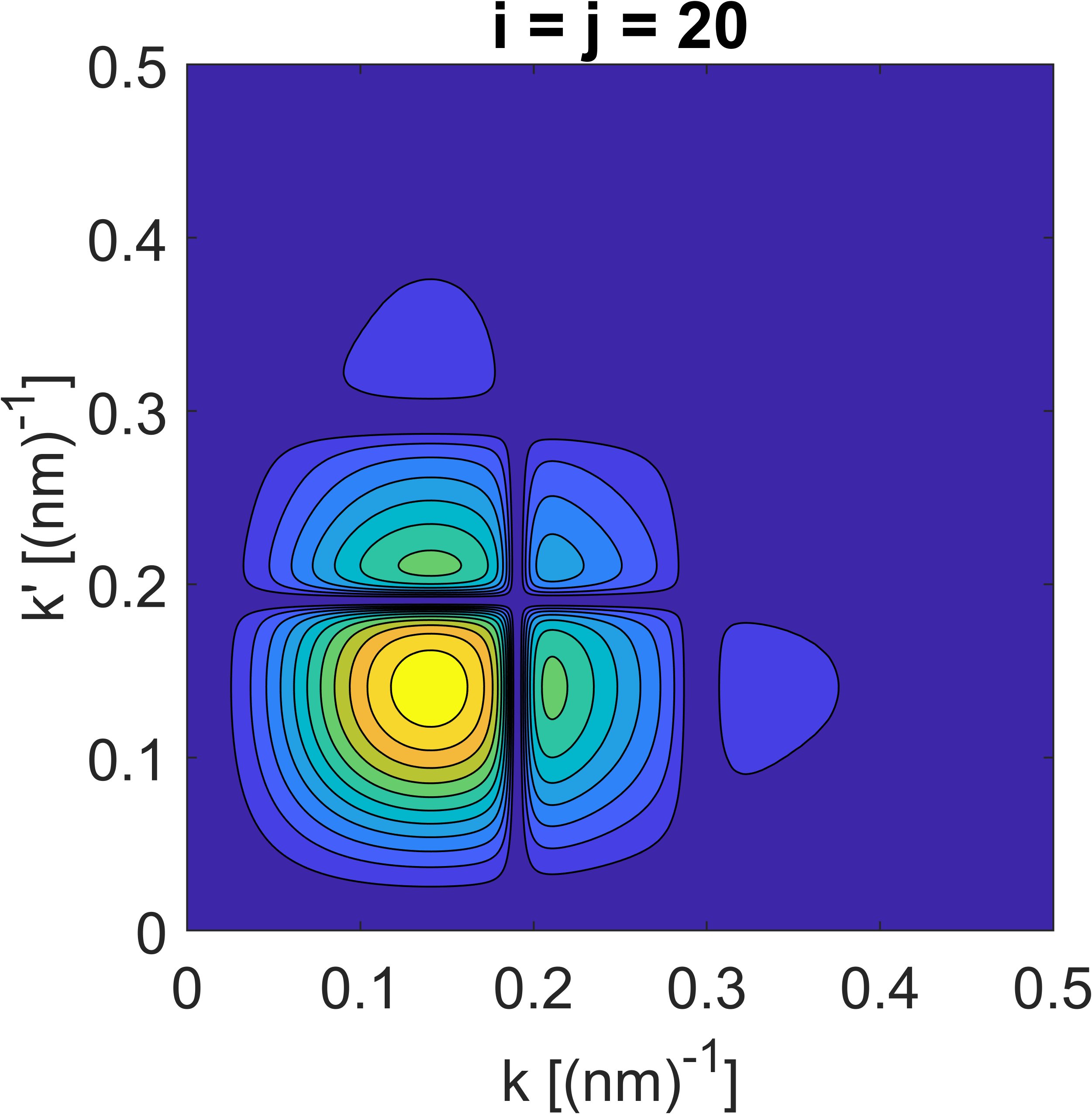}
\end{subfigure}
\caption{Calculation of the quadratic exciton-phonon coupling term $|f_{i,j,\mathbf{k,k'}}|^{2}$ for the first 20 vibration modes, with $i = j$. Here, the nanowire radius $R=25\ \mathrm{nm}$. Modes 1 to 4 are shown with an inset that offers a magnified view in the vicinity of the contour plot's maximum value. Plots are normalized: dark blue denotes 0, and yellow denotes 1 (arbitrary units).} 
\label{fig:coolplots_25nm_1}
\end{figure}

\newpage

\begin{figure}[H]
\ContinuedFloat 
\centering
\begin{subfigure}{.235\textwidth}
  \includegraphics[width=\linewidth]{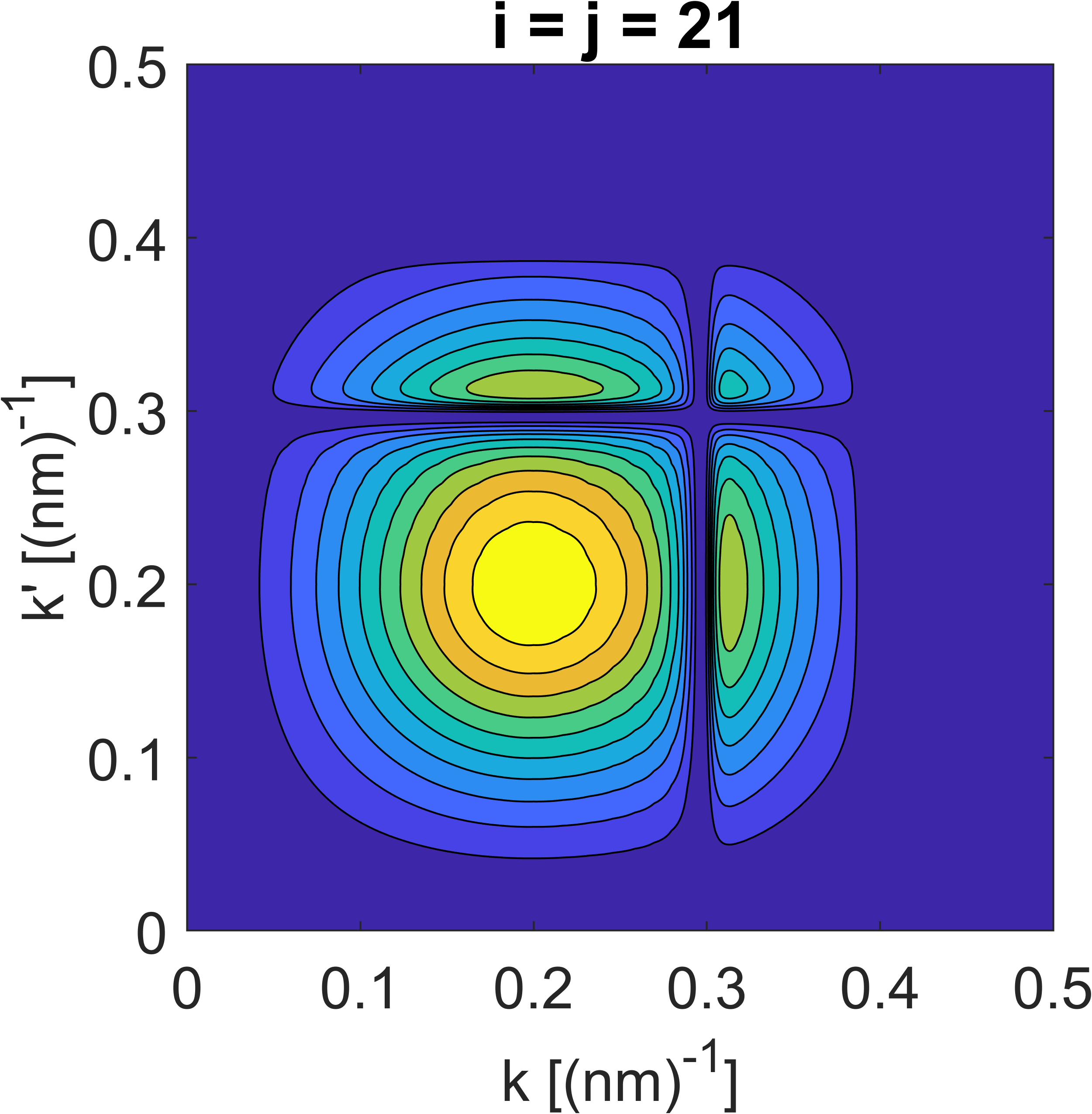}
\end{subfigure}%
\begin{subfigure}{.235\textwidth}
  \includegraphics[width=\linewidth]{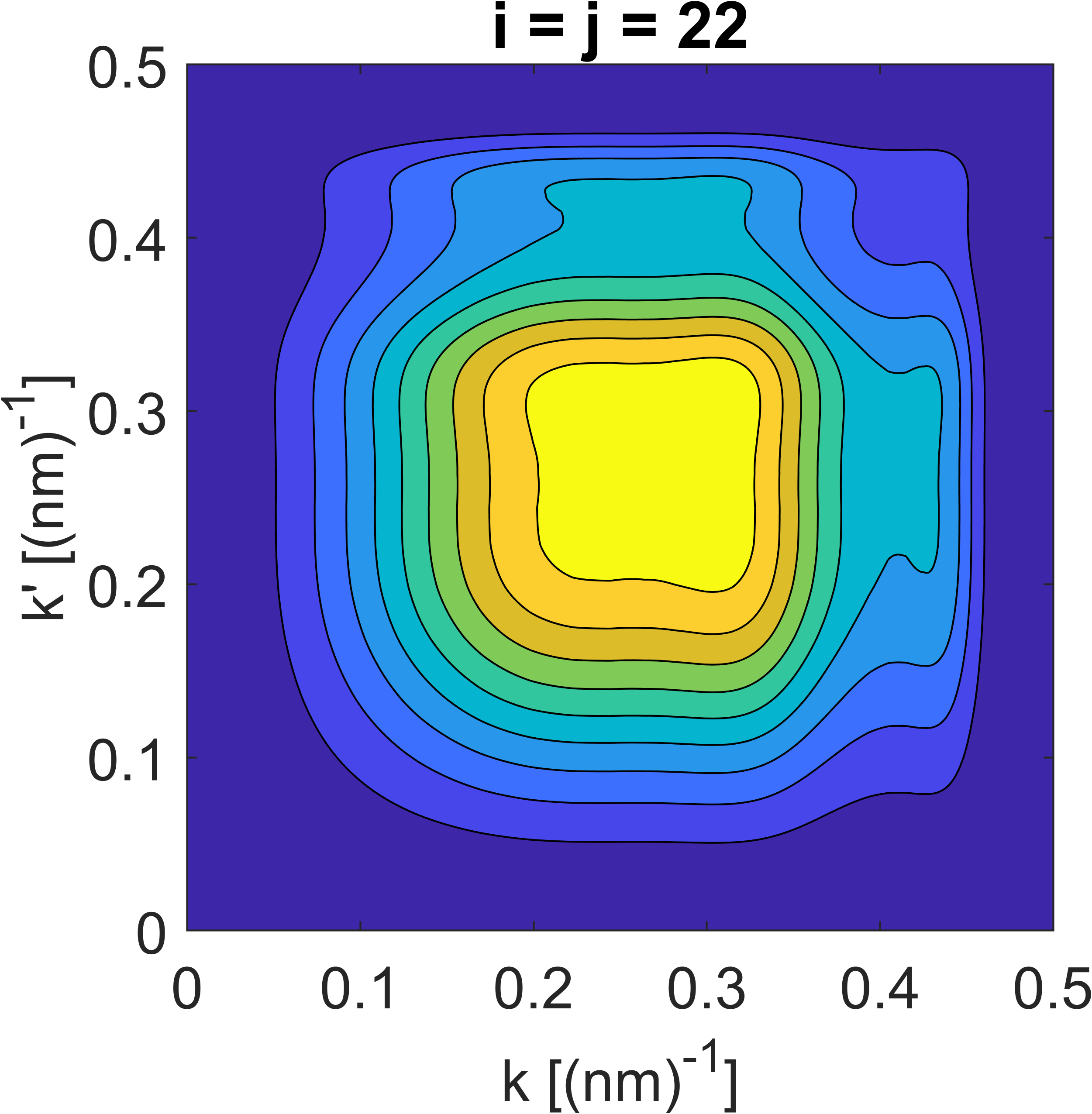}
\end{subfigure}
\begin{subfigure}{.235\textwidth}
  \includegraphics[width=\linewidth]{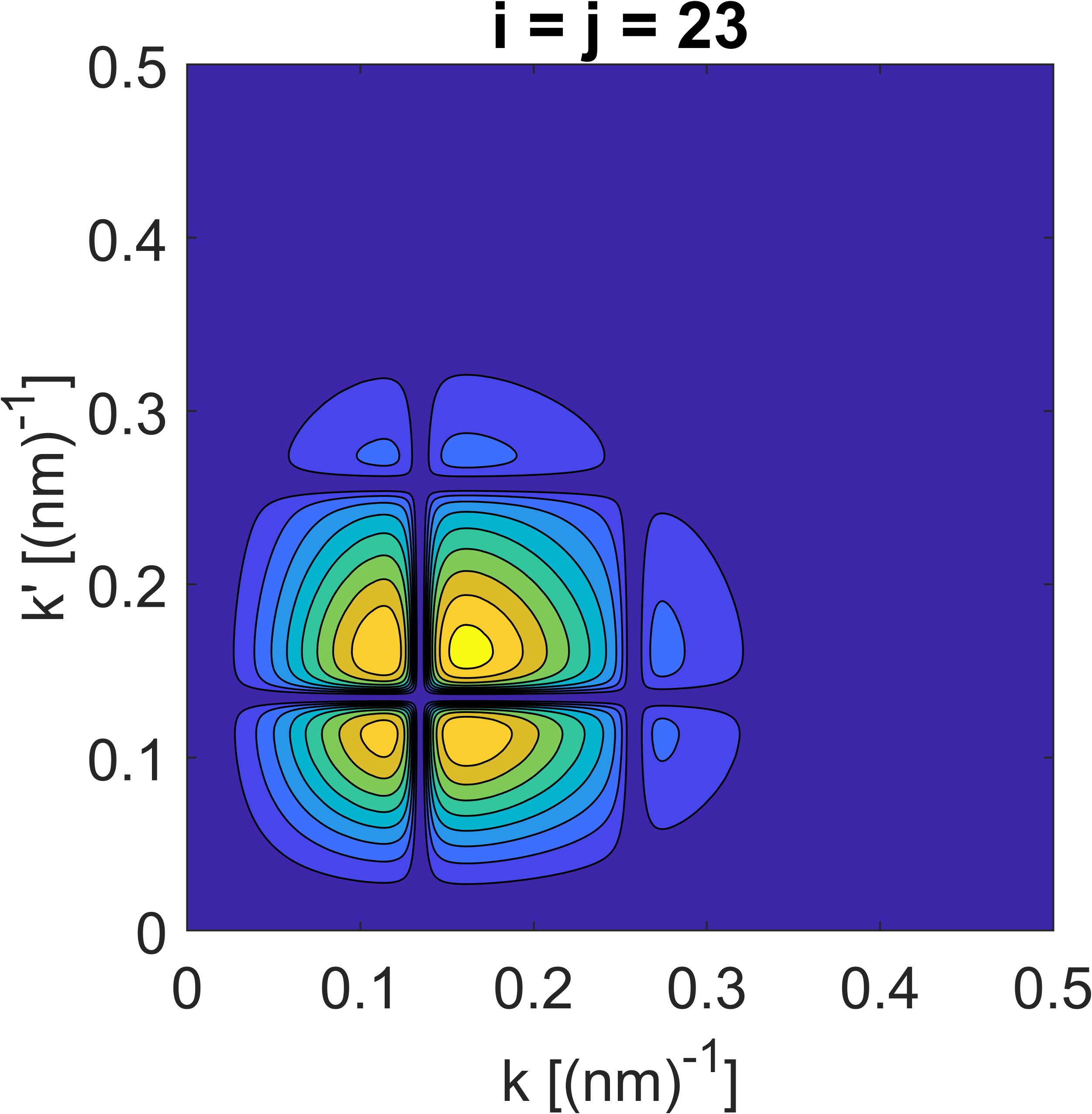}
\end{subfigure}
\begin{subfigure}{.235\textwidth}
  \includegraphics[width=\linewidth]{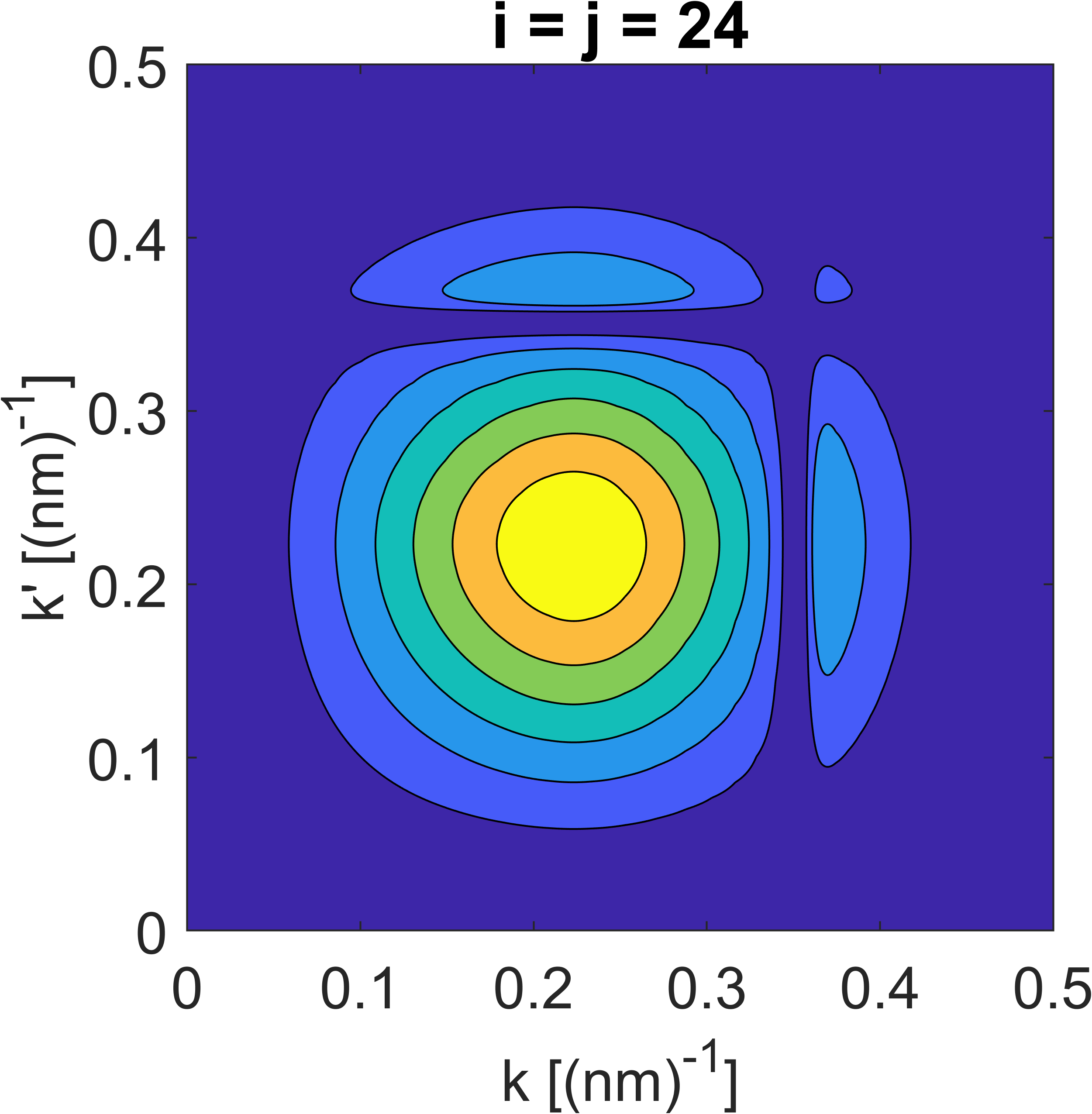}
\end{subfigure}
\newline
\newline
\begin{subfigure}{.235\textwidth}
  \includegraphics[width=\linewidth]{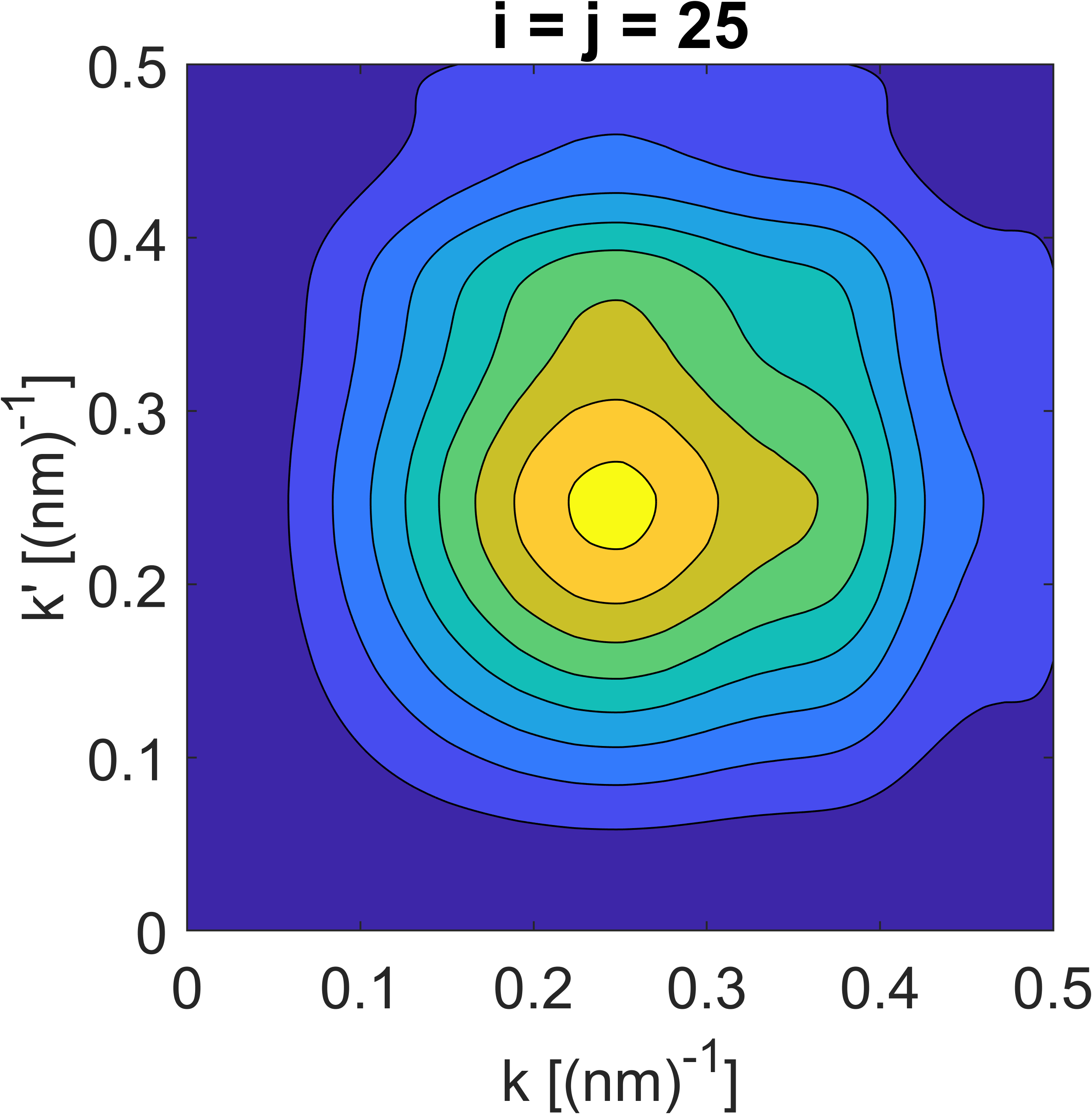}
\end{subfigure}%
\begin{subfigure}{.235\textwidth}
  \includegraphics[width=\linewidth]{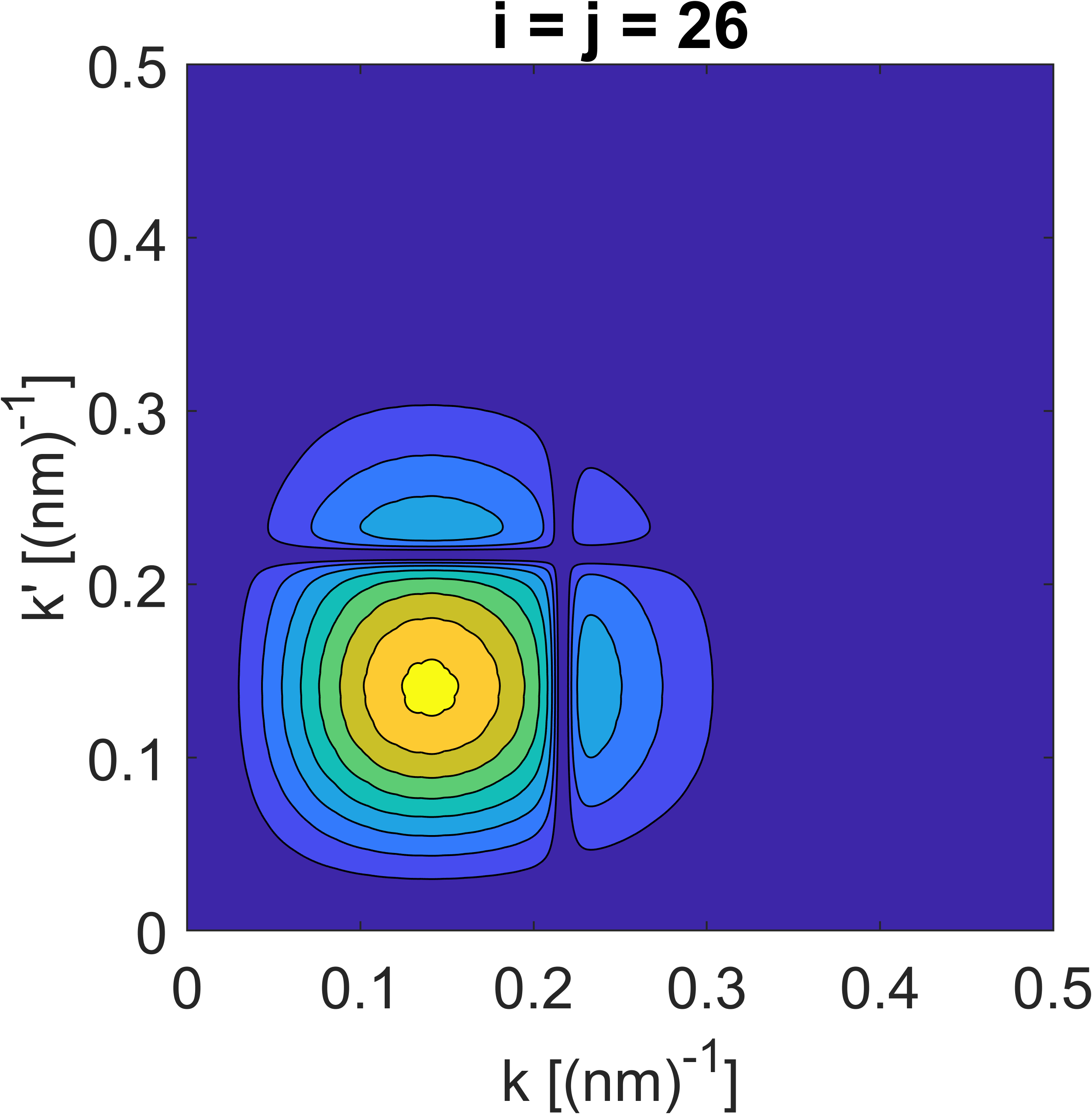}
\end{subfigure}
\begin{subfigure}{.235\textwidth}
  \includegraphics[width=\linewidth]{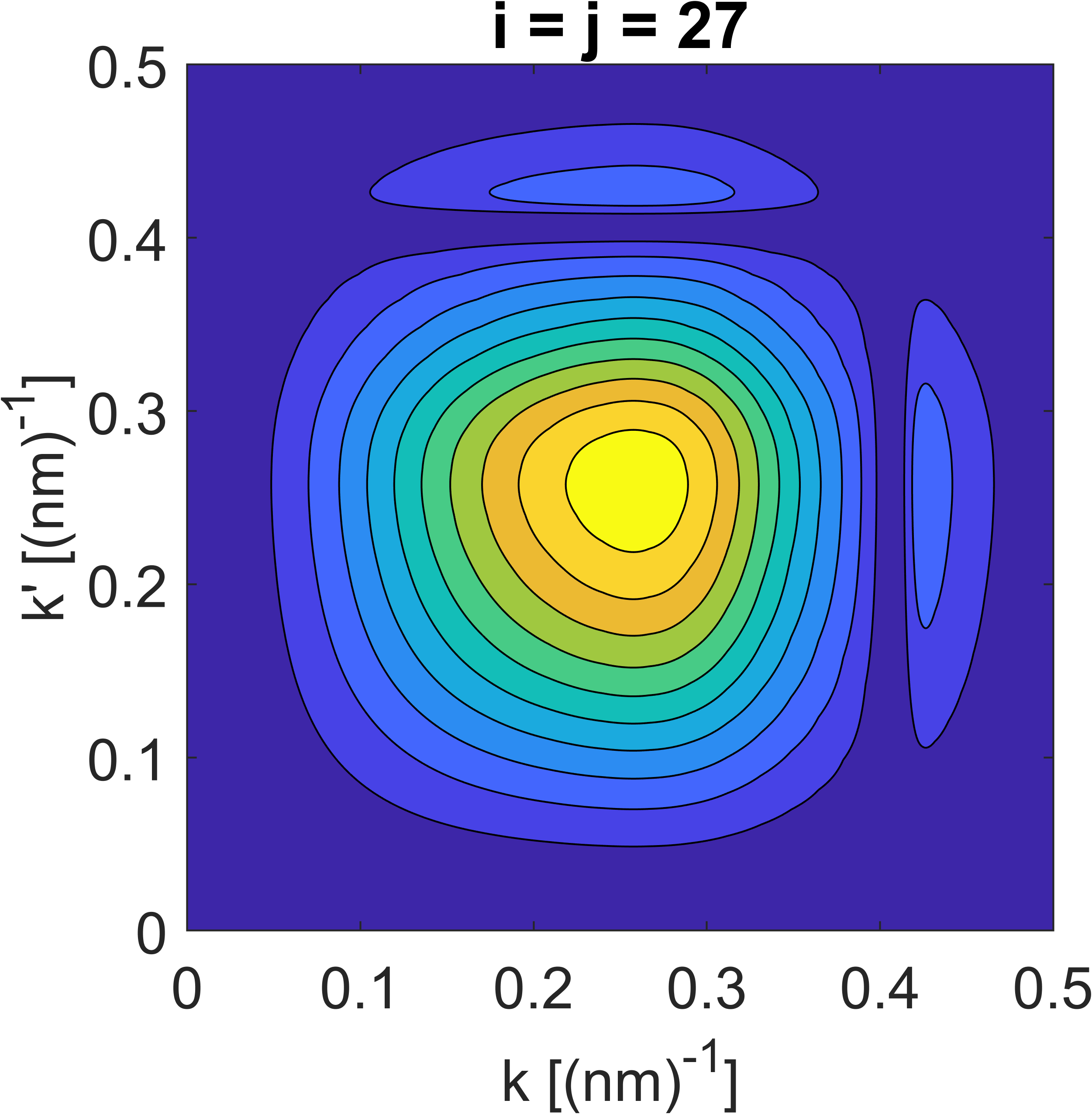}
\end{subfigure}
\begin{subfigure}{.235\textwidth}
  \includegraphics[width=\linewidth]{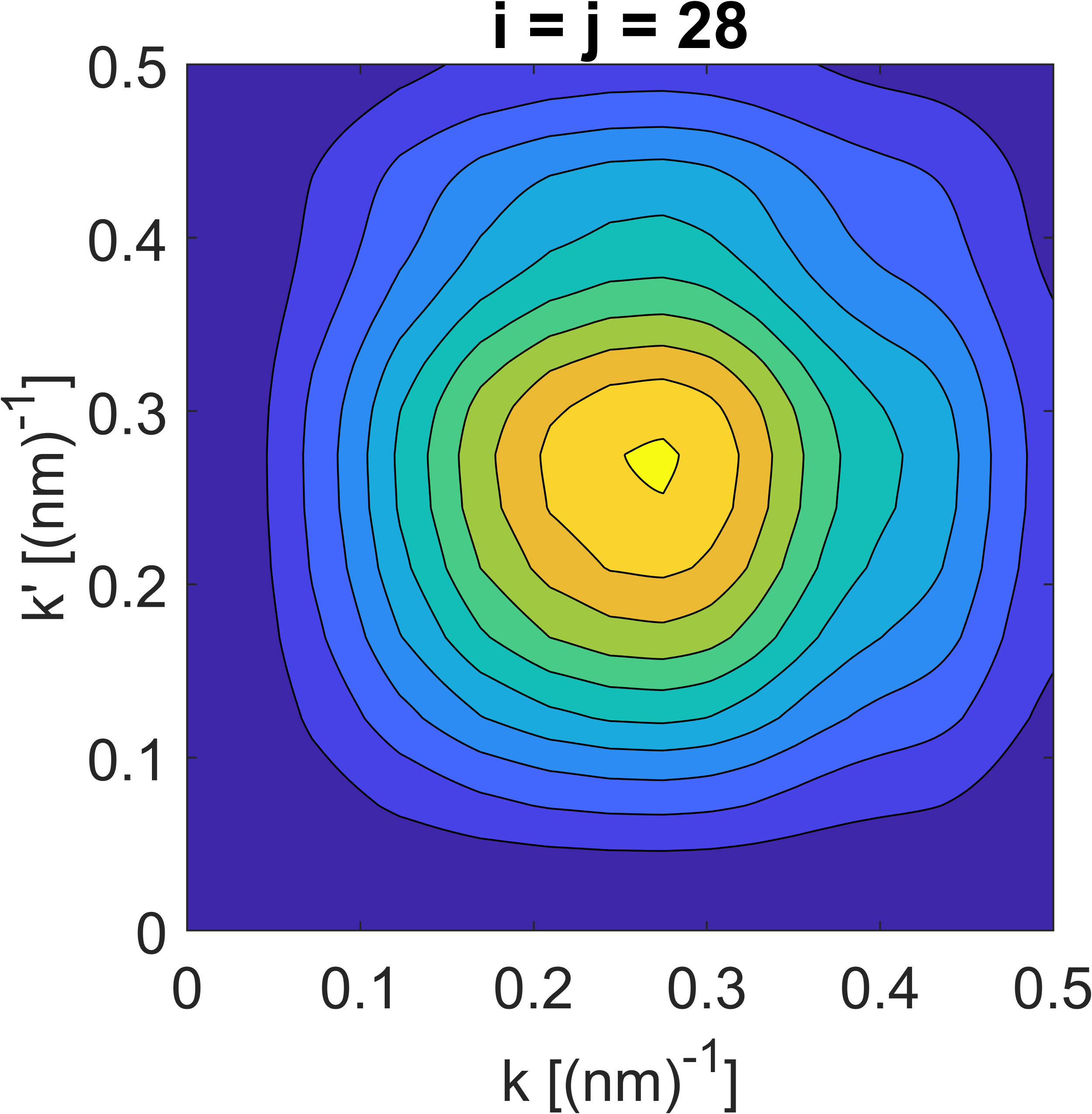}
\end{subfigure}
\newline
\newline
\begin{subfigure}{.235\textwidth}
  \includegraphics[width=\linewidth]{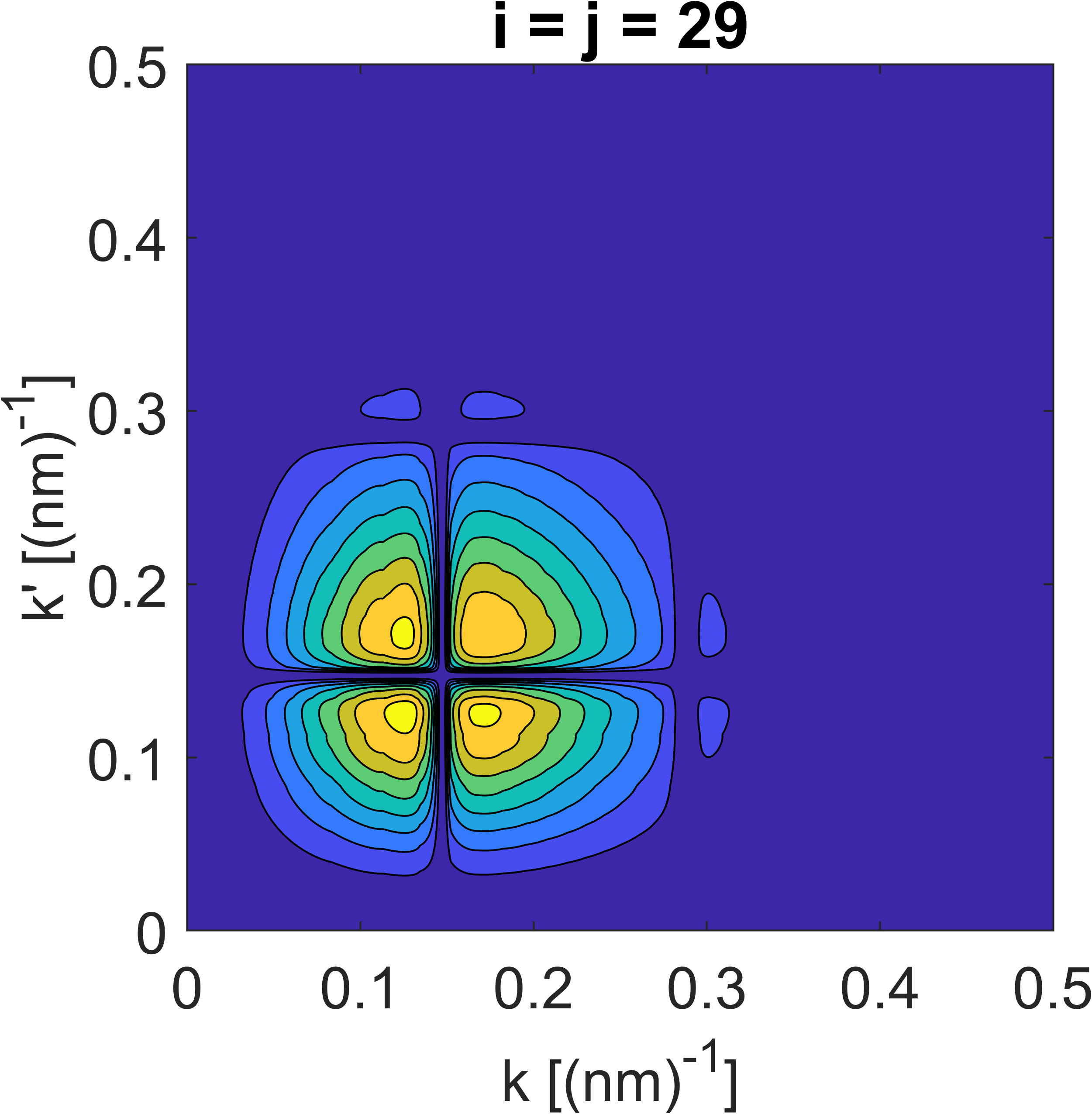}
\end{subfigure}%
\begin{subfigure}{.235\textwidth}
  \includegraphics[width=\linewidth]{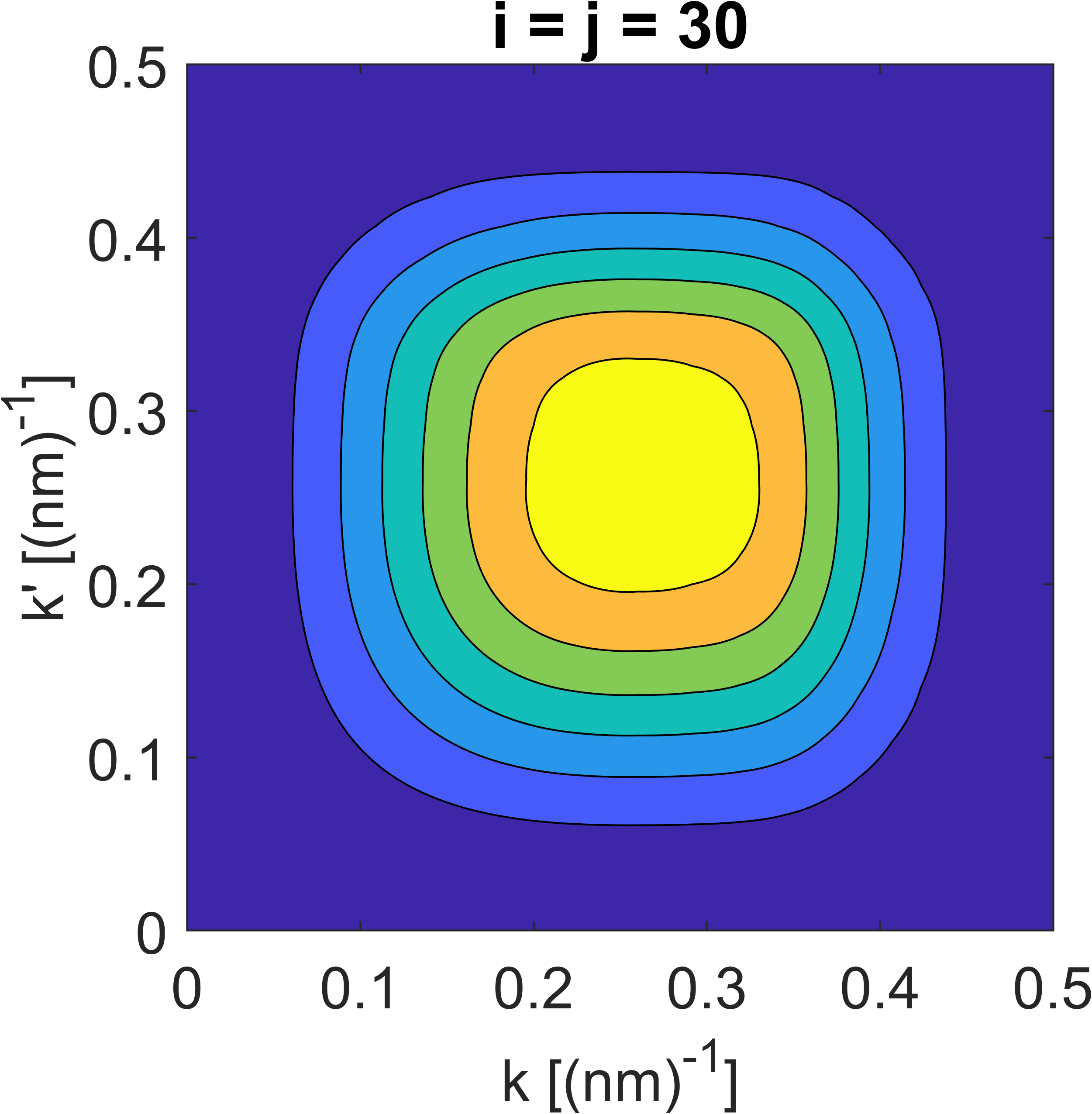}
\end{subfigure}
\begin{subfigure}{.235\textwidth}
  \includegraphics[width=\linewidth]{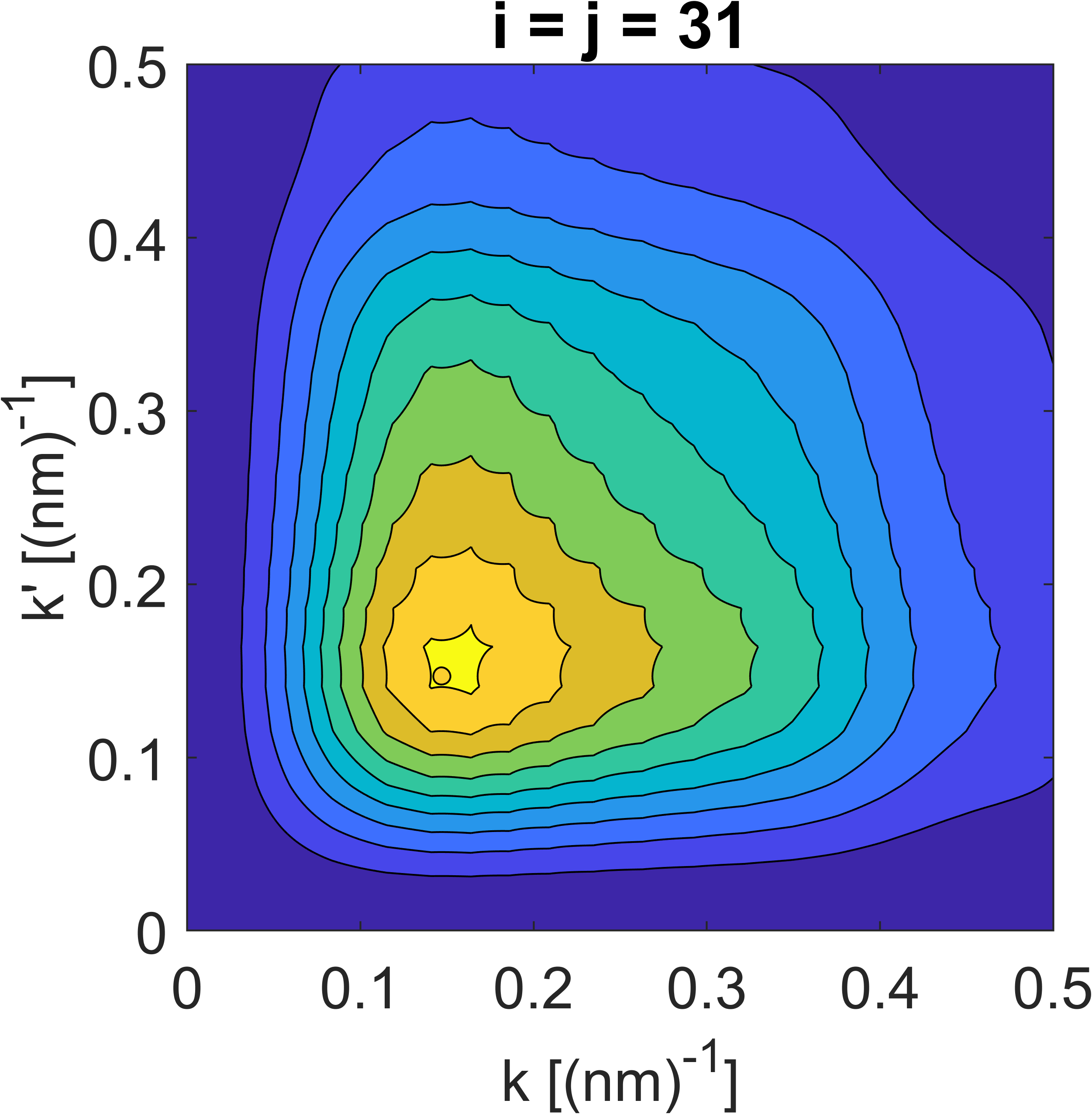}
\end{subfigure}
\begin{subfigure}{.235\textwidth}
  \includegraphics[width=\linewidth]{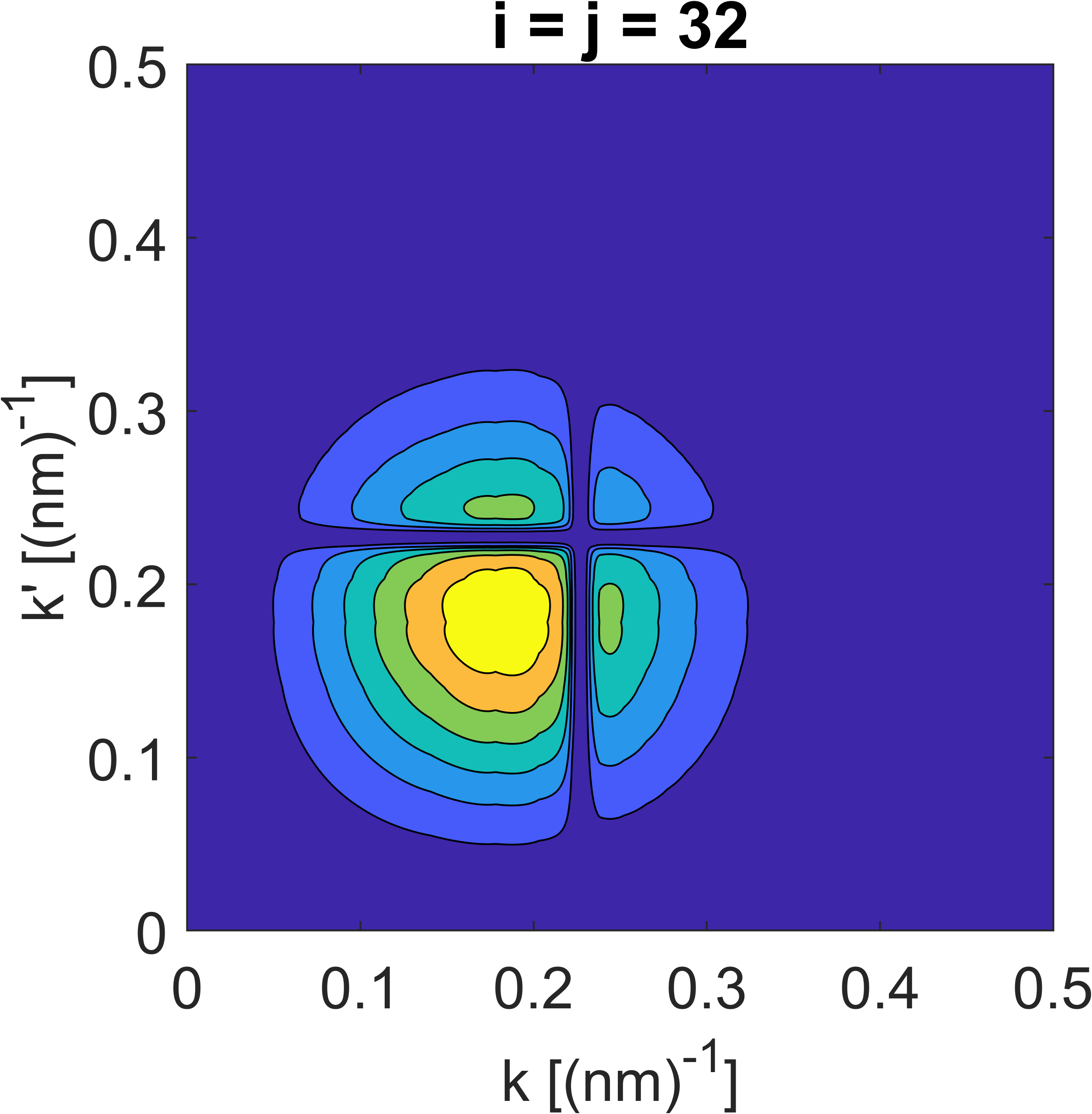}
\end{subfigure}
\newline
\newline
\begin{subfigure}{.235\textwidth}
  \includegraphics[width=\linewidth]{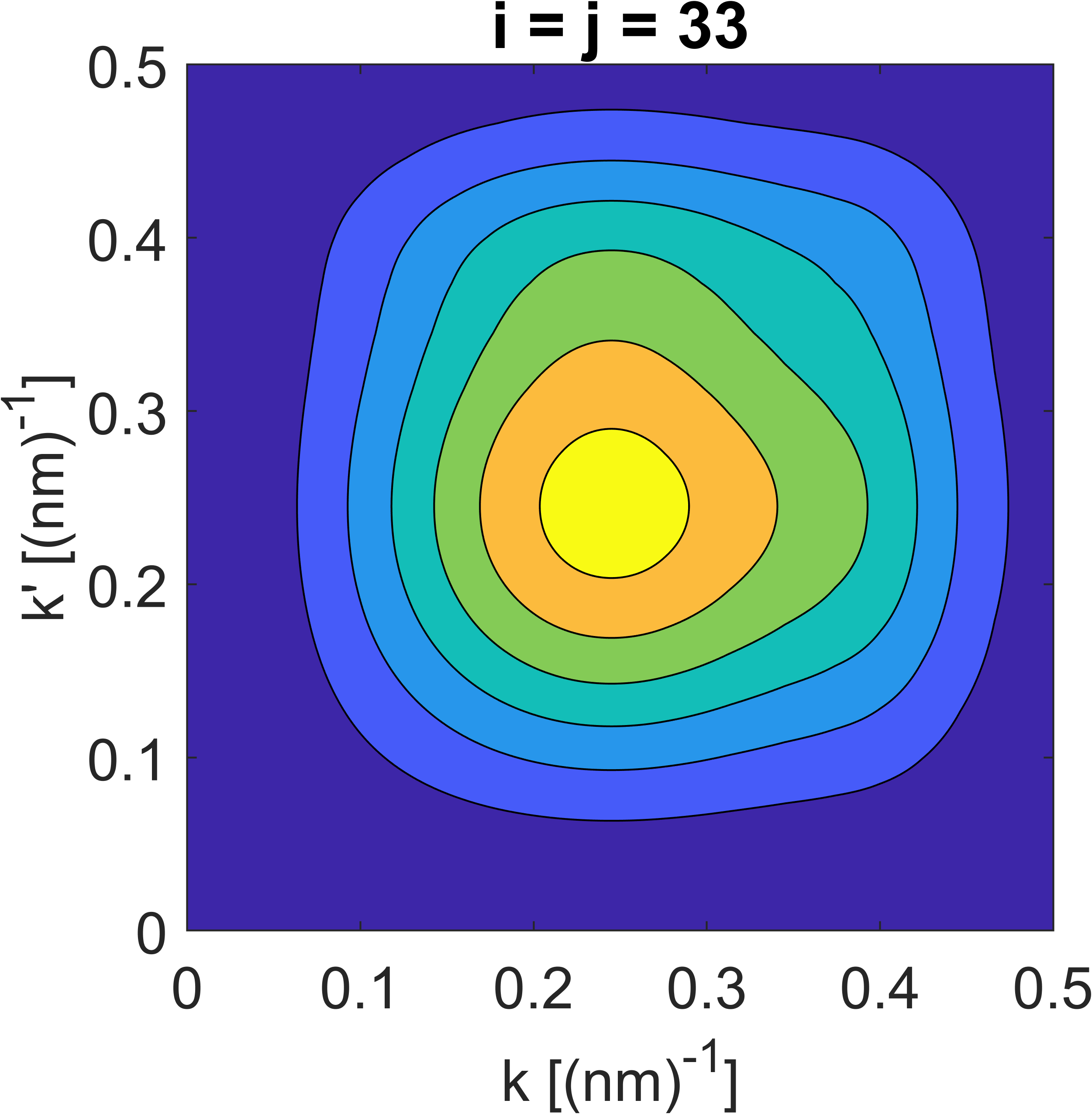}
\end{subfigure}%
\begin{subfigure}{.235\textwidth}
  \includegraphics[width=\linewidth]{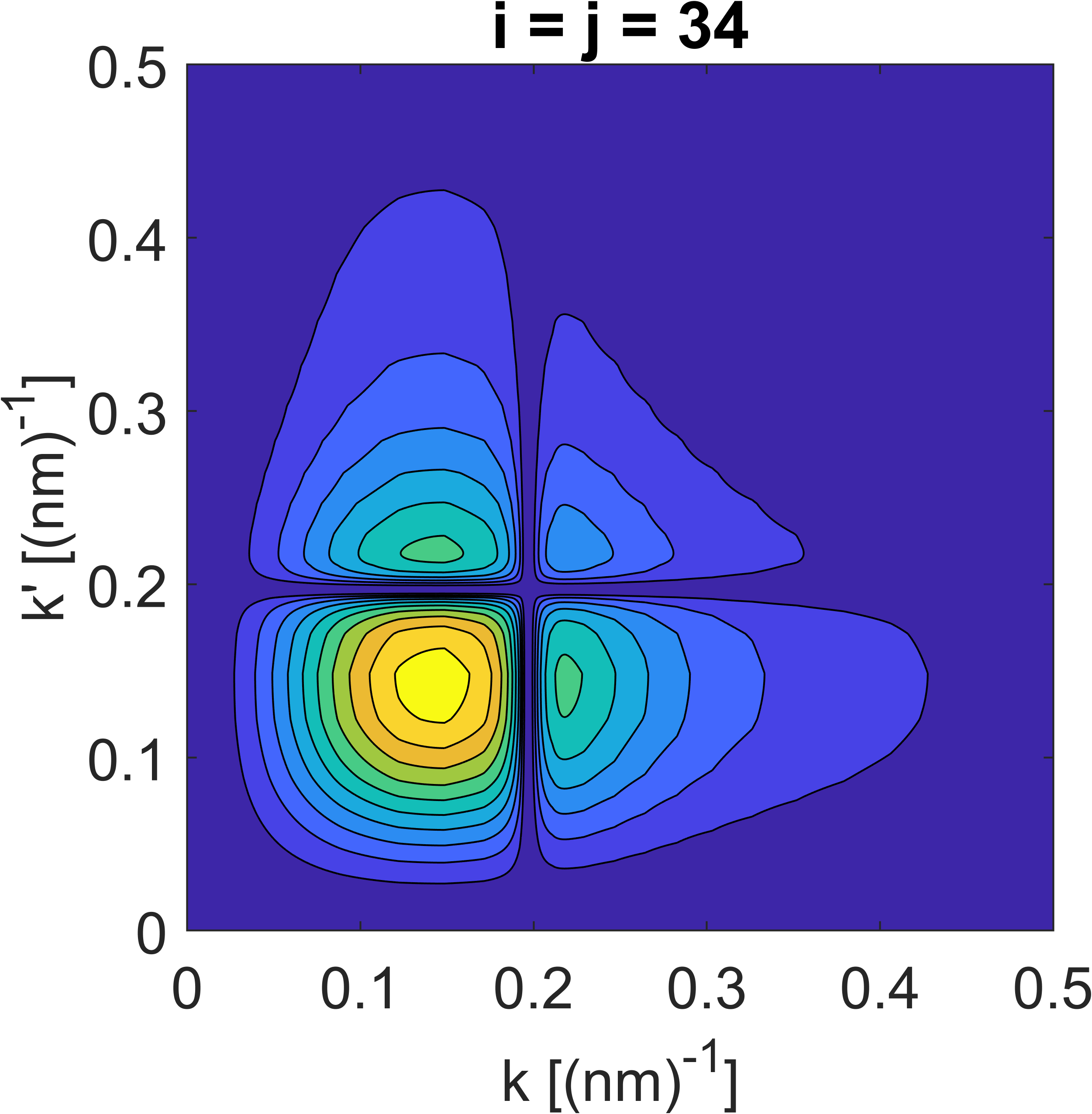}
\end{subfigure}
\begin{subfigure}{.235\textwidth}
  \includegraphics[width=\linewidth]{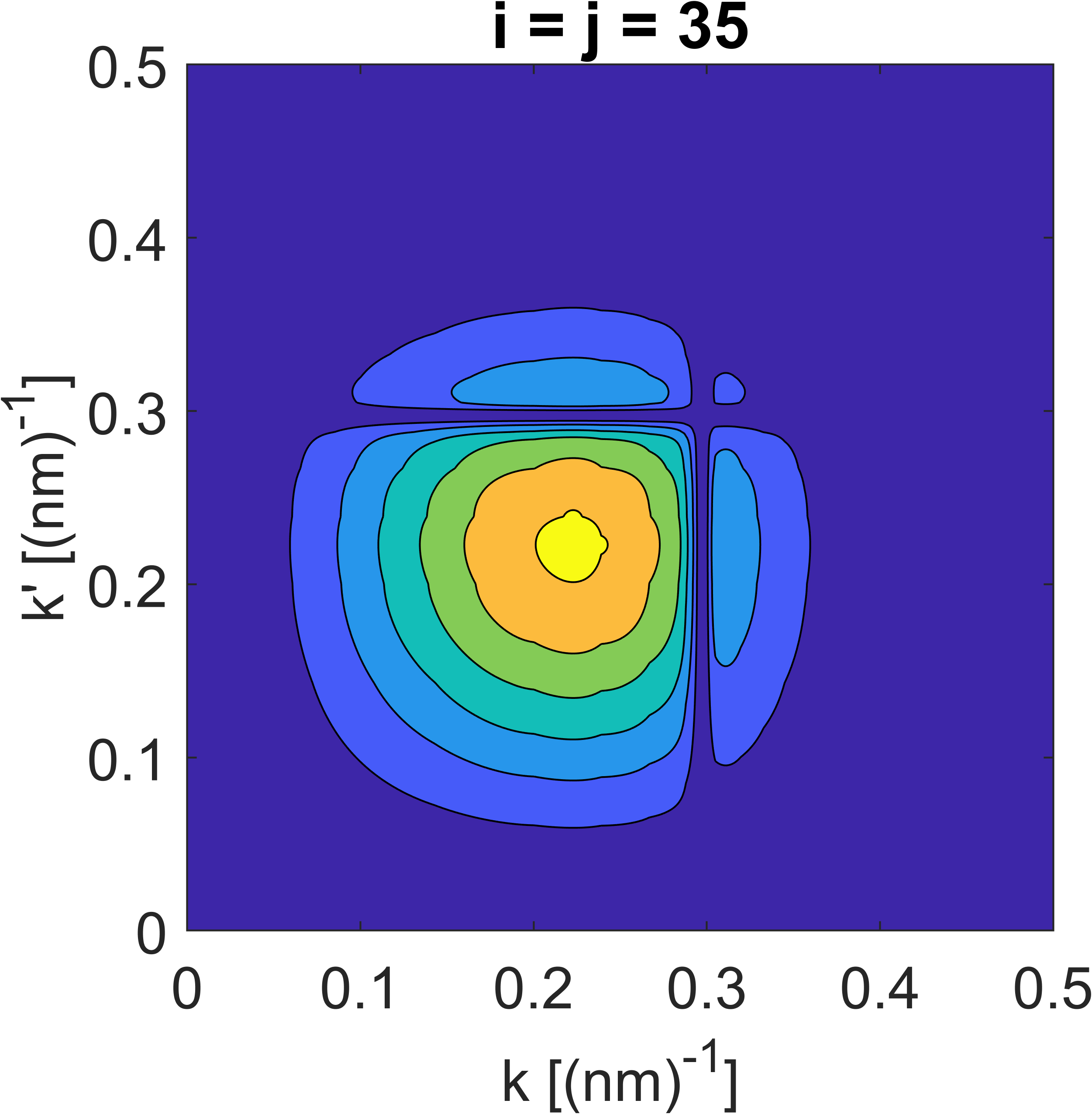}
\end{subfigure}
\begin{subfigure}{.235\textwidth}
  \includegraphics[width=\linewidth]{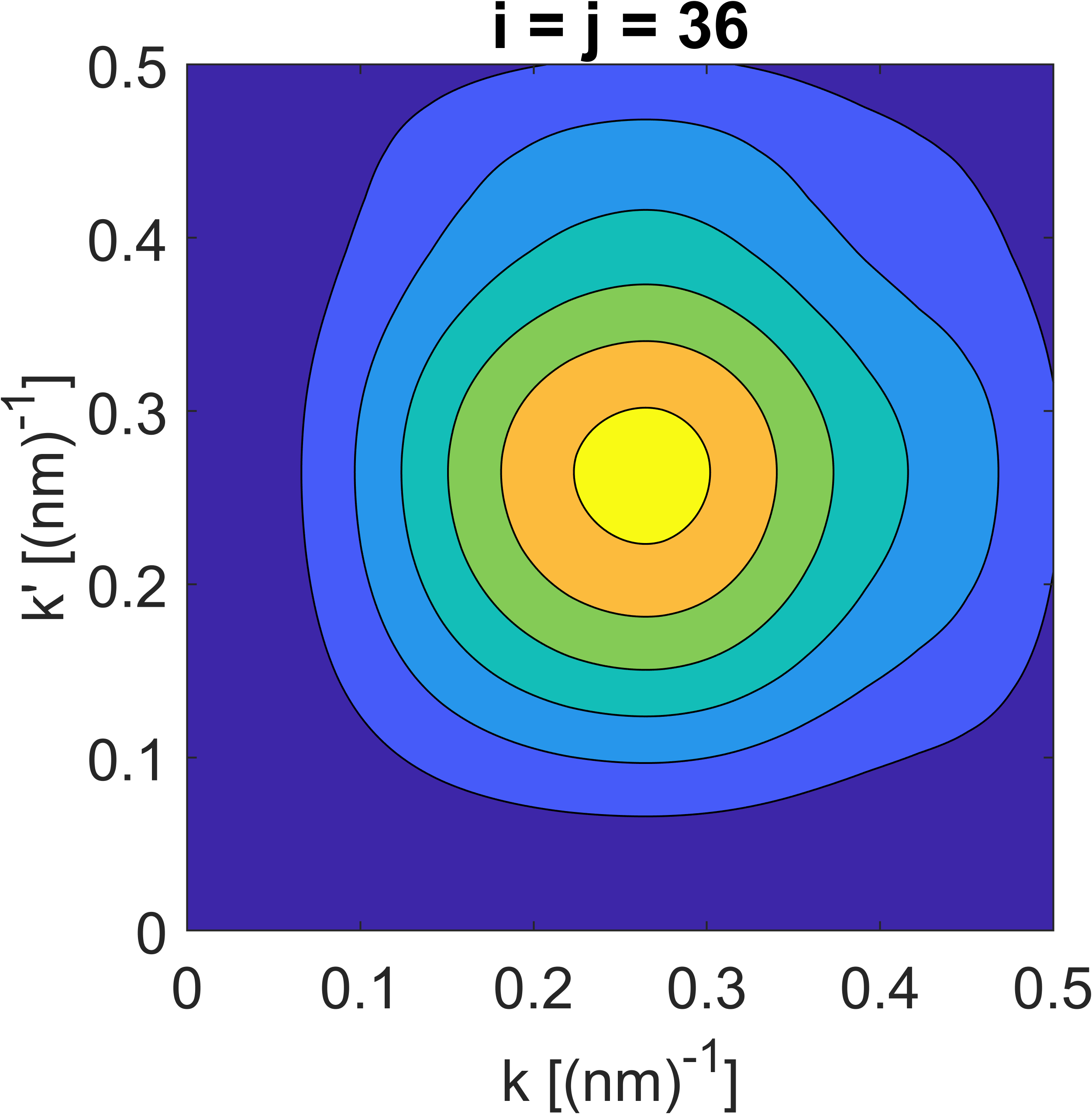}
\end{subfigure}
\newline
\newline
\begin{subfigure}{.235\textwidth}
  \includegraphics[width=\linewidth]{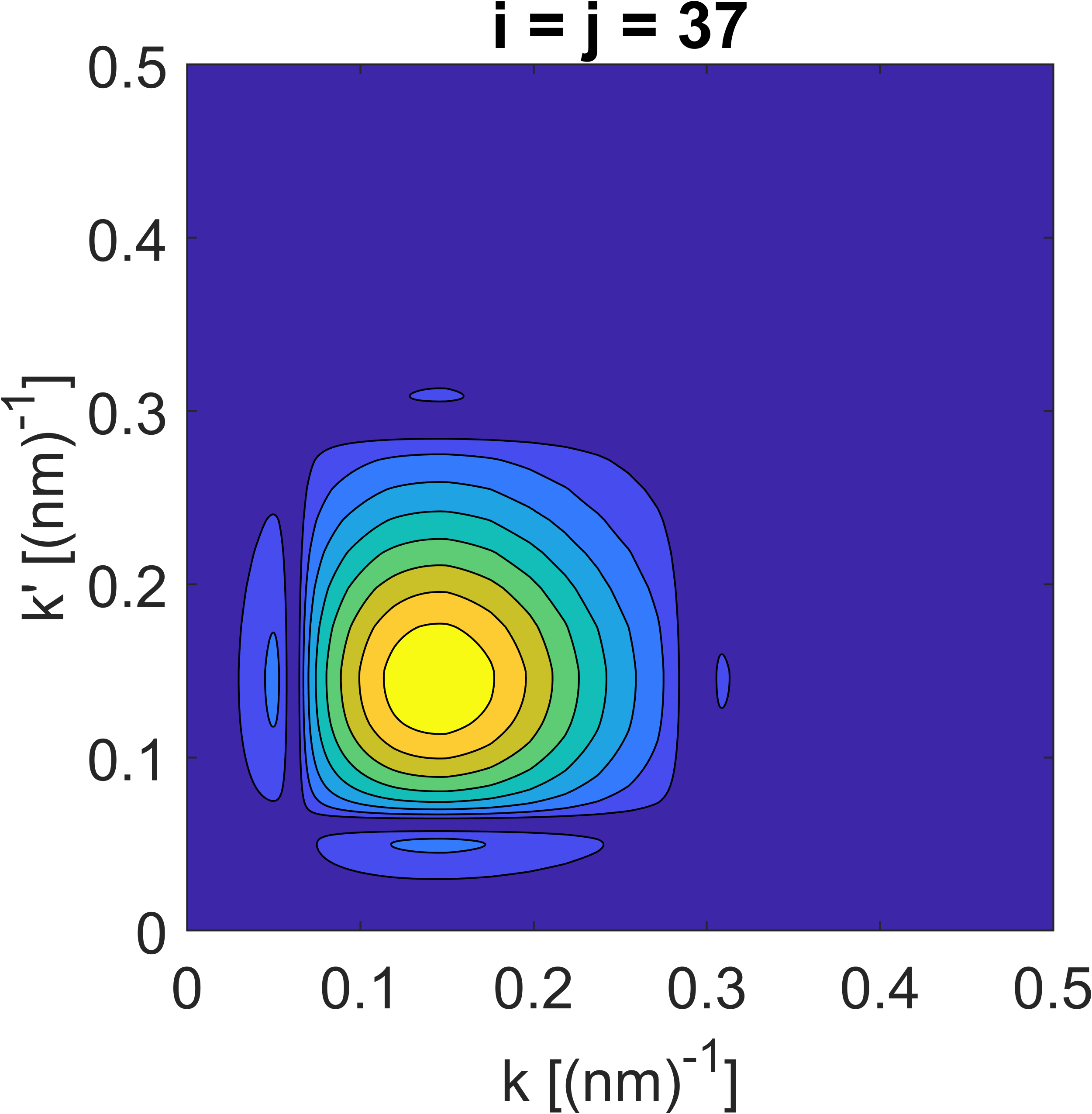}
\end{subfigure}%
\begin{subfigure}{.235\textwidth}
  \includegraphics[width=\linewidth]{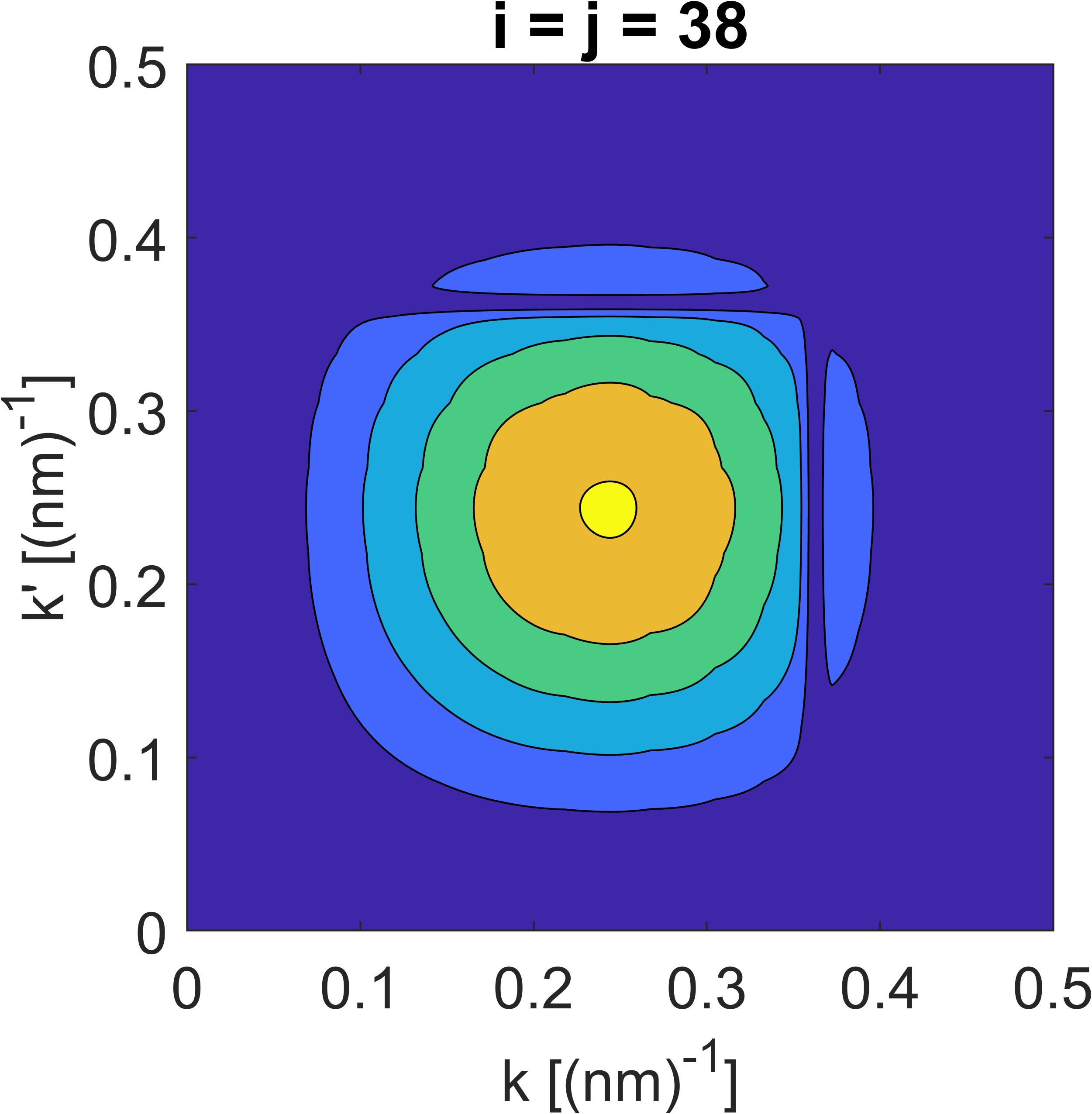}
\end{subfigure}
\begin{subfigure}{.235\textwidth}
  \includegraphics[width=\linewidth]{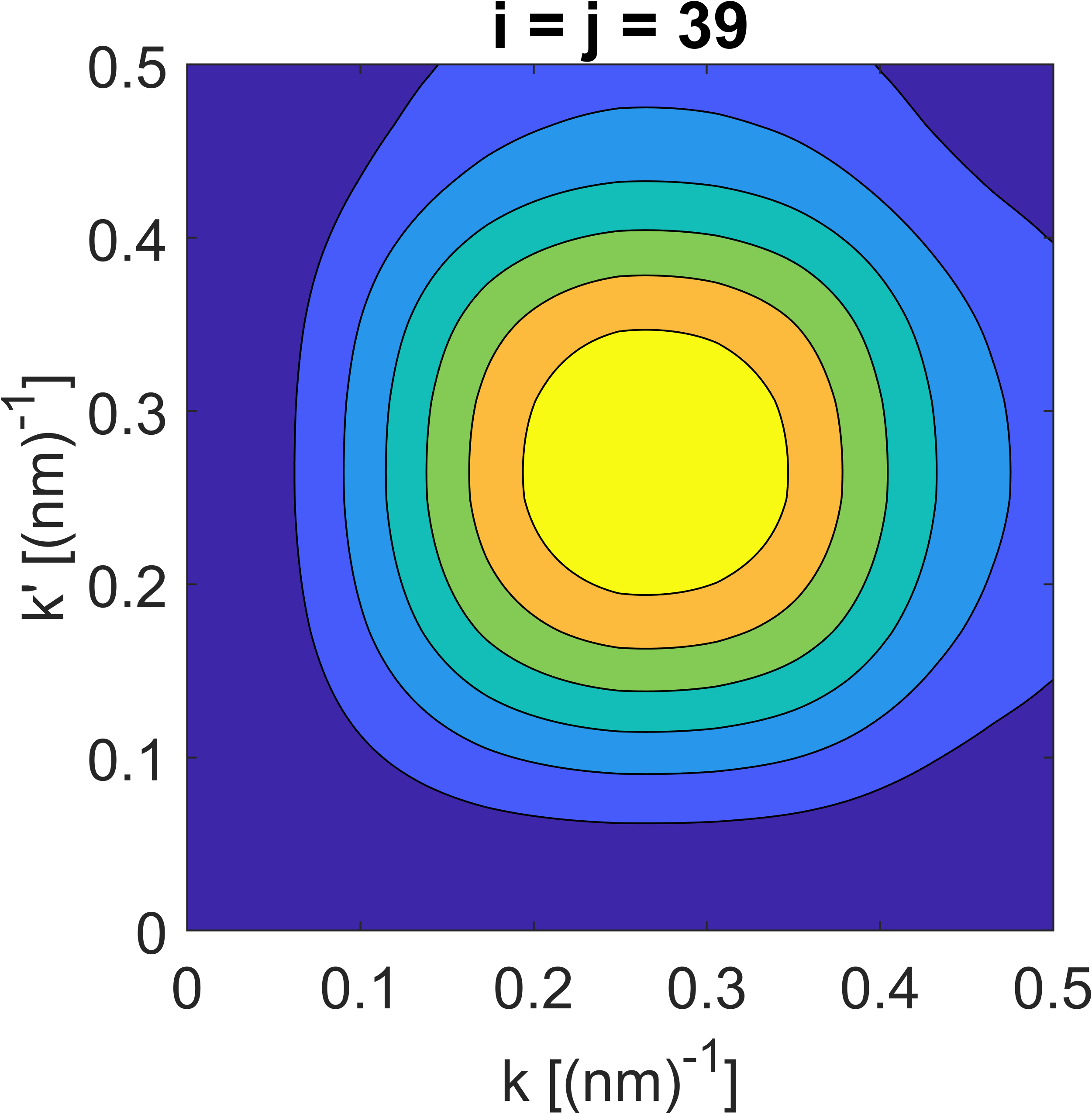}
\end{subfigure}
\begin{subfigure}{.235\textwidth}
  \includegraphics[width=\linewidth]{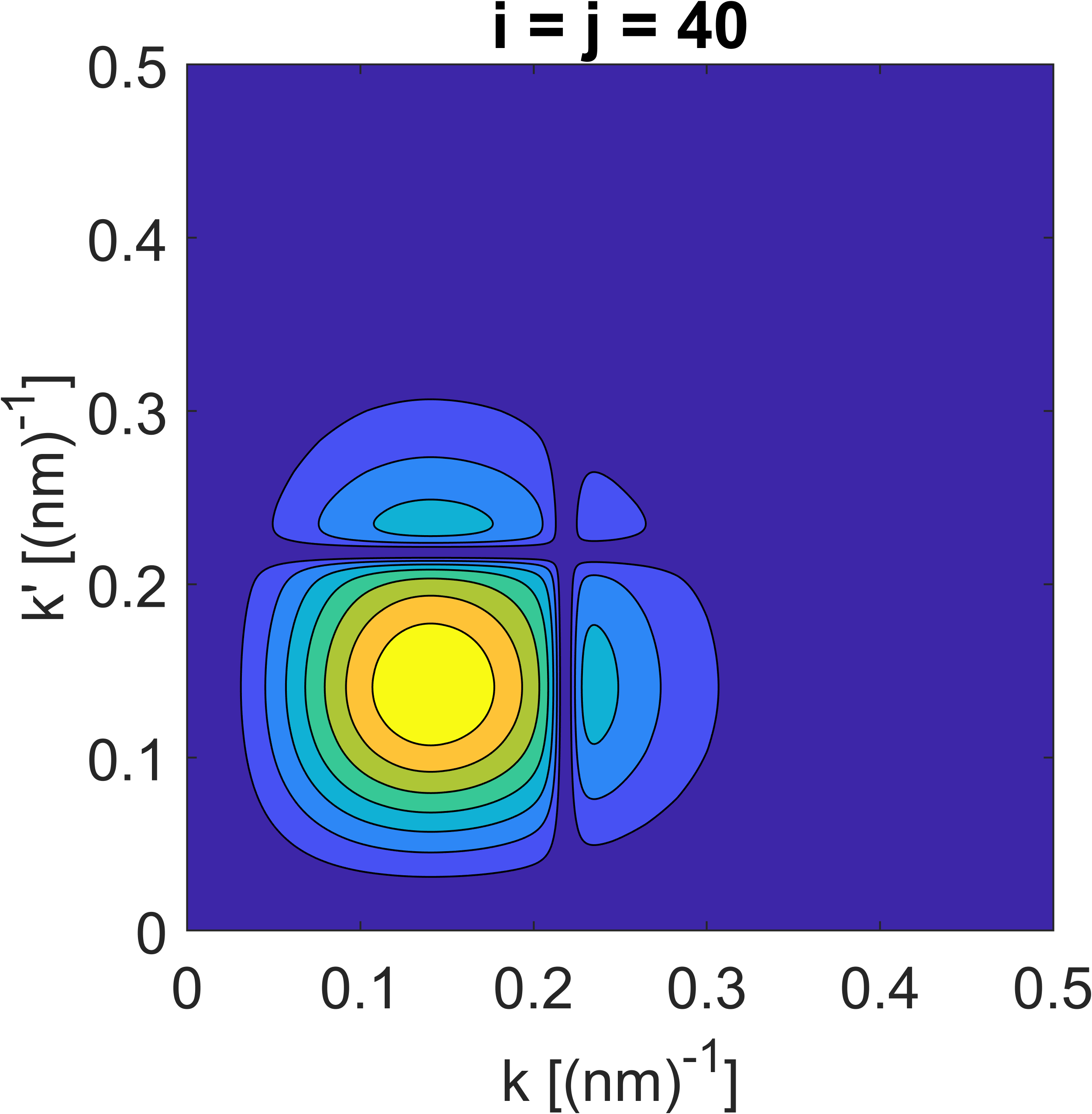}
\end{subfigure}
\caption{Calculation of the quadratic exciton-phonon coupling term $|f_{i,j,\mathbf{k,k'}}|^{2}$ for vibration modes 21 to 40, with $i = j$. Here, the nanowire radius $R=25\ \mathrm{nm}$. The plots above maintain the style introduced in the previous page.} 
\label{fig:coolplots_25nm_1_next}
\end{figure}

\newpage

\begin{figure}[htbp] 
  \includegraphics[width=0.5\linewidth]{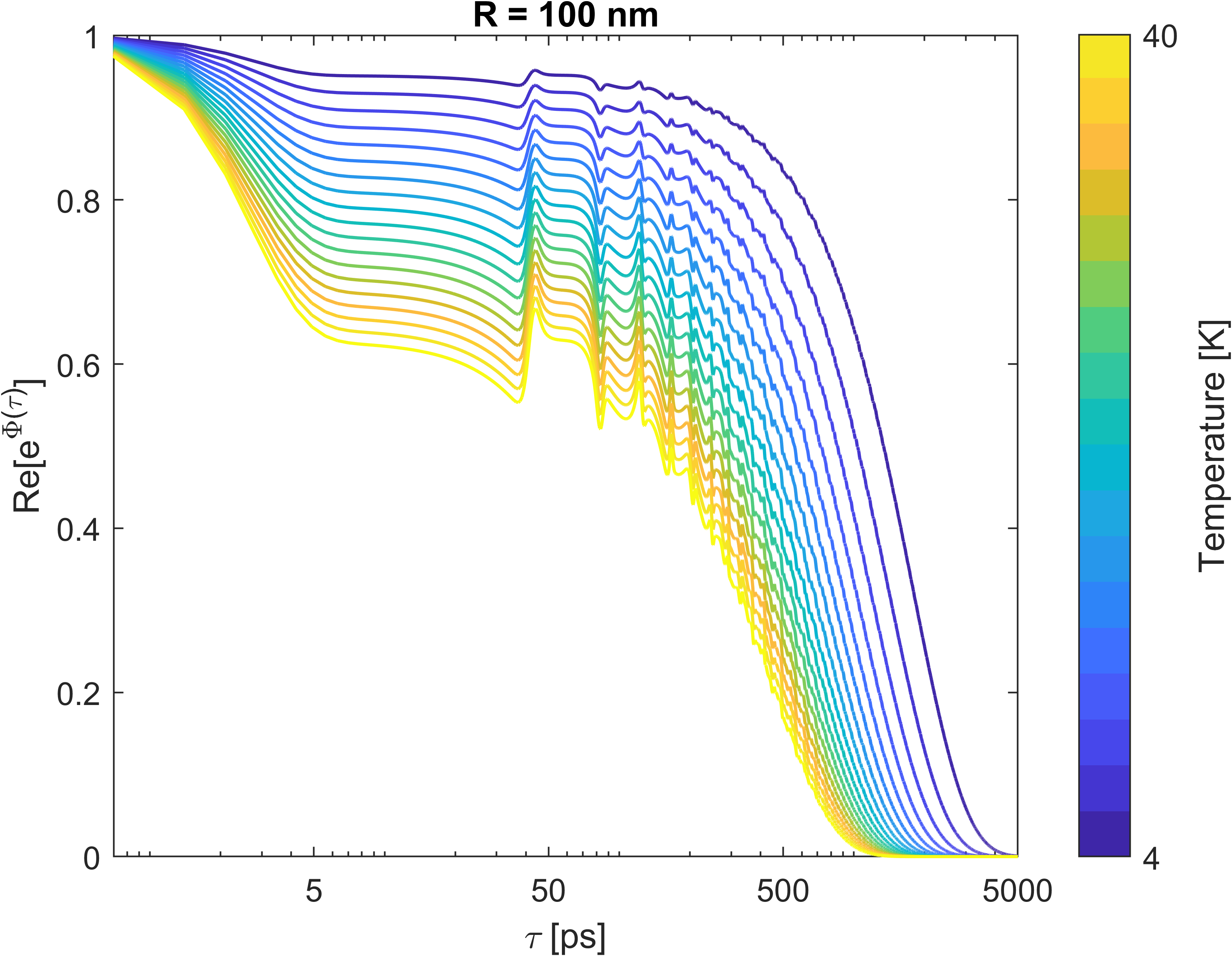}
\caption{Time evolution of the QD coherence for a nanowire radius $R=100\ \mathrm{nm}$ evaluated at different temperatures $T$ $\in$ [\SI{4}{\kelvin}, \SI{40}{\kelvin}] (swept in steps of \SI{2}{\kelvin}).}
\label{fig:coherences_100nm}
\end{figure}

\end{document}